\pdfoutput=1

\documentclass[12pt, reqno]{article}
\usepackage{epsfig}
\usepackage{amssymb}
\usepackage{amsmath}
\usepackage{mathrsfs}

\allowdisplaybreaks[4] 
\begin{document}

\begin{titlepage}

\indent\indent
\begin{flushright}
LU TP 09-33
\end{flushright}
\vspace*{0.2cm}

\begin{center}
{\Large \bf Renormalization and additional degrees of freedom
within\\[0.2cm]
the chiral effective theory for spin-1 resonances}\\[2 cm]

{\bf Karol Kampf$^{\,1,2}$, Jiri Novotny$^{\,2}$ and Jaroslav
Trnka$^{\,2,3}$}
\\[1.2 cm]
$\ ^{1}${\it Department of Theoretical Physics, Lund University, S\"olvegatan 14A, SE 223-62 Lund, Sweden.}\\[0.4cm]
$\ ^{2}${\it Institute of Particle and Nuclear Physics, Faculty of
Mathematics and Physics,\\ Charles University in Prague, V
Hole\v{s}ovi\v{c}k\'ach 2, 18000
Prague, Czech Republic.}\\[0.4cm]
$\ ^{3}${\it Department of Physics, Princeton University, 08540
Princeton, NJ, USA.}
\end{center}

\vspace*{1.0cm}

\begin{abstract}

We study in detail various aspects of the renormalization of the
spin-1 resonance propagator in the effective field theory framework.
First, we briefly review the formalisms for the description of
spin-1 resonances in the path integral formulation with the stress
on the issue of propagating degrees of freedom. Then we calculate
the one-loop $1^{--}$ meson self-energy within the Resonance chiral
theory in the chiral limit using different methods for the
description of spin-one particles, namely the Proca field,
antisymmetric tensor field and the first order formalisms. We
discuss in detail technical aspects of the renormalization procedure
which are inherent to the power-counting non-renormalizable theory
and give a formal prescription for the organization of both the
counterterms and one-particle irreducible graphs. We also construct
the corresponding propagators and investigate their properties. We
show that the additional poles corresponding to the additional
one-particle states are generated by loop corrections, some of which
are negative norm ghosts or tachyons. We count the number of such
additional poles and briefly discuss their physical meaning.

\end{abstract}

\end{titlepage}

\setcounter{footnote}{0} \newpage


\tableofcontents

\newpage

\section{Introduction}

As is well known, in the low energy region the dynamical degrees of
freedom of QCD are not quarks and gluons but the low lying hadronic
states and, as a consequence, a non-perturbative description of the
their dynamics is inevitable. An approach using effective
Lagrangians appears to be very efficient for this purpose and it has
made a considerable progress recently. In the very low energy region
($E\ll \Lambda _H\sim 1{\rm GeV}$), the octet of the lightest
pseudoscalar mesons ($\pi $, $K$, $\eta $) represents the
only relevant part of the QCD spectrum. The Chiral Perturbation Theory ($%
\chi $PT) \cite{Weinberg,Gasser1,Gasser2} based on the spontaneously broken
chiral symmetry $SU(3)_L\times SU(3)_R$ grew into a very successful
model-independent tool for the description of the Green functions (GF) of
the quark currents and related low-energy phenomenology. The pseudoscalar
octet is treated as the octet of pseudo-Goldstone bosons (PGB) and $\chi $PT
is organized according to the Weinberg power-counting formula \cite{Weinberg}
as a rigorously defined simultaneous perturbative expansion in small momenta
and the light quark masses. Recently, the calculations are performed at the
next-to-next-to-leading order $O(p^6)$ (for a comprehensive review and
further references see \cite{Bijnens:2006zp}).

In the intermediate energy region ($\Lambda _H\leq E<2{\rm GeV}$),
where the set of relevant degrees of freedom includes also the low
lying resonances, the situation is less satisfactory. This region is
not separated by a mass gap from the rest of the spectrum and, as a
consequence, there is no appropriate scale playing the role
analogous to that of $\Lambda _H$ in $\chi PT$. Therefore, the
effective theory in this region cannot be constructed as a
straightforward extension of the $\chi $PT low energy expansion by
means of introducing resonances \emph{e.g.} as homogenously (but
nonlinearly)
transformed matter fields in the sense of \cite{Coleman:1969sm}, \cite%
{Callan:1969sn} and pushing the scale $\Lambda _H$ to $2 {\rm
GeV}$.

In order to introduce another type of effective Lagrangian
description, the considerations based on the large $N_C$ expansion
together with the high-energy constraints derived from
perturbative QCD and OPE appear to be
particularly useful. In the limit $N_C\rightarrow \infty $%
, the chiral symmetry is enlarged to $U(3)_L\times U(3)_R$ and the
spectrum relevant for the correlators of the quark bilinears
consists of an infinite tower of free stable mesonic resonaces
exchanged in each channel and classified according to the symmetry
group $U(3)_V$. An appropriate description should therefore require
an infinite number of resonance fields entering the $U(3)_L\times
U(3)_R$ symmetric effective Lagrangian. Because the quasi-classical
expansion is correlated with the large $N_C$ expansion, the
interaction vertices are suppressed by an appropriate power of
$\,N_C^{-1/2}$ according to the number of the meson legs. At the
leading order only the tree graphs have to be taken into account .
An approximation to this general picture where we limit the number
of the resonance fields to one in each channel and matching the
resulting theory in the high energy region with OPE is known as the
Resonance Chiral Theory (R$\chi $T) (it was introduced in seminal
papers \cite{Ecker1,Ecker2}). Integrating out the resonance fields
from the Lagrangian of R$\chi $T in the low energy region and the
subsequent matching with $\chi $PT has become very successful tool
for the
estimates of the resonance contribution to the values of the $O(p^4)$ \cite%
{Ecker1} and $O(p^6)$ \cite{Cirigliano:2006hb,Kampf:2006bn} low energy
constants (LEC) entering the $\chi $PT Lagrangian. Therefore, studying R$%
\chi $T can help us to understand not only the dynamics of resonances but
also the origin of LECs in $\chi $PT.

However, even when restricting to the case of the matter field formalism, it
is known from the very beginning \cite{Ecker2} that the form of the R$\chi $%
T Lagrangian is not determined uniquely. The reason is that the
resonances with a given spin can be described in many ways using
fields with different Lorentz structure. For example, for the
spin-one resonances one can use \emph{\ i.a.} the Proca vector field
or the antisymmetric tensor field or both (within the first order
formalism \cite{Bruns:2004tj,FO1}). Though the theories based on
different types of fields with Lagrangians which contain only finite
number of operators are not strictly equivalent already on the tree
level (in general, it is necessary to include nonlocal interaction
or infinite number of operators and contact terms to ensure the
complete equivalence, see \cite{FO1}), we can always ensure a weak
equivalence of all three formalisms up to a given fixed chiral order
(this was established to $O(p^4)$ in \cite{Ecker2} and enlarged to
$O(p^6)$ in \cite{FO1}).

As we have mentioned above, the lack of the mass gap (which could
provide us with a scale playing the role analogous to $\Lambda
_{H}$) prevents us from using a straightforward extension of the
Weinberg power-counting formula \cite{Weinberg} taking the resonance
masses and momenta of the order $O(p)$ on the same footing as for
PGB. Also the usual chiral power counting which takes the resonance
masses as an additional heavy scale (which is counted as $O(1)$)
fails within the R$\chi $T in a way analogous to the $\chi PT$ with
baryons \cite{Gasser:1987rb}. Nevertheless, it seems to be fully
legitimate to go beyond the tree level R$\chi $T and calculate the
loops \cite{vecform,SS,SS2,VV,SanzCillero:2009pt,
Rosell:2009yb,SanzCillero:2007ib, SanzCillero:2009ap,
Rosell:2005ai}.

Being suppressed by one power of $1/N_{C}$, the loops allow to
encompass such NLO effects in the $1/N_{C}$ expansion as resonance
widths, resonance cuts and the final state interaction and (by means
matching with $\chi $PT) to determine the NLO resonance contribution
to LEC (and their running with renormalization scale).

However, we can expect both technical and conceptual complications
connected with the renormalization of the effective theory for which
no natural organization of the expansion (other than the $1/N_C$
counting) exists. Especially, because there is no natural analog of
the Weinberg power counting in R$\chi $T, we can expect mixing of
the naive chiral orders in the process of the renormalization
(\emph{e.g} the loops renormalize the $O(p^2)$ LEC and also
counterterms of unusually high chiral orders are needed). Also a
straightforward construction of the propagator from the self-energy
using the Dyson re-summation can bring about the appearance of new
poles in the GF. Because the spin-one particles are described using
fields transforming under the reducible representation of the
rotation group and due to the lack of an appropriate protective
symmetry, some of these additional poles can correspond to new
degrees of freedom, which are frozen at the tree level. The latter
might be felt as a pathological artefact of the not carefully enough
formulated theory, particularly because these extra poles might be
negative norm ghosts or tachyons \cite{Slovak}. On the other hand,
however, we could also try to take an advantage of this feature and
to adjust the poles in such a way that they correspond to the well
established resonance states \cite{Kampf:2008xp}.

Let us note, that similar problems are generic for the description
of the higher spin particles in terms of quantum field theory. As an
example we can mention e.g. the problem with the renormalization of
quantum gravity which is trying to be cured by imposing additional
symmetry or by introducing a non-perturbative quantization believing
that UV divergences are only artefact of a perturbative theory. In
the context of the extensions of the $\chi PT$, this has been
studied in connection with introducing of the spin-$3/2$
isospin-$3/2~\ \Delta (1232)$ resonance in the baryonic sector (for
a review see \cite{Pascalutsa:2006up} and references therein). The
Rarita-Schwinger field commonly used for its description contains
along with the spin-$3/2$ sector also spin-$1/2$ sector, which is
frozen at the tree level due to the form of the free equations of
motion. These provides the necessary constraints reducing the
number of propagating spin degrees of freedom to four corresponding to spin $%
3/2$ particles. However, these constraints are generally not present
in the interacting theory and negative norm ghost
\cite{Johnson:1960vt}
 and/or tachyonic \cite{Velo:1969bt}
 poles might
appear beyond the tree level. The appearance of these extra
unphysical degrees of freedom can be avoided by means of the
requirement of additional protective gauge symmetry under which the
interaction Lagrangian has to be invariant. Such a symmetry, which
is also a symmetry of the kinetic term (but not of the mass term),
is an analog of the $U(1)$ gauge symmetry of the electromagnetic
field and its role is also similar. As it has been shown by means of
path integral formalism, it leads to the same constraints as in the
noninteracting theory and prevent therefore the extra spin-$1/2$
states from propagating.

On the other hand, it has been proved, \ that the most  general
interaction Lagrangian at most bilinear in  Rarita-Schwinger field \
(\emph{i.e.} without the protective gauge symmetry) is on shell
equivalent to the gauge invariant one \cite{Krebs:2008zb} .  The
latter is, however, nonlocal (or equivalently it contains an
infinite number of terms). Also the above protective gauge symmetry
is, as a rule, in a conflict with chiral symmetry, and has therefore
to be implemented  with a care. Though there are efficient methods
how to handle this obstacles in concrete loop calculations
\cite{Pascalutsa:2006up}, \cite{Krebs:2008zb}, the problem still has
not been solved completely.

In the following, we would like to discuss these problems in more
detail. As an explicit example we use the one-loop renormalization
of the propagator corresponding to the fields which originally
describe $1^{--}$ vector resonance ($\rho $ meson) at the tree level
within the Proca field, the antisymmetric tensor field and within
the first order formalism in the chiral limit. The situation here is
quite similar to the case of spin-3/2 resonances discussed above. In
addition, to the spin-1 degrees of freedom, there are extra sectors
that are frozen at the tree level. There exists a protective gauge
symmetry which prevents these modes from propagation. The kinetic
term is invariant with respect to this symmetry while the mass term
is not.

By means of an explicit calculation we will show that (unlike the
ordinary $\chi PT$) the one-loop corrections to the self-energy need
counterterms with a number of derivatives ranging from zero up to
six and also that a new kinetic counterterm with two derivatives
(which was not present in the tree level Lagrangian) is necessary.
We will also demonstrate that the corresponding propagator obtained
by means of Dyson re-summation of the one-particle irreducible
self-energy insertions has unavoidably additional poles. Due to the
unusual higher order growth of the self-energy in the UV region some
of them are inevitably pathological (with
 a negative norm or a negative mass squared). Though these
additional poles are decoupled in the limit $N_C\rightarrow \infty
$, \ for reasonable concrete values of the parameters of the
Lagrangian they might appear near or even inside the region for
which $R\chi T$ was originally designed. We also discuss briefly
within the antisymmetric tensor formalism a possible interpretation
of some of the non-pathological poles as a manifestation of the
dynamical generation of various types of additional 1+- states. We
will also show that the appropriate adjustment of coupling constants
in the antisymmetric tensor case allows us (at least in principle)
to generate in this way the one which could be identified
\emph{e.g.} with the $b_1(1235)$ meson \cite{Kampf:2008xp}. Such a
mechanism is analogous to the model \cite{Boglione:2002vv} for the
dynamical generation of the scalar resonances from the bare
quark-antiquark ''seed'', the propagator of which develops (after
dressing with pseudoscalar meson loops) additional poles
identified \emph{e.g.} as $a_0(980)$ (cf. also \cite{Tornqvist:1995kr},\cite%
{Tornqvist:1995ay}).

The paper organized as follows. In Section \ref{Propagators and
poles} we remind the basic facts about the propagators and briefly
discuss the issue of the additional degrees of freedom in all three
formalisms for the description of spin-one resonances. We use the
path integral formulation where the protective symmetry analogous to
the Rarita-Schwinger case is manifest. In Section
\ref{Section_power_counting} we discuss the power counting. We try
to formulate here a formal self-consistent organization of the
counterterms and one-particle irreducible graphs, which sorts the
operators in the Lagrangian according to the number of derivatives
as well as number of the resonance fields and which is useful for
the proof of renormalizability of the $R\chi T$ as an effective
theory. In Section \ref{ch4} we present the results of the explicit
calculation of the self-energies. Then we give a list of
counterterms and briefly discuss the renormalization prescription.
Section \ref{Section_propagators} is devoted to the construction of
the propagators and to the discussion of their poles. Because the
basic ideas are similar
within all three formalisms, we concentrate here on the antisymmetric tensor case. Section %
\ref{Section_summary} contains summary and conclusions. Some of the long
formulae are postponed to the appendices: the explicit form of the
renormalization scale independent parameters of the self-energies are
collected in Appendix \ref{Appendix_alpha_beta}, namely for the Proca field
in \ref{appendix Proca}, for the antisymmetric tensor field in \ref{appendix
tensor} and for the first order formalism in \ref{appendix first order}. In
Appendix \ref{Appendix_positivity} we give a proof of the positivity of the
spectral functions for the antisymmetric tensor propagator.

\section{Propagators and poles\label{Propagators and poles}}

In this section, we collect the basic properties of the propagators and the
corresponding self-energies within the Proca field, the antisymmetric tensor
field and the first order formalisms. The discussion will be as general as
possible without explicit references to $R\chi T$, which can be assumed as
the special example of the general case.

\subsection{Proca formalism}

\subsubsection{General properties of the propagator}

We start our discussion with a standard textbook example of the
interacting Proca field. Let us write the Lagrangian in the form
\begin{equation}
\mathcal{L}=\mathcal{L}_{0}+\mathcal{L}_{int},
\end{equation}
where the free part of the Lagrangian is
\begin{equation}
\mathcal{L}_{0}=-\frac{1}{4}\widehat{V}_{\mu \nu }\widehat{V}^{\mu \nu }+%
\frac{1}{2}M^{2}V_{\mu }V^{\mu }
\end{equation}
with
\begin{equation}
\widehat{V}_{\mu \nu }=\partial _{\mu }V_{\nu }-\partial _{\nu }V_{\mu }.
\end{equation}
Without any additional assumptions on the form and symmetries of
the interaction part of the Lagrangian $\mathcal{L}_{int}$, we can
expect the following general structure of the full two-point
one-particle irreducible (1PI) Green function
\begin{equation}
\Gamma _{\mu \nu }^{(2)}(p)=(M^{2}-p^{2}+\Sigma ^{T}(p^{2}))P_{\mu \nu
}^{T}+(M^{2}+\Sigma ^{L}(p^{2}))P_{\mu \nu }^{L}.  \label{vector_gamma_2}
\end{equation}
Here
\begin{eqnarray}
P_{\mu \nu }^{L} &=&\frac{p_{\mu }p_{\nu }}{p^{2}}  \label{P_L} \\
P_{\mu \nu }^{T} &=&g_{\mu \nu }-\frac{p_{\mu }p_{\nu }}{p^{2}}  \label{P_T}
\end{eqnarray}
are the usual longitudinal and transverse projectors and $\Sigma ^{T,L}$ are
the corresponding transverse and longitudinal self-energies, which vanish in
the free field limit. Inverting (\ref{vector_gamma_2}) we get for the full
propagator
\begin{equation}
\Delta _{\mu \nu }(p)=-\frac{1}{p^{2}-M^{2}-\Sigma ^{T}(p^{2})}P_{\mu \nu
}^{T}+\frac{1}{M^{2}+\Sigma ^{L}(p^{2})}P_{\mu \nu }^{L}.
\end{equation}
The possible (generally complex) poles of such a propagator are of two
types; either at $p^{2}=s_{V}$, where $s_{V}$ is given by the solutions of
\begin{equation}
s_{V}-M^{2}-\Sigma ^{T}(s_{V})=0,  \label{V_pole_spin_1}
\end{equation}
or at $p^{2}=s_{S}$ where $s_{S}$ is the solution of
\begin{equation}
M^{2}+\Sigma ^{L}(s_{S})=0.  \label{V_pole_spin_0}
\end{equation}

Let us first discuss the poles of the first type. Assuming that (\ref%
{V_pole_spin_1}) is satisfied for $s_V=M_V^2>0$, then for $p^2\rightarrow $ $%
M_V^2$
\begin{eqnarray}
\Delta _{\mu \nu }(p) &=&\frac{Z_V}{p^2-M_V^2}\left( -g_{\mu \nu }+\frac{%
p_\mu p_\nu }{M^2}\right) +O(1)  \notag \\
&=&\frac{Z_V}{p^2-M_V^2}\sum_\lambda \varepsilon _\mu ^{(\lambda
)}(p)\varepsilon _\nu ^{(\lambda )*}(p)+O(1)
\end{eqnarray}
where
\begin{equation}
Z_V=\frac 1{1-\Sigma ^{^{\prime }T}(M_V^2)}
\end{equation}
and where $\varepsilon _\mu ^{(\lambda )}(p)$ are the usual spin-one
polarization vectors. Under the condition $Z_V>0$ the poles of this type
correspond to spin-one one particle states $|p,\lambda ,V\rangle $ which
couple to the Proca field as
\begin{equation}
\langle 0|V_\mu (0)|p,\lambda ,V\rangle =Z_V{}^{1/2}\varepsilon _\mu
^{(\lambda )}(p).
\end{equation}
At least one of these states is expected to be perturbative in the sense
that its mass and coupling to $V_\mu $ can be written as
\begin{eqnarray}
M_V^2 &=&M^2+\delta M_V^2 \\
Z_V &=&1+\delta Z_V,
\end{eqnarray}
where $\delta M_V^2$ and $\delta Z_V$ are small corrections vanishing in the
free field limit. This solution corresponds to the original degree of
freedom described by the free part of the Lagrangian $\mathcal{L}_0$. The
additional one particle states corresponding to the other possible
(non-perturbative) solutions of (\ref{V_pole_spin_1}) decouple in the free
field limit.

The second type of poles is given by (intrinsically nonperturbative)
solutions of (\ref{V_pole_spin_0}). Suppose that this condition is satisfied
by $s_{S}=M_{S}^{2}>0$. For $p^{2}\rightarrow $ $M_{S}^{2}$
\begin{equation}
\Delta _{\mu \nu }(p)=\frac{Z_{S}}{p^{2}-M_{S}^{2}}\frac{p_{\mu }p_{\nu }}{%
M_{S}^{2}}+O(1)
\end{equation}
where
\begin{equation}
Z_{S}=\frac{1}{\Sigma ^{^{\prime }L}(M_{S}^{2})}.
\end{equation}
Assuming $Z_{S}>0$ this pole corresponds to the spin-zero one particle state
$|p,S\rangle $ which couples to $V_{\mu }$ as
\begin{equation}
\langle 0|V_{\mu }(0)|p,S\rangle =\mathrm{i}p_{\mu }\frac{Z_{S}{}^{1/2}}{%
M_{S}}.
\end{equation}
For the free field this scalar mode is frozen and does not propagate
according to the special form of the Proca field Lagrangian. Therefore, in
the limit of vanishing interaction the extra scalar state decouples.

Without any additional assumptions on the symmetries of the interaction
Lagrangian we can therefore expect the appearance of additional dynamically
generated degrees of freedom.

The general picture is, however, more subtle. Note that, the
interpretation of the above additional spin-one and spin-zero
poles as physical one-particle asymptotic states depends on the
proper positive sign of the corresponding residues $Z_V,\,Z_S>0$,
otherwise the norm of these states is negative and the poles
correspond to the negative norm ghosts. Similarly, also poles with
$M_{V,S}^2<0$ can be generated, which correspond to the tachyonic
states. Let us illustrate this feature using a toy example.
Suppose, that the only interaction terms are of the form
\begin{equation}
\mathcal{L}_{int}\equiv \mathcal{L}_{ct}=-\frac \alpha 4\widehat{V}_{\mu \nu
}\widehat{V}^{\mu \nu }-\frac \beta 2(\partial _\mu V^\mu )^2+\frac \gamma
{2M^2}(\partial _\mu \widehat{V}^{\mu \nu })(\partial ^\rho \widehat{V}%
_{\rho \nu })+\frac \delta {2M^2}(\partial _\mu \partial _\rho V^\rho
)(\partial ^\mu \partial _\sigma V^\sigma ).  \label{L_V_toy}
\end{equation}
Such a Lagrangian can be typically produced by radiative corrections in an
effective field theory with Proca field, which does not couple to other
fields in a $U(1)$ gauge invariant way, and can provide us with counterterms
necessary to renormalize the loops contributing to the $V$ field
self-energy. $\mathcal{L}_{ct}$ gives rise to the following contributions to
$\Sigma ^T(p^2)$ and $\Sigma ^L(p^2)$
\begin{eqnarray}
\Sigma ^T(p^2) &=&-\alpha p^2+\gamma \frac{p^4}{M^2}  \label{Sigma_T_vector}
\\
\Sigma ^L(p^2) &=&-\beta p^2+\delta \frac{p^4}{M^2}.  \label{Sigma_L_vector}
\end{eqnarray}
As a result, we have two spin-one and two spin-zero one-particle states. The
masses and residue of the spin-one states are then
\begin{eqnarray}
M_{V\pm }^2 &=&M^2\left( 1+\frac{1+\alpha -2\gamma \mp \sqrt{(1+\alpha
)^2-4\gamma }}{2\gamma }\right)  \label{MV} \\
1-\Sigma ^{^{\prime }T}(M_{V\pm }^2) &=&\pm \sqrt{(1+\alpha )^2-4\gamma },
\label{ResV}
\end{eqnarray}
which are real for for $(1+\alpha )^2-4\gamma >0$. In the limit
$\alpha ,\,\gamma \rightarrow 0$, $\alpha /\gamma =const$ we get
either the perturbative solution with mass $M_{V+}$ or (for
$\gamma >0$) an additional spin-one ghost with mass $M_{V-}$ (for
$1+\alpha >0$ and $\gamma <0$ this pole is tachyonic). Similarly
for the spin-zero states
\begin{eqnarray}
M_{S\pm }^2 &=&M^2\left( \frac{\beta \mp \sqrt{\beta ^2-4\delta }}{2\delta }%
\right)  \label{MS} \\
\Sigma ^{^{\prime }L}(M_{S\pm }^2) &=&\mp \sqrt{\beta ^2-4\delta }.
\label{ResS}
\end{eqnarray}
The poles are real for $\beta ^2>4\delta $ and \emph{e.g.} for
$\beta ,\delta >0$ one of the poles is spin-zero ghost. In both
cases for appropriate values of the parameters we can get also two
tachyons or even the complex Lee-Wick pair of ghosts. These features
are of course well known in the connection with the higher
derivative regularization (as well as with the properties of the
gauge-fixing term).

\subsubsection{ Additional degrees of freedom in the path integral formalism$
\label{path integral vector}$}

The additional degrees of freedom discussed in the previous subsection can
be made manifest in the path integral formalism. Let us start with the
generating functional for the interacting Proca field
\begin{equation}
Z[J]=\int \mathcal{D}V\exp \left( \mathrm{i}\int \mathrm{d}^4x\left( -\frac
14\widehat{V}_{\mu \nu }\widehat{V}^{\mu \nu }+\frac 12M^2V_\mu V^\mu +%
\mathcal{L}_{int}(V,J,\ldots )\right) \right) ,
\end{equation}
where the external sources are denoted collectively by $J$. In order to
separate the transverse and longitudinal degrees of freedom of the field $%
V_\mu $ within the path integral we can use the standard Faddeev-Popov trick
with respect to the $U(1)$ gauge transformation of the field $V_\mu $%
\begin{equation}
V_\mu \rightarrow V_\mu +\partial _\mu \Lambda .  \label{V_gauge}
\end{equation}
As a result, we get the generating functional in the form
\begin{equation}
Z[J]=\int \mathcal{D}V_{\perp }\mathcal{D}\Lambda \exp \left( \mathrm{i}\int
\mathrm{d}^4x\left( \frac 12V_{\perp }^\mu \square V_{\perp \mu }+\frac
12M^2V_{\perp }^\mu V_{\perp \mu }+\frac 12M^2\partial _\mu \Lambda \partial
^\mu \Lambda +\mathcal{L}_{int}(V_{\perp }-\partial \Lambda ,J,\ldots
)\right) \right) .  \label{FP_trick}
\end{equation}
Here $\mathcal{D}V_{\perp }=\mathcal{D}V\delta (\partial _\mu V^\mu )$ and
\begin{equation*}
V_{\perp }^\mu =\left( g^{\mu \nu }-\frac{\partial ^\mu \partial ^\nu }{%
\square }\right) V_\nu
\end{equation*}
is the transverse part of the vector field $V^\mu $, the longitudinal part
of which corresponds to the scalar field $\Lambda $,\emph{\ i.e.}
\begin{equation}
V^\mu =V_{\perp }^\mu +\partial ^\mu \Lambda .  \label{V_components}
\end{equation}
The free propagators of the fields $V_{\perp }^\mu $ and $\Lambda $ are
\begin{eqnarray}
\Delta _{\perp }^{\mu \nu }(p) &=&-\frac{P^{T\,\mu \nu }}{p^2-M^2}
\label{propVtrans} \\
\Delta _\Lambda (p) &=&\frac 1{M^2}\frac 1{p^2}.  \label{propLambda}
\end{eqnarray}
Both these propagators have spurious poles at $p^2=0$, however, the only
necessary combination which matters in the Feynman graphs is
\begin{equation}
\Delta _0^{\mu \nu }(p)=\Delta _{\perp }^{\mu \nu }(p)+p^\mu p^\nu \Delta
_\Lambda (p),
\end{equation}
which coincides with the original free propagator of the field $V^\mu $ and
the spurious poles cancel each other.

Note that, provided the interaction Lagrangian $\mathcal{L}_{int}$ is
symmetric under the $U(1)$ gauge transformation (\ref{V_gauge}), the
spin-zero field $\Lambda $ completely decouples and can be integrated out .
The theory can then be formulated solely in terms of the field $V_{\perp
}^{\mu }$. The $U(1)$ invariant form of the interaction allows to simplify
the propagator $\Delta _{\perp }^{\mu \nu }(p)$%
\begin{equation}
\Delta _{\perp }^{\mu \nu }(p)\rightarrow -\frac{g_{\mu \nu }}{p^{2}-M^{2}}
\end{equation}
within the Feynman graphs and the spurious pole $p^{2}=0$ in (\ref%
{propVtrans}) becomes harmless. In this case, the scalar one-particle states
cannot be dynamically generated. On the other hand, in the case when $%
\mathcal{L}_{int}$ is not invariant with respect to (\ref{V_gauge}),
we cannot forget the longitudinal component of $V^{\mu }$ which has
now nontrivial interactions and, as a result, contributions to
$\Sigma ^{L}$ can be generated.

Let us now return to the illustrative example discussed in the previous
subsection. Suppose that the interaction Lagrangian has the form
\begin{equation}
\mathcal{L}_{int}=\mathcal{L}_{ct}+\mathcal{L}_{int}^{^{\prime }}
\end{equation}
where $\mathcal{L}_{ct}$ is the toy interaction Lagrangian (\ref{L_V_toy})
and we assume $\alpha >-1\,$and $\delta >0$ in what follows. Then it is
possible to transform $Z[J]$ to the form of the path integral with all the
additional degrees of freedom represented explicitly in the Lagrangian and
the integration measure. In terms of the transverse and longitudinal degrees
of freedom we get
\begin{eqnarray}
\mathcal{L}_{int}(V_{\perp }-\partial \Lambda ,J,\ldots ) &=&\mathcal{L}%
_{ct}(V_{\perp }-\partial \Lambda ,J,\ldots )+\mathcal{L}_{int}^{^{\prime
}}(V_{\perp }-\partial \Lambda ,J,\ldots )  \notag \\
&=&\frac \alpha 2V_{\perp }^\mu \Box V_{\perp \mu }-\frac \beta 2(\Box
\Lambda )^2+\frac \gamma {2M^2}(\Box V_{\perp }^\mu )(\Box V_{\perp \mu
})+\frac \delta {2M^2}(\partial _\mu \square \Lambda )(\partial ^\mu \Box
\Lambda )  \notag \\
&&+\mathcal{L}_{int}^{^{\prime }}(V_{\perp }-\partial \Lambda ,J\ldots ).
\end{eqnarray}
In order to lower the number of derivatives in the kinetic terms we
integrate in auxiliary scalar fields $\chi $, $\rho $, $\pi $, $\sigma $ and
auxiliary transverse vector field $B_{\perp \mu }$ writing \emph{e.g.}
\begin{equation}
\exp \left( -i\int \mathrm{d}^4x\frac \beta 2(\Box \Lambda )^2\right) =\int
\mathcal{D}\chi \exp \left( i\int \mathrm{d}^4x\left( \frac 1{2\beta }\chi
^2-\partial _\mu \chi \partial ^\mu \Lambda \right) \right)
\end{equation}
and similarly for other higher derivative terms. After the superfluous
degrees of freedom are identified and integrated out, the fields are
re-scaled and then the resulting mass matrix can be diagonalized by means of
\ two symplectic rotations with angles $\theta _V$ and $\theta _S$ (the
technical details are postponed to the Appendix \ref{PI_Proca}). Finally we
get (under the conditions $(1+\alpha )^2>4\gamma $ and $\beta ^2>4\delta $)
\begin{equation}
Z[J]=\int \mathcal{D}V_{\perp }\mathcal{D}B_{\perp }\mathcal{D}\Lambda
\mathcal{D}\chi \mathcal{D}\sigma \exp \left( i\int \mathrm{d}^4x\mathcal{L}%
(V_{\perp },B_{\perp },\Lambda ,\chi ,\sigma ,J,\ldots )\right)
\end{equation}
where
\begin{eqnarray}
\mathcal{L}(V_{\perp },B_{\perp },\Lambda ,\chi ,\sigma ,J,\ldots ) &=&\frac
12V_{\perp }^\mu \square V_{\perp \mu }+\frac 12M_{V+}^2V_{\perp }^\mu
V_{\perp \mu }-\frac 12B_{\perp }^\mu \square B_{\perp }^\mu +\frac
12M_{V-}^2B_{\perp }^\mu B_{\perp \mu }  \notag \\
&&+\frac 12\partial _\mu \sigma \partial ^\mu \sigma -\frac 12M_{S+}^2\sigma
^2-\frac 12\partial _\mu \chi \partial ^\mu \chi -\frac 12M_{S-}^2\chi
^2+\frac 12M^2\partial _\mu \Lambda \partial ^\mu \Lambda  \notag \\
&&+\mathcal{L}_{int}^{^{\prime }}(\overline{V}^{(\theta )},J,\ldots ).
\notag \\
&&
\end{eqnarray}
and
\begin{equation}
\overline{V}^{(\theta )}=\frac{\exp \theta _V}{(1+\alpha )^{1/2}}(V_{\perp
}+B_{\perp })-\partial \chi \cosh \theta _S-\partial \sigma \sinh \theta
_S-\partial \Lambda
\end{equation}
and where $M_{V\pm }^2$, $M_{S\pm }^2$ are the mass eigenvalues
(\ref{MV}) and (\ref{MS}). The theory is now formulated in terms of
two spin one and
two spin zero fields, whereas two of them, namely $B_{\perp }^\mu $ and $%
\chi $ have a wrong sign of the kinetic terms and are therefore
negative norm ghosts. As above, the field $\Lambda $ does not
correspond to any dynamical degree of freedom, its role is merely to
cancel the spurious poles of the free propagators of the transverse
fields $V_{\perp }$ and $B_{\perp } $ at $p^2=0$.

\subsection{Antisymmetric tensor formalism}

For the antisymmetric tensor field in the formalism \cite{Ecker1,Ecker2} the
situation is quite analogous to the Proca field case so our discussion will
be parallel to the previous subsection. Let us write the Lagrangian in the
form
\begin{equation}
\mathcal{L}=\mathcal{L}_0+\mathcal{L}_{int}.
\end{equation}
where the free part is
\begin{equation}
\mathcal{L}_0=-\frac 12(\partial _\mu R^{\mu \nu })(\partial ^\rho R_{\rho
\nu })+\frac 14M^2R_{\mu \nu }R^{\mu \nu },
\end{equation}
and introduce the transverse and longitudinal projectors
\begin{eqnarray}
\Pi _{\mu \nu \alpha \beta }^T &=&\frac 12\left( P_{\mu \alpha }^TP_{\nu
\beta }^T-P_{\nu \alpha }^TP_{\mu \beta }^T\right) \\
\Pi _{\mu \nu \alpha \beta }^L &=&\frac 12\left( g_{\mu \alpha }g_{\nu \beta
}-g_{\nu \alpha }g_{\mu \beta }\right) -\Pi _{\mu \nu \alpha \beta }^T
\end{eqnarray}
with $P_{\mu \alpha }^T$ given by (\ref{P_T}). Again, in analogy with (\ref%
{vector_gamma_2}), for completely general $\mathcal{L}_{int}$ we can expect
the following general form of the full two-point 1PI Green function

\begin{equation}
\Gamma _{\mu \nu \alpha \beta }^{(2)}(p)=\frac 12(M^2+\Sigma ^T(p^2))\Pi
_{\mu \nu \alpha \beta }^T+\frac 12(M^2-p^2+\Sigma ^L(p^2))\Pi _{\mu \nu
\alpha \beta }^L  \label{tensor_gamma_2}
\end{equation}
where $\Sigma ^{T,L}$ are the corresponding self-energies. The full
propagator is then obtained by means of the inversion of $\Gamma _{\mu \nu
\alpha \beta }^{(2)}$ in the form
\begin{equation}
\Delta _{\mu \nu \alpha \beta }(p)=-\frac 2{p^2-M^2-\Sigma ^L(p^2)}\Pi _{\mu
\nu \alpha \beta }^L+\frac 2{M^2+\Sigma ^T(p^2)}\Pi _{\mu \nu \alpha \beta
}^T.  \label{tensor_Delta_2}
\end{equation}
This propagator has two types of poles analogous to (\ref{V_pole_spin_1})
and (\ref{V_pole_spin_0}), either at $p^2=s_V,$ satisfying
\begin{equation}
s_V-M^2-\Sigma ^L(s_V)=0,  \label{T_pole_parity+}
\end{equation}
or at $p^2=s_{\widetilde{V}}$ where
\begin{equation}
M^2+\Sigma ^T(s_{\widetilde{V}})=0.  \label{T_pole_parity-}
\end{equation}
Assuming that the solution of (\ref{T_pole_parity+}) satisfies $s_V=$ $%
M_V^2>0,$ the propagator behaves at this pole as
\begin{eqnarray}
\Delta _{\mu \nu \alpha \beta }(p) &=&\frac{Z_V}{p^2-M_V^2}\frac{p_\mu
g_{\nu \alpha }p_\beta -p_\nu g_{\mu \alpha }p_\beta -(\alpha
\leftrightarrow \beta )}{M_V^2}+O(1)  \notag \\
&=&\frac{Z_V}{p^2-M_V^2}\sum_\lambda u_{\mu \nu }^{(\lambda )}(p)u_{\alpha
\beta }^{(\lambda )}(p)^{*}+O(1)
\end{eqnarray}
where
\begin{equation}
Z_V=\frac 1{1-\Sigma ^{^{\prime }L}(M_V^2)}
\end{equation}
and the wave function $u_{\mu \nu }^{(\lambda )}(p)$ can be expressed in
terms of the spin-one polarization vectors $\varepsilon _\nu ^{(\lambda
)}(p) $ as
\begin{equation}
u_{\mu \nu }^{(\lambda )}(p)=\frac{\mathrm{i}}{M_V}\left( p_\mu \varepsilon
_\nu ^{(\lambda )}(p)-p_\nu \varepsilon _\mu ^{(\lambda )}(p)\right) .
\label{wave_function}
\end{equation}
For $Z_V>0$ the pole of this type corresponds therefore to the spin-one
state $|p,\lambda ,V\rangle $ which couples to $R_{\mu \nu }$ as
\begin{equation}
\langle 0|R_{\mu \nu }(0)|p,\lambda ,V\rangle =Z_V{}^{1/2}u_{\mu \nu
}^{(\lambda )}(p).  \label{u_function}
\end{equation}
Analogously to the Proca case, at least one of these poles is expected to be
perturbative and corresponds to the original degree of freedom described by
the free Lagrangian $\mathcal{L}_0$. This means
\begin{eqnarray}
M_V^2 &=&M^2+\delta M_V^2 \\
Z_V &=&1+\delta Z_V
\end{eqnarray}
with small corrections $\delta M_V^2$ and $\delta Z_V$ vanishing in the free
field limit. The other possible nonperturbative solutions of (\ref%
{T_pole_parity+}) decouple in this limit.

Provided there exists a solution of (\ref{T_pole_parity-}) for which $s_{%
\widetilde{V}}=M_{\widetilde{V}}^2>0$, we get at this pole
\begin{eqnarray}
\Delta _{\mu \nu \alpha \beta }(p) &=&\frac{Z_{\widetilde{V}}}{p^2-M_{%
\widetilde{V}}^2}\left( g_{\mu \alpha }g_{\nu \beta }+\frac{p_\mu g_{\nu
\alpha }p_\beta -p_\mu g_{\nu \beta }p_\alpha }{M_A^2}-(\mu \leftrightarrow
\nu )\right) +O(1)  \notag \\
&=&\frac{Z_{\widetilde{V}}}{p^2-M_{\widetilde{V}}^2}\sum_\lambda w_{\mu \nu
}^{(\lambda )}(p)w_{\alpha \beta }^{(\lambda )}(p)^{*}+O(1)
\end{eqnarray}
where
\begin{equation}
Z_{\widetilde{V}}=\frac 1{\Sigma ^{^{\prime }T}(M_{\widetilde{V}}^2)}
\end{equation}
and the wave function is dual to the wave function (\ref{wave_function})
\begin{equation}
w_{\mu \nu }^{(\lambda )}(p)=\widetilde{u}_{\mu \nu }^{(\lambda )}(p)=\frac
12\varepsilon _{\mu \nu \alpha \beta }u^{(\lambda )\alpha \beta }(p).
\end{equation}
Provided $Z_{\widetilde{V}}>0$, the poles of this type correspond to
the spin-one particle states $|p,\lambda ,\widetilde{V}\rangle $
with the opposite intrinsic parity in comparison with $|p,\lambda
,V\rangle $, which couple to the antisymmetric tensor field as
\begin{equation}
\langle 0|R_{\mu \nu }(0)|p,\lambda ,\widetilde{V}\rangle =Z_{\widetilde{V}%
}{}^{1/2}w_{\mu \nu }^{(\lambda )}(p).  \label{w_function}
\end{equation}
This degree of freedom is frozen in the free propagator due to the specific
form of the free Lagrangian and it decouples in the limit of the vanishing
interaction.

As in the Proca field case, we can therefore generally expect dynamically
generated additional degrees of freedom, which can be either regular
asymptotic states ($M_{V,\widetilde{V}}^{2},\,Z_{V,\widetilde{V}}>0$) or
negative norm ghosts ($M_{V,\widetilde{V}}^{2}>0,\,Z_{V,\widetilde{V}}<0$)
or tachyons ($M_{V,\widetilde{V}}^{2}<0$). Complex poles on the unphysical
sheets can be then interpreted as resonances.

As the toy illustration of these possibilities, let us take the interaction
Lagrangian similar to (\ref{L_V_toy}) in the Proca field case \emph{e.g. }in
the form
\begin{eqnarray}
\mathcal{L}_{int} &\equiv &\mathcal{L}_{ct}=-\frac{\alpha -\beta }2(\partial
_\mu R^{\mu \nu })(\partial ^\rho R_{\rho \nu })-\frac \beta 4(\partial _\mu
R^{\alpha \beta })(\partial ^\mu R_{\alpha \beta })  \notag \\
&&+\frac{\gamma -\delta }{2M^2}(\partial _\alpha \partial _\mu R^{\mu \nu
})(\partial ^\alpha \partial ^\rho R_{\rho \nu })+\frac \delta
{4M^2}(\partial _\rho \partial _\mu R^{\alpha \beta })(\partial ^\rho
\partial ^\mu R_{\alpha \beta }).  \label{L_R_toy}
\end{eqnarray}
We get then the following contributions to the longitudinal and transverse
self-energies
\begin{eqnarray}
\Sigma ^L(p^2) &=&-\alpha p^2+\gamma \frac{p^4}{M^2} \\
\Sigma ^T(p^2) &=&-\beta p^2+\delta \frac{p^4}{M^2}.
\end{eqnarray}
These are exactly the same as (\ref{Sigma_L_vector}) and (\ref%
{Sigma_T_vector}) (with the identification $\Sigma ^{T,L}\leftrightarrow
\Sigma ^{L,T}$). Therefore, provided we further identify $M_{S\pm
}^2\leftrightarrow M_{\widetilde{V}\pm }^2$, the properties of the poles and
residues are the same as in the previous subsection (see the discussion
after (\ref{Sigma_L_vector}) and (\ref{Sigma_T_vector})), with the only
exception that instead of the extra spin-zero states with the mass (\ref{MS}%
) we have now extra spin-one states with the same mass (\ref{MS}) but with
the opposite parity in comparison with the original degrees of freedom
described by the free lagrangian $\mathcal{L}_0$.

\subsubsection{Path integral formulation\label{path integral tensor}}

We can again made the additional degrees of freedom manifest within the path
integral approach in the way parallel to subsection \ref{path integral
vector}. An analog of the $U(1)$ gauge symmetry used in the case of the
Proca field formalism in order to separate the transverse and longitudinal
components of the field $V_\mu $ is here the following transformation with a
pseudovector\footnote{%
This is of course true only in the case of the proper tensor field $R_{\mu
\nu }$. Provided $R_{\mu \nu }$ is a pseudotensor, the parameter of the
transformation is vectorial.} parameter $\Lambda _\alpha $
\begin{equation}
R^{\mu \nu }\rightarrow R^{\mu \nu }+\frac 12\varepsilon ^{\mu \nu \alpha
\beta }\widehat{\Lambda }_{\alpha \beta },  \label{R_gauge}
\end{equation}
where
\begin{equation}
\widehat{\Lambda }_{\alpha \beta }=\partial _\alpha \Lambda _\beta -\partial
_\beta \Lambda _\alpha .
\end{equation}
This leaves the kinetic term invariant, while the mass term is changed.
Note, that the transformation with the parameters $\Lambda _\alpha $ and $%
\Lambda _\alpha ^\lambda $ where
\begin{equation}
\Lambda _\alpha ^\lambda =\Lambda _\alpha +\partial _\alpha \lambda
\label{Lambda_gauge}
\end{equation}
are the same. This residual gauge invariance has to be taken into account
when using the Faddeev-Popov trick in order to isolate the longitudinal and
transverse degrees of freedom of the field $R_{\mu \nu }$. Analog of the
formula (\ref{V_components}) is now
\begin{equation}
R^{\mu \nu }=R_{\parallel }^{\mu \nu }+\frac 12\varepsilon ^{\mu \nu \alpha
\beta }\widehat{\Lambda }_{\alpha \beta }
\end{equation}
where $R_{\parallel }^{\mu \nu }$ is the longitudinal component of $R_{\mu
\nu }$. Its transverse component is described with the transverse component $%
\Lambda _{\perp }^\mu $ of the field $\Lambda ^\mu $ where
\begin{equation}
\Lambda ^\mu =\Lambda _{\perp }^\mu +\partial ^\mu \lambda .
\end{equation}
Starting with the path integral representation of the generating functional%
\footnote{%
Here $J$ are the external sources, cf. previous subsecrion.}
\begin{equation}
Z[J]=\int \mathcal{D}R\exp \left( \mathrm{i}\int \mathrm{d}^4x\left( -\frac
12(\partial _\mu R^{\mu \nu })(\partial ^\rho R_{\rho \nu })+\frac
14M^2R_{\mu \nu }R^{\mu \nu }+\mathcal{L}_{int}(R^{\mu \nu },J,\ldots
)\right) \right)
\end{equation}
and using the Faddeev-Popov trick twice with respect to the transformations (%
\ref{R_gauge}) and (\ref{Lambda_gauge}) we finally find for $Z[J]$ the
following representation
\begin{equation}
Z[J]=\int \mathcal{D}R_{\parallel }\mathcal{D}\Lambda _{\perp }\exp \left(
\mathrm{i}\int \mathrm{d}^4x\mathcal{L}(R_{\parallel }^{\mu \nu },\Lambda
_{\perp }^\mu ,\ldots )\right)
\end{equation}
where the integral measure is
\begin{equation}
\mathcal{D}R_{\parallel }\mathcal{D}\Lambda _{\perp }=\mathcal{D}R\mathcal{D}%
\Lambda \delta (\partial _\alpha R_{\mu \nu }+\partial _\nu R_{\alpha \mu
}+\partial _\mu R_{\nu \alpha })\delta (\partial _\mu \Lambda ^\mu )
\end{equation}
and
\begin{eqnarray}
R_{\parallel }^{\mu \nu } &=&-\frac 1{2\square }(\partial ^\mu g^{\nu \alpha
}\partial ^\beta +\partial ^\nu g^{\mu \beta }\partial ^\alpha -(\mu
\leftrightarrow \nu ))R_{\alpha \beta } \\
\Lambda _{\perp }^\mu &=&\left( g^{\mu \nu }-\frac{\partial ^\mu \partial
^\nu }{\square }\right) \Lambda _\nu .
\end{eqnarray}
are the longitudinal part of the tensor field $R^{\mu \nu }$and the
transverse part of the vector field $\Lambda ^\mu $ (describing the
transverse part of the tensor field $R^{\mu \nu }$) respectively\footnote{%
Note again that, the field $\Lambda _{\mu }$ has opposite parity than the
field $R_{\mu \nu }$ (being pseudovector for proper tensor field $R_{\mu \nu
}$ and vice versa).}. The Lagrangian expressed in these variables reads
\begin{equation}
\mathcal{L}(R_{\parallel }^{\mu \nu },\Lambda _{\perp }^\mu ,J,\ldots
)=\frac 14R_{\parallel }^{\mu \nu }\square R_{\parallel \,\mu \nu }+\frac
14M^2R_{\parallel }^{\mu \nu }R_{\parallel \,\mu \nu }+\frac 12M^2\Lambda
_{\perp }^\mu \square \Lambda _{\perp \mu }+\mathcal{L}_{int}(R_{\parallel
}^{\mu \nu }-\frac 12\varepsilon ^{\mu \nu \alpha \beta }\widehat{\Lambda }%
_{\alpha \beta },J,\ldots ).
\end{equation}
The free propagators of the fields $R_{\parallel }^{\mu \nu }$ and $\Lambda
_{\perp }^\mu $ are therefore
\begin{eqnarray}
\Delta _{\parallel }^{\mu \nu \alpha \beta }(p) &=&-\frac 2{p^2-M^2}\Pi
^{L\,\mu \nu \alpha \beta } \\
\Delta _{\perp }^{\mu \nu }(p) &=&-\frac 1{M^2}\frac 1{p^2}P^{T\,\mu \nu }
\end{eqnarray}
and, similarly to the case of the Proca field, they have spurious poles at $%
p^2=0$. Due to the form of the interaction, however, only the combination
\begin{eqnarray}
\Delta _0^{\mu \nu \alpha \beta }(p) &=&\Delta _{\parallel }^{\mu \nu \alpha
\beta }(p)+\varepsilon ^{\mu \nu \rho \sigma }\varepsilon ^{\alpha \beta
\kappa \lambda }p_\rho p_\kappa \Delta _{\perp \,\sigma \lambda }(p)  \notag
\\
&=&-\frac 2{p^2-M^2}\Pi ^{L\,\mu \nu \alpha \beta }+\frac 2{M^2}\Pi ^{T\,\mu
\nu \alpha \beta }
\end{eqnarray}
corresponding to the free propagator of the original tensor field $R^{\mu
\nu }$ is relevant within the Feynman graphs and the spurious poles cancel.
By analogy with the Proca field case, for the interaction Lagrangian
invariant with respect to the transformation (\ref{R_gauge}) the field $%
\Lambda _{\perp }^\mu $ completely decouples and can be integrated out. Such
a form of the interaction also allows to modify the propagator $\Delta
_{\parallel }^{\mu \nu \alpha \beta }(p)$ within the Feynman graphs
\begin{equation}
\Delta _{\parallel }^{\mu \nu \alpha \beta }(p)\rightarrow -\frac{g_{\mu
\alpha }g_{\nu \beta }-g_{\mu \beta }g_{\nu \alpha }}{p^2-M^2}
\end{equation}
and no spurious pole at $p^2=0\,$effectively appears. In this case the
opposite parity spin-one states discussed in the previous subsection cannot
be dynamically generated.

In order to illustrate the appearance of the additional degrees of freedom
connected with the interaction Lagrangian (\ref{L_R_toy}) within the path
integral formalism, we can make the same exercise with the interaction
Lagrangian (\ref{L_R_toy}) as we did in the previous subsection with (\ref%
{L_V_toy}). Our aim is again to make the additional degrees of freedom
explicit in the path integral representation of $Z[J]$. The procedure is
almost one-to-one to the case of the Proca fields so that we will be more
concise. The technical details can be found in the Appendix \ref%
{PI_antisymmetric}.

We assume the interaction Lagrangian to be of the form
\begin{equation}
\mathcal{L}_{int}=\mathcal{L}_{ct}+\mathcal{L}_{int}^{^{\prime }},
\end{equation}
where $\mathcal{L}_{ct}$ is given by (\ref{L_R_toy}) and we assume $\alpha
>-1\,$and $\delta >0$ as above. $\mathcal{L}_{int}$ can be then re-express
it in terms of the longitudinal and transverse components of the original
field $R_{\mu \nu }$
\begin{equation}
\mathcal{L}_{int}(R_{\parallel }^{\mu \nu }-\frac 12\varepsilon ^{\mu \nu
\alpha \beta }\widehat{\Lambda }_{\alpha \beta },J,\ldots )=\mathcal{L}%
_{ct}(R_{\parallel }^{\mu \nu }-\frac 12\varepsilon ^{\mu \nu \alpha \beta }%
\widehat{\Lambda }_{\alpha \beta },J,\ldots )+\mathcal{L}_{int}^{^{\prime
}}(R_{\parallel }^{\mu \nu }-\frac 12\varepsilon ^{\mu \nu \alpha \beta }%
\widehat{\Lambda }_{\alpha \beta },J,\ldots )
\end{equation}
where
\begin{eqnarray}
\mathcal{L}_{ct}(R^{\mu \nu }-\frac 12\varepsilon ^{\mu \nu \alpha \beta }%
\widehat{\Lambda }_{\alpha \beta }J,\ldots ) &=&\frac \alpha 4R_{\parallel
}^{\mu \nu }\square R_{\parallel \,\mu \nu }+\frac \gamma {4M^2}(\square
R_{\parallel }^{\mu \nu })(\square R_{\parallel \,\mu \nu })  \notag \\
&&+\frac \beta 2(\square \Lambda _{\perp }^\mu )(\square \Lambda _{\perp \mu
})-\frac \delta {2M^2}(\partial ^\alpha \square \Lambda _{\perp }^\mu
)(\partial _\alpha \square \Lambda _{\perp \mu }).
\end{eqnarray}
We then introduce the auxiliary (longitudinal) antisymmetric tensor field $%
B_{\parallel }^{\mu \nu }$ and (transverse) vector fields $\chi _{\perp
}^\mu $, $\rho _{\perp }^\mu $, $\sigma _{\perp }^\mu $ and $\pi _{\perp
}^\mu $ in order to avoid the higher derivative terms in a complete analogy
with the Proca field case. Again, not all the fields correspond to
propagating degrees of freedom and such redundant fields can be integrated
out. After rescaling the fields and diagonalization of the resulting mass
terms by means of two symplectic rotations with angles $\theta _V$ and $%
\theta _{\widetilde{V}}$ exactly as in the case of the Proca fields (see the
Appendix \ref{PI_antisymmetric} for details) we end up with
\begin{equation}
Z[J]=\int \mathcal{D}R_{\parallel }\mathcal{D}B_{\parallel }\mathcal{D}%
\Lambda _{\perp }\mathcal{D}\chi _{\perp }\mathcal{D}\rho _{\perp }\mathcal{D%
}\sigma _{\perp }\mathcal{D}\pi _{\perp }\exp \left( \mathrm{i}\int \mathrm{d%
}^4x\mathcal{L}(R_{\parallel },B_{\parallel },\Lambda _{\perp },\chi _{\perp
},\rho _{\perp },\sigma _{\perp },\pi _{\perp },J,\ldots )\right)
\end{equation}
with (cf. (\ref{integrated_L_V}))
\begin{eqnarray}
\mathcal{L} &=&\frac 14R_{\parallel }^{\mu \nu }\square R_{\parallel \,\mu
\nu }+\frac 14M_{V+}^2R_{\parallel }^{\mu \nu }R_{\parallel \,\mu \nu }
\notag \\
&&-\frac 14B_{\parallel }^{\mu \nu }\square B_{\parallel \,\mu \nu }+\frac
14M_{V-}^2B_{\parallel }^{\mu \nu }B_{\parallel \,\mu \nu }  \notag \\
&&+\frac 12M^2\Lambda _{\perp }^\mu \square \Lambda _{\perp \mu }  \notag \\
&&-\frac 12\chi _{\perp }^\mu \square \chi _{\perp \mu }+\frac 12M_{%
\widetilde{V}-}^2\chi _{\perp }^\mu \chi _{\perp \mu }+\frac 12\sigma
_{\perp }^\mu \square \sigma _{\perp \mu }+\frac 12M_{\widetilde{V}%
+}^2\sigma _{\perp }^\mu \sigma _{\perp \mu }  \notag \\
&&+\mathcal{L}_{int}(\overline{R}^{(\theta )},J,\ldots )
\end{eqnarray}
where
\begin{equation*}
\overline{R}^{(\theta )\mu \nu }=\frac{\exp \theta _V}{(1+\alpha )^{1/2}}%
(R_{\parallel }^{\mu \nu }+B_{\parallel }^{\mu \nu })-\frac 12\varepsilon
^{\mu \nu \alpha \beta }\left( \widehat{\Lambda }_{\alpha \beta }+\widehat{%
\sigma }_{\perp \alpha \beta }\sinh \theta _{\widetilde{V}}+\widehat{\chi }%
_{\perp \alpha \beta }\cosh \theta _{\widetilde{V}}\right)
\end{equation*}
and with the diagonal mass terms corresponding to the eigenvalues (\ref{MV}, %
\ref{MS}) (with identification $M_{\widetilde{V}\pm }^2\rightarrow M_{S\pm
}^2$). Again we have two pairs of fields with the opposite signs of the
kinetic terms, namely $(R_{\parallel }^{\mu \nu },B_{\parallel }^{\mu \nu })$
and $(\chi _{\perp }^\mu ,\sigma _{\perp }^\mu )$ respectively. As a result
we have found four spin-one states, two of them being negative norm ghosts,
namely $B_{\parallel }^{\mu \nu }$ and $\sigma _{\perp }^\mu $ and two of
them with the opposite parity, namely $\chi _{\perp }^\mu $ and $\sigma
_{\perp }^\mu $. As in the Proca field case, the field $\Lambda _{\perp
}^\mu $ effectively compensates the spurious $p^2=0$ poles in the $%
R_{\parallel }^{\mu \nu }$ and $B_{\parallel }^{\mu \nu }$ propagators
within Feynman graphs.\qquad \qquad

\subsection{First order formalism}

The first order formalism is a natural alternative to the previous two (for
the motivation and details of the quantization see \cite{FO1}, cf. also \cite%
{Bruns:2004tj}). It introduces both vector and antisymmetric tensor fields
into the Lagrangian, therefore the analysis is a little bit more complex in
comparison with previous two cases. In this case, the Lagrangian is of the
form
\begin{equation}
\mathcal{L}=\mathcal{L}_0+\mathcal{L}_{int}
\end{equation}
where now the free part is
\begin{equation}
\mathcal{L}_0=MV_\nu \partial _\mu R^{\mu \nu }+\frac 12M^2V_\mu V^\mu
+\frac 14M^2R_{\mu \nu }R^{\mu \nu }.
\end{equation}
Instead of just one one-particle irreducible two point Green function we
have a matrix
\begin{equation}
\Gamma ^{(2)}(p)=\left(
\begin{array}{ll}
\Gamma _{VV}^{(2)}(p)_{\mu \nu } & \Gamma _{VR}^{(2)}(p)_{\alpha \mu \nu }
\\
\Gamma _{RV}^{(2)}(p)_{\mu \nu \alpha } & \Gamma _{RR}^{(2)}(p)_{\mu \nu
\alpha \beta }%
\end{array}
\right)  \label{matrix_gamma}
\end{equation}
where (without any additional assumptions on the form of $\mathcal{L}_{int}$%
) the matrix elements have the following general form (cf. (\ref%
{vector_gamma_2}) and (\ref{tensor_gamma_2}))
\begin{eqnarray}
\Gamma _{RR}^{(2)}(p)_{\mu \nu \alpha \beta } &=&\frac 12(M^2+\Sigma
_{RR}^T(p^2))\Pi _{\mu \nu \alpha \beta }^T+\frac 12(M^2+\Sigma
_{RR}^L(p^2))\Pi _{\mu \nu \alpha \beta }^L \\
\Gamma _{VV}^{(2)}(p)_{\mu \nu } &=&(M^2+\Sigma _{VV}^T(p^2))P_{\mu \nu
}^T+(M^2+\Sigma _{VV}^L(p^2))P_{\mu \nu }^L \\
\Gamma _{RV}^{(2)}(p)_{\mu \nu \alpha } &=&\frac{\mathrm{i}}2\left( M+\Sigma
_{RV}(p^2)\right) \Lambda _{\mu \nu \alpha } \\
\Gamma _{VR}^{(2)}(p)_{\alpha \mu \nu } &=&\frac{\mathrm{i}}2\left( M+\Sigma
_{VR}(p^2)\right) \Lambda _{\alpha \mu \nu }^t.
\end{eqnarray}
Here $\Sigma _{RR}^{T,L}(p^2)$, $\Sigma _{VV}^{T,L}(p^2)$ and $\Sigma
_{RV}(p^2)=\Sigma _{VR}(p^2)$ are corresponding self-energies and the
off-diagonal tensor structures are
\begin{equation}
\Lambda _{\mu \nu \alpha }=-\Lambda _{\alpha \mu \nu }^t=p_\mu g_{\nu \alpha
}-p_\nu g_{\mu \alpha }.
\end{equation}
This matrix of propagators
\begin{equation}
\Delta (p)=\left(
\begin{array}{ll}
\Delta _{VV}(p)_{\mu \nu } & \Delta _{VR}(p)_{\alpha \mu \nu } \\
\Delta _{RV}(p)_{\mu \nu \alpha } & \Delta _{RR}(p)_{\mu \nu \alpha \beta }%
\end{array}
\right)
\end{equation}
can be obtained by means of the inversion of the matrix (\ref{matrix_gamma})
with the result
\begin{eqnarray}
\Delta _{RR}(p)_{\mu \nu \alpha \beta } &=&\frac 2{M^2+\Sigma
_{RR}^T(p^2)}\Pi _{\mu \nu \alpha \beta }^T+2\frac{M^2+\Sigma _{VV}^T(p^2)}{%
D(p^2)}\Pi _{\mu \nu \alpha \beta }^L \\
\Delta _{VV}(p)_{\mu \nu } &=&\frac 1{M^2+\Sigma _{VV}^L(p^2)}P_{\mu \nu }^L+%
\frac{M^2+\Sigma _{RR}^L(p^2)}{D(p^2)}P_{\mu \nu }^T \\
\Delta _{RV}(p)_{\mu \nu \alpha } &=&-\mathrm{i}\frac{M+\Sigma _{RV}(p^2)}{%
D(p^2)}\Lambda _{\mu \nu \alpha } \\
\Delta _{VR}(p)_{\alpha \mu \nu } &=&-\mathrm{i}\frac{M+\Sigma _{VR}(p^2)}{%
D(p^2)}\Lambda _{\alpha \mu \nu }^t,
\end{eqnarray}
where
\begin{equation}
D(p^2)=(M^2+\Sigma _{RR}^L(p^2))(M^2+\Sigma _{VV}^T(p^2))-p^2(M+\Sigma
_{RV}(p^2))(M+\Sigma _{VR}(p^2)).
\end{equation}
Let us now discuss the structure of the poles, which is now richer than in
previous two cases. We have three possible types of poles, namely $s_V$, $s_{%
\widetilde{V}}$ and $s_S$, being solutions of
\begin{eqnarray}
D(s_V) &=&0  \notag \\
M^2+\Sigma _{RR}^T(s_{\widetilde{V}}) &=&0  \notag \\
M^2+\Sigma _{VV}^L(s_S) &=&0  \label{RV_zeros}
\end{eqnarray}
respectively. As far as the pole $s_V$ is concerned, let us assume $%
s_V=M_V^2>0$. We get then at this pole (see also previous two subsections)
\begin{eqnarray}
\Delta _{RR}(p)_{\mu \nu \alpha \beta } &=&\frac{Z_{RR}}{p^2-M_V^2}%
\sum_\lambda u_{\mu \nu }^{(\lambda )}(p)u_{\alpha \beta }^{(\lambda
)}(p)^{*}+O(1) \\
\Delta _{VV}(p)_{\mu \nu } &=&\frac{Z_{VV}}{p^2-M_V^2}\sum_\lambda
\varepsilon _\mu ^{(\lambda )}(p)\varepsilon _\nu ^{(\lambda )*}(p)+O(1) \\
\Delta _{RV}(p)_{\mu \nu \alpha } &=&\frac{Z_{RV}}{p^2-M_V^2}\sum_\lambda
u_{\mu \nu }^{(\lambda )}(p)\varepsilon _\alpha ^{(\lambda )}(p)^{*}+O(1) \\
\Delta _{VR}(p)_{\alpha \mu \nu } &=&\frac{Z_{VR}}{p^2-M_V^2}\sum_\lambda
\varepsilon _\alpha ^{(\lambda )}(p)u_{\mu \nu }^{(\lambda )*}(p)+O(1)
\end{eqnarray}
where $u_{\mu \nu }^{(\lambda )}(p)$ is given by (\ref{wave_function}) and
the residue are
\begin{eqnarray}
Z_{RR} &=&\frac{M^2+\Sigma _{VV}^T(M_V^2)}{D^{^{\prime }}(M_V^2)} \\
Z_{VV} &=&\frac{M^2+\Sigma _{RR}^L(M_V^2)}{D^{^{\prime }}(M_V^2)} \\
Z_{RV} &=&\frac{M+\Sigma _{RV}(M_V^2)}{D^{^{\prime }}(M_V^2)}M_V=Z_{VR}=%
\frac{M+\Sigma _{VR}(M_V^2)}{D^{^{\prime }}(M_V^2)}M_V.
\end{eqnarray}
Note that, as a consequence of (\ref{RV_zeros}) we get the following
relation
\begin{equation}
Z_{RR}Z_{VV}=Z_{RV}^2=Z_{VR}^2,
\end{equation}
(remember $\Sigma _{RV}(p^2)=\Sigma _{VR}(p^2)$), therefore assuming $%
Z_{RR},Z_{VV}>0$ the pole $p^2=M_V^2>0$ corresponds to the spin-one
one-particle state $|p,\lambda ,V\rangle $ which couples to the fields as
\begin{eqnarray}
\langle 0|R_{\mu \nu }(0)|p,\lambda ,V\rangle &=&Z_{RR}{}^{1/2}u_{\mu \nu
}^{(\lambda )}(p) \\
\langle 0|V_\mu (0)|p,\lambda ,V\rangle &=&Z_{VV}{}^{1/2}\varepsilon _\mu
^{(\lambda )}(p).
\end{eqnarray}
Again at least one of such states is expected to be perturbative as above
and it correspond to the original degree of freedom described by $\mathcal{L}%
_0$; the others decouple when the interactions is switched off. The other
possible poles, $s_S=M_S^2$ and $s_{\widetilde{V}}=M_{\widetilde{V}}^2$ are
analogical to the spin-zero and spin-one (opposite parity) states discussed
in detail in the previous two subsections; they correspond to the modes
which are frozen at the leading order and decouple in the free field limit.
As we have already discussed, without further restriction on the form of the
interaction, all the additional states can be also negative norm ghosts or
tachyons.

Let us illustrate the general case using a toy interaction Lagrangian of the
form
\begin{eqnarray}
\mathcal{L}_{ct} &=&-\frac{\alpha _{V}}{4}\widehat{V}_{\mu \nu }\widehat{V}%
^{\mu \nu }-\frac{\beta _{V}}{2}(\partial _{\mu }V^{\mu })^{2}  \notag \\
&&-\frac{\alpha _{R}-\beta _{R}}{2}(\partial _{\mu }R^{\mu \nu })(\partial
^{\rho }R_{\rho \nu })-\frac{\beta _{R}}{4}(\partial _{\mu }R^{\alpha \beta
})(\partial ^{\mu }R_{\alpha \beta }).  \label{L_RV_toy}
\end{eqnarray}%
This gives
\begin{eqnarray}
\Sigma _{RR}^{L}(p^{2}) &=&-\alpha _{R}p^{2}  \notag \\
\Sigma _{RR}^{T}(p^{2}) &=&-\beta _{R}p^{2}  \notag \\
\Sigma _{VV}^{T}(p^{2}) &=&-\alpha _{V}p^{2}  \notag \\
\Sigma _{VV}^{L}(p^{2}) &=&-\beta _{V}p^{2}  \notag \\
\Sigma _{RV}(p^{2}) &=&\Sigma _{VR}(p^{2})=0
\end{eqnarray}%
and for $\beta _{V,R}>0$ the spectrum of one-particle states consists of one
spin-zero ghost, one spin-one ghost with opposite parity. Their masses and
residue are
\begin{eqnarray}
M_{S}^{2} &=&\frac{M^{2}}{\beta _{V}},\,\,\,\,\,Z_{S}=-\frac{1}{\beta _{V}}
\\
M_{\widetilde{V}}^{2} &=&\frac{M^{2}}{\beta _{R}},\,\,\,\,Z_{\widetilde{V}}=-%
\frac{1}{\beta _{R}}
\end{eqnarray}%
(provided $\beta _{R}<0$ or $\beta _{V}<0$ the corresponding states are
tachyons) and two spin-one states with masses
\begin{eqnarray}
M_{V\pm }^{2} &=&M^{2}\frac{1+\alpha _{R}+\alpha _{V}\pm \sqrt{\mathcal{D}}}{%
2\alpha _{R}\alpha _{V}}  \notag \\
\mathcal{D} &=&(1+\alpha _{R}+\alpha _{V})^{2}-4\alpha _{R}\alpha _{V}.
\end{eqnarray}%
To get both $M_{V\pm }^{2}>0$ we need $\mathcal{D}>0$, $\alpha _{V}\alpha
_{R}>0$ and $1+\alpha _{R}+\alpha _{V}>0$; in this case we get for the
residue $Z_{RR}^{(\pm )}$ and $Z_{VV}^{(\pm )}$ at poles $M_{V\pm }^{2}$
\begin{equation}
\alpha _{R}Z_{RR}^{(+)}Z_{RR}^{(-)}=\alpha _{V}Z_{VV}^{(+)}Z_{VV}^{(-)}=%
\frac{1}{\mathcal{D}}>0
\end{equation}%
Assuming $Z_{RR}^{(-)},\,Z_{VV}^{(-)}>0$ (note that, for small couplings $%
M_{V-}^{2}=M^{2}(1+O(\alpha _{R},\alpha _{V}))$ with $%
Z_{RR}^{(-)},Z_{VV}^{(-)}=1+O(\alpha _{R},\alpha _{V})$ corresponds to the
perturbative solution), the additional spin one-state is either positive
norm state for $\alpha _{V,R}>0$ or ghost for $\alpha _{V,R}<0$ (in this
latter case the extra kinetic terms in $\mathcal{L}_{ct}$ have wrong signs).

Also in this case the propagating degrees of freedom can be made manifest
within the path integral formalism. The corresponding discussion is in a
sense synthesis of subsections \ref{path integral vector} and \ref{path
integral tensor} and is postponed to Appendix \ref{PI_first_order}.

\section{Organization of the counterterms\label{Section_power_counting}}

Let us now return to the concrete case of $R\chi T$. Our aim is to
calculate the one loop self-energies defined in the previous
section in all three formalisms discussed there. In the process of
the loop calculation we are lead to the problem of performing a
classification of the countertems, which have to be introduced in
order to renormalize infinities. For this purpose, it is
convenient to have a scheme, which allows us to assign to each
operator in the Lagrangian and to each Feynman graph an
appropriate expansion index. Indices of the counterterms, which
are necessary in order to cancel the divergences of the given
Feynman graph, should be then correlated with the indices of the
vertices of the graph as well as with the number of the loops.
When we restrict ourselves to the (one-particle irreducible)
graphs with a given index, the number of the allowed operators
contributing to the graph as well as that of necessary
counterterms should be finite.

There are several possibilities how to do it, some of them being quite
efficient but purely formal and unphysical, some of them having good
physical meaning, but not very useful in practise.

In the literature, several attempts to organize the individual terms of the $%
R\chi T$ Lagrangian can be found. Let us briefly comment on some of them
from the point of view of its applicability to our purpose.

The first one is intimately connected with the effective chiral Lagrangian $%
\mathcal{L}_{\chi ,\mathrm{res}}$ which appears as a result of the
(tree-level) integrating out of the resonances from the $R\chi T$. Such a
counting assigns to each operator of the resonance part of the $R\chi T$
Lagrangian $\mathcal{L}_{\mathrm{res}}$ a chiral order according to the
minimal chiral order of the coupling (LEC) of the effective chiral
Lagrangian $\mathcal{L}_{\chi ,\mathrm{res}}$ to which the corresponding
operator contributes \cite{Kampf:2006bn}, \cite{Cirigliano:2006hb}. More
generally, in this scheme the chiral order of the operators from $\mathcal{L}%
_{\mathrm{res}}$ refers to the minimal chiral order of its contribution to
the generating functional of the currents $Z[v,a,p,s]=%
\sum_nZ^{(2n)}[v,a,p,s] $. The loop expansion of $Z[v,a,p,s]$ formally
corresponds to the expansion around the classical fields which are solutions
of the classical equation of motion. The formal chiral order of the
resonance fields corresponds then to the chiral order of the leading term of
the expansion of the classical resonance fields in powers of $p$ and
external sources according to the standard chiral power counting,\emph{\ i.e.%
}
\begin{equation}
V^\mu =O(p^3),\,\,\,\,\,R^{\mu \nu }=O(p^2).\,  \label{fields order I}
\end{equation}
At the same time, for the resonance mass (which plays a role of the hadronic
scale within the standard power counting) we take
\begin{equation}
\,M=O(1),  \label{mass order I}
\end{equation}
and for the external sources as usual
\begin{equation}
v^\mu ,a^\mu =O(p),\,\,\,\,\,\chi ,\chi ^{+}=O(p^2).  \label{source order
I}
\end{equation}
The resonance propagators are then of the (minimal) order $O(1)$ and the
order of the operators which contain the resonance fields is at least $%
O(p^4) $. This formal power counting therefore restricts both the number of
the resonance fields in the generic operator as well as the number of the
derivatives. When combined with the large $N_C$ arguments, it allows for the
construction of the complete operator basis necessary for the saturation of
the LEC's in the chiral Lagrangian at a given chiral order and a leading
order in the $1/N_C$ expansion \cite{Cirigliano:2006hb}.

Originally this type of power counting was designed for the leading order
(tree-level) matching of $R\chi T$ and $\chi PT$ within the large $N_C$
expansion and there is no straightforward extension to the general graph $%
\Gamma $ with $L$ loops. The reason is that the above power
counting of the resonance propagators inside the loops does not
reproduce correctly the standard chiral order of the graph. As a
result, the loop graphs violate the naive chiral power counting in
a way analogous to the $\chi PT$ with baryons \cite{Gasser:1987rb}
.

The second possibility applicable to loops is to generalize the Weinberg
\cite{Weinberg} power counting scheme and \emph{formally} arrange the
computation as an expansion in the power of the momenta \emph{and} the
resonance masses \cite{Lutz:2008km} (though there is no mass gap and no
natural scale which would give to such a formal power counting a reasonable
physical meaning\footnote{%
Sometimes it is argued \cite{Lutz:2008km},\cite{Harada:2003jx},
that such a counting can be used within the large $N_{C}$ limit,
due to the fact that the natural $\chi PT$ scale $\Lambda _{\chi
PT}=4\pi F=O(\sqrt{N_{C}})$ grows with $N_{C}$ while the masses of
the resonances behave as $O(1)$. In fact this results only in the
suppression of the loops but generally not in the suppression of
the counterterm contributions. In the latter case the expansion is
rather controlled by the scale $\Lambda _{H}\sim M_{R}=O(1)$,
where $M_{R}$ is the typical mass of the higher resonance in the
considered channel not included in truncated Lagrangian
corresponding to minimal hadronic ansatz.}). Nevertheless,
provided we make a following assignment to the resonance field and
to the resonance mass
$M$%
\begin{equation}
V^\mu ,R^{\mu \nu }=O(1),\,\,\,M=O(p)
\end{equation}
we get for the kinetic and mass term of the resonance field
\begin{equation}
\mathcal{L}_{kin},\mathcal{L}_{mass}=O(p^2)
\end{equation}
\emph{i.e.} the same order as for the lowest order chiral
Lagrangian, which allows the same power counting of the resonance
propagators as for PGB within the pure $\chi PT$. As a result, the
Weinberg formula for the order $D_\Gamma $ of a given graph $\Gamma
$ with $L$ loops built from the vertices with the order $D_V$,
\begin{equation}
D_\Gamma =2+2L+\sum_V(D_V-2),  \label{weinberg}
\end{equation}
remains valid also within $R\chi T$. Note however, that now $p^2/M^2=O(1)$
and therefore the counterterms needed for renormalization of the graph with
chiral order $D_\Gamma $ might contain more than $D_\Gamma $ derivatives
(this feature is typical for graphs with resonances inside the loops because
of the nontrivial numerator of the resonance propagator). Therefore this
type of power counting is less useful for the classification of the
counterterms than in the case of the pure $\chi PT$, where $D_\Gamma $ gives
an upper bound on the number of derivatives of the counterterms needed to
renormalize $\Gamma $.

There are also some other complications, which depreciate this counting in
the case of $R\chi T$. First note that the interaction vertices with the
resonance fields can carry a chiral order smaller than two. This applies
\emph{e.g.} to the trilinear vertex in the antisymmetric tensor
representation
\begin{equation}
\mathcal{O}^{RRR}=\mathrm{i}g_{\rho \sigma }\langle R_{\mu \nu }R^{\mu \rho
}R^{\nu \sigma }\rangle  \label{RRR_vertex}
\end{equation}
or to the odd intrinsic parity vertex mixing the vector and rge
antisymmetric tensor field in the first order formalism
\begin{equation}
\mathcal{O}^{RV}=\varepsilon _{\alpha \beta \mu \nu }\langle \{V^\alpha
,R^{\mu \nu }\}u^\beta \rangle .  \label{VRGB_vertex}
\end{equation}
Therefore, increasing number of such vertices will decrease the formal
chiral order causing again a mismatch between the chiral counting and the
loop expansion. Furthermore, such a naive scheme unlike the previous one
does not restrict the number of the resonance fields in a general operator
because only the number of derivatives, the resonance masses and the
external sources score.

The former drawback can be \emph{formally} cured by adding an artificial
power of $M$ in front of such operators\footnote{%
In the case of $\mathcal{O}^{RV}$ it seems to be natural from the
dimensional reason.} (or equivalently counting the corresponding
couplings as $O(p^2)$ and $O(p)$ respectively) in order to increase
artificially their chiral order and preserve the validity of the
Weinberg formula, which now can serve as a \emph{formal} tool for
the classification of the counterterms. How to treat the latter
drawback we will discuss further bellow. Let us, however, stress
once again, that there is \emph{no} physical content in such a
classification scheme, though it might be technically useful.

Third possibility how to assign an index to the given interaction
terms and to the general graphs, independent of the previous two,
is offered by the large $N_{C}$ expansion. In the
$N_{C}\rightarrow \infty $ limit, the amplitude of the interaction
of the $n$ mesonic resonances is suppressed at least by the factor
$O(N_{C}^{1-n/2})$ and, more generally, the matrix element of
arbitrary number of quark currents and $n$ mesons in the initial
and final states has the same leading order behavior; \emph{e.g.}
for the GB decay constant we get $F=O(N_{C}^{1/2})$. Because
within the chiral building blocks the GB fields always go with the
factor $1/F$, we can
treat the coupling $c_{\mathcal{O}}\,$corresponding to the operator $%
\mathcal{O}$ of the $R\chi T$ Lagrangian as $c_{\mathcal{O}}=O(N_{C}^{\omega
_{\mathcal{O}}})$, where
\begin{equation}
\omega _{\mathcal{O}}=1-\frac{n_{R}^{\mathcal{O}}}{2}-s_{\mathcal{O}},
\label{NC_index}
\end{equation}%
$n_{R}^{\mathcal{O}}$ is the number of the resonance fields contained in $%
\mathcal{O}$ and $s_{\mathcal{O}}$ is a possible additional suppression
coming \emph{e.g.} from multiple flavor traces or from the fact, that this
coupling appears as a counterterm renomalizing the loop divergences\footnote{%
Note that, each additional mesonic loop yields a further suppression
$1/N_{C} $, see also bellow.}. From such an operator, generally the
infinite number of vertices $V\,$with increasing number $n_{GB}^{V}$
of GB legs can be derived, each accompanied with a factor
$c_{\mathcal{O}}F^{-n_{GB}/2}$ and therefore, suppressed as
$O(N_{C}^{\omega _{{V}}})$, where the index $\omega _{V}$ is
given by\footnote{%
Here and in what follows we use subscript $\mathcal{O}$ when referring to
the operator, while the superscript $V$ corresponds to the concrete vertex
derived from the operator $\mathcal{O}$.}
\begin{equation}
\omega _{V}=1-\frac{n_{R}^{\mathcal{O}}}{2}-\frac{n_{GB}^{V}}{2}-s_{\mathcal{%
O}}.
\end{equation}%
For a given graph, we have the large $N_{C}$ behavior $O(N_{C}^{\omega _{{%
\Gamma }}})$ where\footnote{%
Here and in what follows, the sum over $\mathcal{O}$ include all the
operators from which the individual vertices entering the graph $\Gamma $
are derived with necessary multiplicity.\label{note}}
\begin{equation}
\omega _{\Gamma }=\sum_{V}\omega _{V}=1-\frac{1}{2}E-L-\sum_{\mathcal{O}}s_{%
\mathcal{O}},  \label{large_N_C}
\end{equation}%
where $L$ is number of the loops, $E$ is the number of external mesonic
lines and we have used the identities
\begin{eqnarray}
\sum_{V}(n_{R}^{V}+n_{GB}^{V}) &=&2I_{R}+2I_{GB}+E  \notag \\
I_{R}+I_{GB} &=&L+V-1
\end{eqnarray}%
relating $L$ and $E$ with the number of resonance and GB internal lines $%
I_{R}$ and $I_{GB}$. The loop expansion is therefore correlated with the
large $N_{C}$ expansion; higher loops need additionally $N_{C}-$suppressed
counterterms $\mathcal{O}_{ct}$ with higher $s_{\mathcal{O}_{ct}}$:
\begin{equation}
s_{\mathcal{O}_{ct}}=\left( 1-\frac{1}{2}E\right) -\omega _{\Gamma }=L+\sum_{%
\mathcal{O}}s_{\mathcal{O}}  \label{large_N_C_s}
\end{equation}%
Though the formula (\ref{large_N_C}) refers seemingly to individual
vertices, reformulated in in the form (\ref{large_N_C_s}) it points to the
members of the chiral symmetric operator basis of the $R\chi T$ Lagrangian.
However, as it stays, it does not suit for our purpose because the large $%
N_{C}$ counting rules give no restriction for the number of derivatives as
well as to the number of resonance fields (once the couplings respect the
leading order large $N_{C}$ behavior described above). The formula (\ref%
{large_N_C_s}) expresses merely the the fact that the large $N_{C}$
expansion coincide with the loop one.

Let us now describe another useful technical way how to classify the
couterterms, which could overcome the problems with the above schemes and is
in a sense a combination of them. Let us start with the familiar formula for
the degree of superficial divergence $d_\Gamma \,$of a given \emph{one
particle irreducible} graph $\Gamma $, which provides us with the upper
bound on the number of derivatives $d_{\mathcal{O}_{ct}}$ in a counterterm $%
\mathcal{O}_{ct}$ needed for the renormalization of $\Gamma $. Because in
the Proca and antisymmetric tensor formalisms the spin $1$ resonance
propagator behaves as\footnote{%
In the case of the first order formalism, the mixed propagator behaves as $%
O(p^{-1})$. In this case, $d_{\Gamma }=4L-2I_{GB}-I_{RV}+\sum_{\mathcal{O}%
}d_{\mathcal{O}}$ where $I_{RV}$ is number of the internal mixed lines. In
the following considerations we can take the r.h.s. of (\ref{d_ct}) as an
upper bound on $d_{\Gamma }$ with the conclusions unchanged.} $O(1)$ for $%
p\rightarrow \infty $, we get
\begin{equation}
d_{\mathcal{O}_{ct}}\leq d_\Gamma =4L-2I_{GB}+\sum_{\mathcal{O}}d_{\mathcal{O%
}}  \label{d_ct}
\end{equation}
where $d_{\mathcal{O}}$ means the number of derivatives of the vertex $V$
derived from the operator $\mathcal{O}$. Eliminating $I_{GB}$ in favour of $%
L $ and $I_R$ and using the identity
\begin{equation*}
\sum_{\mathcal{O}}n_R^{\mathcal{O}}=2I_R+E_R,
\end{equation*}
relating $I_R$ with the number of external resonance lines $E_R$, we get
eventually
\begin{equation*}
d_{\mathcal{O}_{ct}}\leq d_\Gamma =2+2L+\sum_{\mathcal{O}}(d_{\mathcal{O}%
}+n_R^{\mathcal{O}}-2)-E_R.
\end{equation*}
Adding further to both sides $\sum_{\mathcal{O}}(2n_s^{\mathcal{O}}+2n_p^{%
\mathcal{O}}+n_v^{\mathcal{O}}+n_a^{\mathcal{O}})$, the total number of
insertions of the external $v$, $a$, $p$ and $s$ sources weighted with its
chiral order, we have
\begin{equation*}
D_{\mathcal{O}_{ct}}+n_R^{ct}-2\leq 2L+\sum_{\mathcal{O}}(D_{\mathcal{O}%
}+n_R^{\mathcal{O}}-2)
\end{equation*}
where $D_{\mathcal{O}}$ is the usual chiral order (as in pure $\chi PT$) of
generic operator $\mathcal{O}$. Therefore, introducing an index $i_{\mathcal{%
O}}$ of a general operator $\mathcal{O}$ as follows\footnote{%
Analogous assignment of the chiral order to the interaction terms
with at least two resonance fields is proposed in
\cite{Lutz:2008km}, note however, that
in this reference it is used by means of substitution $D_V\rightarrow i_{%
\mathcal{O}}$ in the Weinberg formula (\ref{weinberg}) with counting $M=O(p)$%
.}
\begin{equation}
i_{\mathcal{O}}=D_{\mathcal{O}}+n_R^{\mathcal{O}}-2  \label{index_O}
\end{equation}
we get analog of the Weinberg formula\footnote{%
This can be recovered for $n_R^{\mathcal{O}}=0$, when the inequality changes
to the equality.}, now in the form of an upper bound
\begin{equation}
i_{\mathcal{O}_{ct}}\leq i_\Gamma =2L+\sum_{\mathcal{O}}i_{\mathcal{O}}.
\label{power_counting}
\end{equation}
Let us now discuss its properties more closely. First, the number of
operators with given $i_{\mathcal{O}}\leq i_{\max }$ is finite, because this
requirement limits both the number of derivatives as well as the number of
resonance fields. Second, note that, for general operator $\mathcal{O}$ the
index $i_{\mathcal{O}}\geq 0$. We have $i_{\mathcal{O}}=0$ for the leading
order $\chi PT$ Lagrangian, for the resonance mass (counter)terms as well as
for the resonance-GB mixing term $\langle A^\mu u_\mu \rangle $ possible for
$1^{+-}$ resonances in the Proca field formalism\footnote{%
Note however, that this term can be removed by means of the field
redefinition.}. The usual interaction terms with one resonance field and $%
O(p^2)$ building blocks correspond to the sector $i_{\mathcal{O}}=1$, the
same is true for the trilinear resonance vertex (\ref{RRR_vertex}) as well
as for the ``mixed''\ vertex (\ref{VRGB_vertex}), while the two resonance
vertices with $O(p^2)$ building blocks correspond to the sector $i_{\mathcal{%
O}}=2$, \emph{etc}.

Therefore, according to the formula (\ref{power_counting}), the loop
expansion is correlated with the organization of the operators and
loop graphs according to the indices $i_{\mathcal{O}}$ and $i_\Gamma
$ respectively analogously to the pure $\chi PT$, with the only
exception that also lower sectors of the Lagrangian w.r.t.
$i_{\mathcal{O}}$ are renormalized at each step. Therefore, we get
the renormalizability provided we limit ourselves to the graphs
composed from one-particle ireducible building blocs for which the
RHS of (\ref{power_counting}) is smaller or equal to $i_{\max }$.

The counting rules can be summarized as follows
\begin{equation}
R_{\mu \nu },\,V_\mu =O(p),\,M=O(1)
\end{equation}
and for the external sources as usual
\begin{equation}
v^\mu ,a^\mu =O(p),\,\,\chi ,\chi ^{+}=O(p^2).
\end{equation}
Note also that, the index $i_{\mathcal{O}}$ can be rewritten as
\begin{equation}
i_{\mathcal{O}}=D_{\mathcal{O}}-2\left( 1-\frac{n_R^{\mathcal{O}}}2\right)
\end{equation}
and in the last bracket we recognize the exponent controlling the leading
large $N_C\,$ behavior of the coupling constant in front of the operator $%
\mathcal{O}$. Remember, however, that the loop induced counterterms have an
additional $1/N_C$ suppression for each loop (cf. (\ref{large_N_C})).
Therefore it is natural to modify the index $i_{\mathcal{O}}$ and $i_\Gamma $
as follows (the coefficient $1/2$ is a matter of convenience, see bellow)
\begin{eqnarray}
\widehat{i}_{\mathcal{O}} &=&\frac{i_{\mathcal{O}}}2+s_{\mathcal{O}}=\frac
12D_{\mathcal{O}}-\left( 1-\frac{n_R^{\mathcal{O}}}2-s_{\mathcal{O}}\right)
=\frac 12D_{\mathcal{O}}-\omega _{\mathcal{O}}  \notag \\
\widehat{i}_\Gamma &=&\frac{i_\Gamma }2+s_\Gamma =L+\sum_{\mathcal{O}}\frac{%
i_{\mathcal{O}}}2+s_\Gamma =2L+\sum_{\mathcal{O}}\widehat{i}_{\mathcal{O}}
\end{eqnarray}
where $\omega _{\mathcal{O}}$ is given by (\ref{NC_index}) and we have used (%
\ref{large_N_C_s}) in the last line. With such a modified indices $\widehat{i%
}_{\mathcal{O}}$, $\widehat{i}_\Gamma $ the formula (\ref{power_counting})
has the form
\begin{equation}
\widehat{i}_{{\mathcal{O}_{ct}}}\leq \widehat{i}_\Gamma =2L+\sum_{\mathcal{O}%
}\widehat{i}_{\mathcal{O}}  \label{power_counting_hat}
\end{equation}
The content of this redefinition of $i_{\mathcal{O}}$ is evident: the
operators are now classified according to the combined derivative and large $%
N_C$ expansion according to the counting rules (for pure $\chi PT$
introduced in \cite{Moussallam:1994xp}, \cite{Kaiser:1998ds}, \cite%
{Kaiser:2000gs})
\begin{equation}
p=O(\delta ^{1/2}),\,\,v,a=O(\delta ^{1/2}),\,\,\chi ,\chi ^{+}=O(\delta
),\,\,\frac 1{N_C}=O(\delta )
\end{equation}

In what follows we shall use for the classification of the counterterms and
for the organization of our calculation the index $i_{\mathcal{O}}$ given by
(\ref{index_O}) and (\ref{power_counting}). Note however, that these
formulae similarly to the previous cases, do not have much of physical
content and serve only as a \emph{formal} tool for the proof of the
renormalizability and for the ordering of the counterterms. Namely, the
index $i_\Gamma $ which is by construction related to the superficial degree
of the divergence (and which applies to one-particle irreducible graphs
only) does not reflect the infrared behavior of the (one-particle
irreducible) graph $\Gamma $, rather it refers to its ultraviolet properties.

Note also, that the hierarchy of the contributions to the GF by means of
fixing $i_\Gamma $ for \emph{one-particle irreducible building blocks}%
\footnote{%
That means at a given level $i_{\mathrm{max}}$ we allow for all the graphs
with one-particle irreducible building blocks satisfying $i_\Gamma \leq i_{%
\mathrm{max}}$. This point of view is crucial in order to preserve the
symmetric properties of the corresponding GF.} might appear to be unusual.
For instance, let us assume the antisymmetric tensor formalism. Taking then $%
i_\Gamma =0$ allows only the tree graphs with vertices from pure $O(p^2)$
chiral Lagrangian with resonaces completely decoupled (the only $i_{\mathcal{%
O}}=0$ relevant term with resonance fields is the resonance mass terms) and
such a case is therefore equivalent to the LO $\chi PT$. When fixing $%
i_\Gamma \leq 1$, also the terms linear in the resonance fields (at least in
the antisymmetric tensor formalism, where the linear sources start at $%
O(p^2) $) can be used as the one-particle irreducible building blocks and
again only the tree graphs are in \ the game. However, the resonance
propagator is still derived from the mass terms only. Therefore, summing up
all the tree graphs with resonance internal lines leads then effectively to
the contributions equivalent to those of the pure $O(p^4)$ $\chi PT$
operators with $O(p^4)$ LEC saturated with the resonances in the usual way%
\footnote{%
Here we tacitly assume that the trilinear term without derivatives has been
removed by means of field redefinition, cf. \cite%
{Cirigliano:2006hb,Kampf:2006bn}.}. Because the resonance kinetic term has $%
i_{\mathcal{O}}=2$, the resonances start to propagate only when we take $%
i_\Gamma \leq 2$. At this level we recover the complete NLO $\chi PT$ as a
part of the theory (including the loop graphs) supplemented with tree graphs
built from the free resonance propagators and vertices with $i_{\mathcal{O}%
}\leq 2$. As far as the resonance part of the Lagrangian is
concerned, these vertices coincide with the $O(p^6)$ vertices in
the first type of power counting we have considered in the
beginning of this section (where we assumed $R_{\mu \nu }=O(p^2)$,
see (\ref{fields order I})) but also the four resonance term
without derivatives is allowed. The resonance loops start to
contribute at $i_\Gamma \leq 3$ (with the resonance tadpoles) and
$i_\Gamma \leq 4$ (with the pure resonance bubbles). In order to
renormalize the corresponding divergences, plethora of new
counterterms with increasing number of resonances as well as
increasing order of the chiral building blocks is needed. In what
follows we will encounter graphs with $i_\Gamma =6$ (the mixed GB
and resonance bubbles) for which we will need counterterms up to
the index $i_{\mathcal{O}}\leq 6$.

\section{The self-energies at one loop\label{ch4}}

In this section we present the main result of our paper, namely the one-loop
self-energies within all three formalisms discussed in the Section \ref%
{Propagators and poles} in the chiral limit. In what follows, the
loops are calculated within the dimensional regularization scheme.
In order to avoid complications with the $d-$dimensional Levi-Civita
tensor, we use its simplest variant known as Dimensional reduction,
\emph{i.e.} we perform the four-dimensional tensor algebra first in
order to reduce the tensor integrals to scalar ones and only then we
continue to $d$ dimensions.

\subsection{The Proca field case}

\begin{figure}[tbp]
\par
\centering\includegraphics[scale=1]{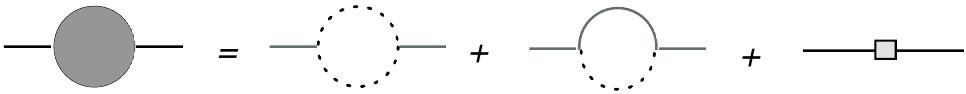} \caption{The
one-loop graphs contributing to the self-energy of the Proca field.
The dotted and full lines corresponds to the Goldstone boson and
resonance propagators respectively. Both one-loop graphs have
$i_\Gamma=6$}\label{vectorg}
\end{figure}

Our starting point is the following Lagrangian for $1^{--}$ resonances \cite%
{Prades:1993ys} (see also \cite{Knecht:2001xc})
\begin{eqnarray}
\mathcal{L}_{V} &=&-\frac{1}{4}\langle \widehat{V}_{\mu \nu }\widehat{V}%
^{\mu \nu }\rangle +\frac{1}{2}M^{2}\langle V_{\mu }V^{\mu }\rangle   \notag
\\
&&-\frac{\mathrm{i}}{2\sqrt{2}}g_{V}\langle \widehat{V}^{\mu \nu }[u_{\mu
},u_{\nu }]\rangle +\frac{1}{2}\sigma _{V}\varepsilon _{\alpha \beta \mu \nu
}\langle \{V^{\alpha },\widehat{V}^{\mu \nu }\}u^{\beta }\rangle +\ldots
\label{Lagrangian V}
\end{eqnarray}%
where we have written down explicitly only the terms contributing to the
self-energy. Originally it was constructed to encompass terms up to the
order $O(p^{6})$ within the chiral power-counting (\ref{fields order I}, \ref%
{mass order I}). In the large $N_{C}$ limit the couplings behave as $%
g_{V}=O(N_{C}^{1/2})$ and $\sigma _{V}=O(N_{C}^{-1/2})$. This
suggests that the odd intrinsic parity terms are of higher order,
however the vertices relevant for our calculations have the same
order $O(N_{C}^{-1})$ in both cases due to the presence of the
factor $1/F=O(N_{C}^{-1/2})$ which accompanies each Goldstone bosons
field. In the above Lagrangian the operators shown explicitly have
no more than two derivatives and two resonance fields. Therefore,
because the interaction terms are $O(p^{2})$ we would expect (by
analogy with the $\chi $PT power counting) the counterterms
necessary to cancel the divergencies of the one-loop graphs to have
four derivatives at most. However, the nontrivial structure of the
free resonance propagator (namely the presence of the $P_{L}$ part)
results in the failure
of this naive expectation. In fact, according to (\ref{index_O}) and (\ref%
{power_counting}), the operators in (\ref{Lagrangian V}) have index up to $%
i_{\mathcal{O}}\leq 2$, whereas the Feynman graphs corresponding to
the self-energies $\Sigma _{L,T}$ (depicted in Fig. \ref{vectorg})
 $i_{\Gamma }=6$. In order to cancel the
infinite part of the loops we have therefore to introduce a set of
counterterms with two resonance fields and indices\footnote{%
Note that, for these counterterms the index $i_{\mathcal{O}}$ coincides with
the usual chiral order $D_{\mathcal{O}}$.} $i_{\mathcal{O}}\leq 6$, namely
\begin{eqnarray}
\mathcal{L}_{V}^{ct} &=&\frac{1}{2}M^{2}Z_{M}\langle V_{\mu }V^{\mu }\rangle
\mathcal{+}\frac{Z_{V}}{4}\langle \hat{V}_{\mu \nu }\hat{V}^{\mu \nu
}\rangle -\frac{Y_{V}}{2}\langle (D_{\mu }V^{\mu })^{2}\rangle   \notag \\
&&+\frac{X_{V1}}{4}\langle \{D_{\alpha },D_{\beta }\}V_{\mu }\{D^{\alpha
},D^{\beta }\}V^{\mu }\rangle +\frac{X_{V2}}{4}\langle \{D_{\alpha
},D_{\beta }\}V_{\mu }\{D^{\alpha },D^{\mu }\}V^{\beta }\rangle   \notag \\
&&+\frac{X_{V3}}{4}\langle \{D_{\alpha },D_{\beta }\}V^{\beta }\{D^{\alpha
},D^{\mu }\}V_{\mu }\rangle +\frac{X_{V4}}{2}\langle D^{2}V_{\mu }\{D^{\mu
},D^{\beta }\}V_{\beta }\rangle +X_{V5}\langle D^{2}V_{\mu }D^{2}V^{\mu
}\rangle   \notag \\
&&+\mathcal{L}_{V}^{ct(6)}.  \label{ctV}
\end{eqnarray}%

Here the last term accumulates the operators with six derivatives ($i_{%
\mathcal{O}}=6$), which we do not write down explicitly. The bare couplings
are split into a finite part renormalized at a scale ${\mu }$ and a
divergent part. The infinite parts of the bare couplings are fixed according
to
\begin{eqnarray*}
Z_{M} &=&Z_{M}^{r}(\mu ) \\
Z_{V} &=&Z_{V}^{r}(\mu )+\frac{80}{3}\left( \frac{M}{F}\right) ^{2}\sigma
_{V}^{2}\lambda _{\infty } \\
X_{V} &=&X_{V}^{r}(\mu )-\frac{80}{9}\left( \frac{M}{F}\right) ^{2}\sigma
_{V}^{2}\frac{1}{M^{2}}\lambda _{\infty } \\
Y_{V} &=&Y_{V}^{r}(\mu ) \\
X_{V}^{^{\prime }} &=&X_{V}^{^{\prime }r}(\mu )
\end{eqnarray*}%
where
\begin{eqnarray*}
X_{V}^{r}(\mu ) &=&X_{V1}^{r}(\mu )+X_{V5}^{r}(\mu ) \\
X_{V}^{^{\prime }r}(\mu ) &=&X_{V1}^{r}(\mu )+X_{V2}^{r}(\mu
)+X_{V3}^{r}(\mu )+X_{V4}^{r}(\mu )+X_{V5}^{r}(\mu ),
\end{eqnarray*}%
and
\begin{equation*}
\lambda _{\infty }=\frac{{\mu }^{d-4}}{(4\pi )^{2}}\left( \frac{1}{d-4}-%
\frac{1}{2}(\ln 4\pi -\gamma +1)\right) .
\end{equation*}%
The result can be written in the form (in the following formulae $x=s/M^{2}$%
)
\begin{eqnarray*}
\Sigma _{T}^{r}(s) &=&M^{2}\left( \frac{M}{4\pi F}\right) ^{2}\left[
\sum_{i=0}^{3}{\alpha }_{i}x^{i}-\frac{1}{2}g_{V}^{2}\left( \frac{M}{F}%
\right) ^{2}x^{3}\widehat{B}(x)-\frac{40}{9}\sigma _{V}^{2}(x-1)^{2}x%
\widehat{J}(x)\right]  \\
\Sigma _{L}^{r}(s) &=&M^{2}\left( \frac{M}{4\pi F}\right) ^{2}\sum_{i=0}^{3}{%
\beta }_{i}x^{i}
\end{eqnarray*}%
In the above formulae ${\alpha }_{i}$ and ${\beta }_{i}$ can be expressed in
terms of the renormalization scale independent combinations of the
counterterm couplings and $\chi $logs. The explicit formulae are collected
in the Appendix \ref{appendix Proca}. The functions $\widehat{B}(x)$ and $%
\widehat{J}(x)$ correspond to the vacuum bubbles with two Goldstone boson
lines or with one Goldstone boson and one resonance line respectively. On
the first (physical) sheet,
\begin{eqnarray}
\widehat{B}(x) &=&\widehat{B}^{I}(x)=1-\ln (-x)  \notag \\
\widehat{J}(x) &=&\widehat{J}^{I}(x)=\frac{1}{x}\left[ 1-\left( 1-\frac{1}{x}%
\right) \ln (1-x)\right],  \label{loop_functions}
\end{eqnarray}%
where we take the principal branch of the logarithm ($-\pi <\mathrm{{Im}\ln
x\leq \pi }$) with cut for $x<0$. On the second sheet we have then $\widehat{%
B}^{II}(x-\mathrm{i}0)=\widehat{B}^{I}(x+\mathrm{i}0)=\widehat{B}^{I}(x-%
\mathrm{i}0)+2\pi \mathrm{i}$ and similarly for $\widehat{J}(x)$, therefore
\begin{eqnarray}
\widehat{B}^{II}(x) &=&\widehat{B}^{I}(x)+2\pi \mathrm{i}  \notag \\
\widehat{J}^{II}(x) &=&\widehat{J}^{I}(x)+\frac{2\pi \mathrm{i}}{x}\left( 1-%
\frac{1}{x}\right) .  \label{loop_functions_II}
\end{eqnarray}%
The equation for the pole in the $1^{--}$ channel
\begin{equation*}
s-M^{2}-\Sigma _{T}(s)=0
\end{equation*}%
has a perturbative solution corresponding to the original $1^{--}$ vector
resonance, which develops a mass correction and a finite width of the order $%
O(1/N_{C})$ due to the loops. This solution can be written in the form $%
\overline{s}=M_{\mathrm{phys}}^{2}-\mathrm{i}M_{\mathrm{phys}}\Gamma _{%
\mathrm{phys}}$ where
\begin{eqnarray*}
M_{\mathrm{phys}}^{2} &=&M^{2}+\mathrm{Re}\Sigma _{T}(M^{2})=M^{2}\left[
1+\left( \frac{M}{4\pi F}\right) ^{2}\left( \sum_{i=0}^{3}{\alpha }_{i}-%
\frac{1}{2}g_{V}^{2}\left( \frac{M}{F}\right) ^{2}\right) \right]  \\
M_{\mathrm{phys}}\Gamma _{\mathrm{phys}} &=&-\mathrm{Im}\Sigma
_{T}(M^{2})=M^{2}\left( \frac{M}{4\pi F}\right) ^{2}\frac{1}{2}%
g_{V}^{2}\left( \frac{M}{F}\right) ^{2}\pi
\end{eqnarray*}%
which gives a constraint on the values of ${\alpha }_{i}$'s
\begin{equation*}
M_{\mathrm{phys}}^{2}+\frac{1}{\pi }M_{\mathrm{phys}}\Gamma _{\mathrm{phys}%
}=M^{2}\left[ 1+\left( \frac{M}{4\pi F}\right) ^{2}\sum_{i=0}^{3}{\alpha }%
_{i}\right]
\end{equation*}%
and in terms of the physical mass and the width we have then
\begin{eqnarray*}
\Sigma _{T}^{r}(s) &=&M_{\mathrm{phys}}^{2}\left( \frac{M_{\mathrm{phys}}}{%
4\pi F}\right) ^{2}\left[ \sum_{i=0}^{3}{\alpha }_{i}x^{i}-\frac{40}{9}%
\sigma _{V}^{2}(x-1)^{2}x\widehat{J}(x)\right] -\frac{1}{\pi }M_{\mathrm{phys%
}}\Gamma _{\mathrm{phys}}x^{3}\widehat{B}(x) \\
\Sigma _{L}^{r}(s) &=&M_{\mathrm{phys}}^{2}\left( \frac{M_{\mathrm{phys}}}{%
4\pi F}\right) ^{2}\sum_{i=0}^{3}{\beta }_{i}x^{i}.
\end{eqnarray*}%
For further numerical estimates it is convenient to adopt the on shell
renormalization prescription demanding $M^{2}=M_{\mathrm{phys}}^{2}$ and
also to identify $F$ with $F_{\pi }$ (because $F=F_{\pi }$ at the leading
order). This gives
\begin{equation*}
\frac{1}{\pi }\frac{\Gamma _{\mathrm{phys}}}{M_{\mathrm{phys}}}=\left( \frac{%
M}{4\pi F}\right) ^{2}\sum_{i=0}^{3}{\alpha }_{i}
\end{equation*}%
and, introducing parameters $a_{i}$, $b_{i}$ with natural size $O(1)$%
\begin{eqnarray*}
a_{i} &=&\pi \frac{M_{\mathrm{phys}}}{\Gamma _{\mathrm{phys}}}\left( \frac{%
M_{\mathrm{phys}}}{4\pi F_{\pi }}\right) ^{2}{\alpha }_{i}\sim O(1) \\
b_{i} &=&\pi \frac{M_{\mathrm{phys}}}{\Gamma _{\mathrm{phys}}}\left( \frac{%
M_{\mathrm{phys}}}{4\pi F_{\pi }}\right) ^{2}{\beta }_{i}\sim O(1)
\end{eqnarray*}%
we get in this scheme for $\sigma _{T,L}^{r}(x)=M_{\mathrm{phys}}^{-2}\Sigma
_{T,L}^{r}(M_{\mathrm{phys}}^{2}x)$
\begin{eqnarray*}
\sigma _{T}^{r}(x) &=&\frac{1}{\pi }\frac{\Gamma _{\mathrm{phys}}}{M_{%
\mathrm{phys}}}\left( 1+\sum_{i=1}^{3}a_{i}(x^{i}-1)-x^{3}\widehat{B}%
(x)\right) -\frac{40}{9}\left( \frac{M_{\mathrm{phys}}}{4\pi F_{\pi }}%
\right) ^{2}\sigma _{V}^{2}(x-1)^{2}x\widehat{J}(x) \\
\sigma _{L}^{r}(x) &=&\frac{1}{\pi }\frac{\Gamma _{\mathrm{phys}}}{M_{%
\mathrm{phys}}}\sum_{i=0}^{3}b_{i}x^{i}.
\end{eqnarray*}

\subsection{The antisymmetric tensor case}

We start with the following Lagrangian for $1^{--}$ resonances (here only
the terms relevant for the one-loop selfenergy are shown explicitly)
\begin{eqnarray}
\mathcal{L}_{R} &=&-\frac{1}{2}\langle D_{\mu }R^{\mu \nu }D^{\alpha
}R_{\alpha \nu }\rangle +\frac{1}{4}M^{2}\langle R^{\mu \nu }R_{\mu \nu
}\rangle  \notag \\
&+&\frac{iG_{V}}{2\sqrt{2}}\langle R^{\mu \nu }[u_{\mu },u_{\nu }]\rangle
+d_{1}\varepsilon _{\mu \nu \alpha \sigma }\langle D_{\beta }u^{\sigma
}\{R^{\mu \nu },R^{\alpha \beta }\}\rangle  \notag \\
&+&d_{3}\varepsilon _{\rho \sigma \mu \lambda }\langle u^{\lambda }\{D_{\nu
}R^{\mu \nu },R^{\rho \sigma }\}\rangle +d_{4}\varepsilon _{\rho \sigma \mu
\alpha }\langle u_{\nu }\{D^{\alpha }R^{\mu \nu },R^{\rho \sigma }\}\rangle
\notag \\
&+&\mathrm{i}\lambda ^{VVV}\langle R_{\mu \nu }R^{\mu \rho }R^{\nu \sigma
}\rangle +\ldots  \label{Lagrangian T}
\end{eqnarray}%
Note that, in the large $N_{C}$ limit the coupling $G_{V}$ behaves as $%
G_{V}=O(N_{C}^{1/2})$, whereas $d_{i}=O(1)$ and $\lambda
^{VVV}=O(N_{C}^{-1/2}) $. Apparently the intrinsic parity odd part
and the trilinear resonance coupling are thus of higher order.
However, the
trilinear vertices contributing to the one-loop self-energies are $%
O(N_{C}^{-1/2})$ in both cases due to the appropriate power of $%
1/F=O(N_{C}^{-1/2})$ accompanying $u_{\alpha }$. Therefore, the operators
with two and three resonance fields cannot be got rid of using the large $%
N_{C}$ arguments. Also nonzero $d_{i}$ are required in order to satisfy the
OPE constraints for VVP GF at the LO; especially for $d_{3}$ we get \cite%
{VVP}
\begin{equation}
d_{3}=-\frac{N_{C}}{64\pi ^{2}}\left( \frac{M}{F_{V}}\right) ^{2}+\frac{1}{8}%
\left( \frac{F}{F_{V}}\right) ^{2}  \label{d_3}
\end{equation}%
where $F_{V}$ is the strength of the resonance coupling to the vector
current.

\begin{figure}[tbp]
\par
\begin{center}
\epsfig{width=1\textwidth,file=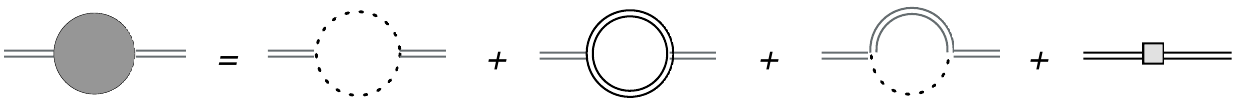}
\end{center}
\caption{The one-loop graphs contributing to the self-energy of the
antisymmetric tensor field. The dotted and double lines correspond
to the Goldstone boson and resonance propagators respectively. The
GB and pure resonance bubbles have $i_\Gamma=4$, while the ``mixed''
one has $i_\Gamma=6$}\label{tensor_graphs}
\end{figure}

The Lagrangian (\ref{Lagrangian T}) includes terms up to the index $i_{%
\mathcal{O}}\leq 2$. The one-loop Feynman graphs contributing to
the self-energy are depicted in Fig. \ref{tensor_graphs}. The
first two bubbles include only interaction vertices with
$i_{\cal{O}}=1$ and therefore they have indices $i_{\Gamma }=4$
while the third one is built from vertices with $i_{\cal{O}}=2$
and has the index $i_{\Gamma }=6$.
In order to cancel the infinite part of the loops we have then to
add counterterms with indices $i_{\mathcal{O}}\leq 6$, namely the
following set
\begin{eqnarray}
\mathcal{L}_{R}^{ct} &=&\frac{1}{4}M^{2}Z_{M}\langle R^{\mu \nu }R_{\mu \nu
}\rangle +\frac{1}{2}Z_{R}\langle D_{\alpha }R^{{\alpha }{\mu }}D^{\beta }R_{%
{\beta }{\mu }}\rangle +\frac{1}{4}Y_{R}\langle D_{\alpha }R^{{\mu }{\nu }%
}D^{\alpha }R_{{\mu }{\nu }}\rangle  \notag \\
&&+\frac{1}{4}X_{R1}\langle D^{2}R^{{\mu }{\nu }}\{D_{\nu },D^{\sigma }\}R_{{%
\mu }{\sigma }}\rangle +\frac{1}{8}X_{R2}\langle \{D_{\nu },D_{\alpha }\}R^{{%
\mu }{\nu }}\{D^{\sigma },D^{\alpha }\}R_{{\mu }{\sigma }}\rangle  \notag \\
&&+\frac{1}{8}X_{R3}\langle \{D^{\sigma },D^{\alpha }\}R^{{\mu }{\nu }%
}\{D_{\nu },D_{\alpha }\}R_{{\mu }{\sigma }}\rangle  \notag \\
&&+\frac{1}{4}W_{R1}\langle D^{2}R^{{\mu }{\nu }}D^{2}R_{{\mu }{\nu }%
}\rangle +\frac{1}{16}W_{R2}\langle \{D^{{\alpha }},D^{\beta }\}R^{{\mu }{%
\nu }}\{D_{{\alpha }},D_{\beta }\}R_{{\mu }{\nu }}\rangle  \notag \\
&&+\mathcal{L}_{R}^{ct(6)},  \label{ctR}
\end{eqnarray}%
where the last term accumulates the operators with six derivatives ($i_{%
\mathcal{O}}=6)$, which we do not write down explicitly. The infinite parts
of the bare couplings are fixed as
\begin{eqnarray*}
Z_{M} &=&Z_{M}^{r}({\mu })+\frac{80}{3}\left( \frac{M}{F}\right)
^{2}d_{1}^{2}\lambda _{\infty }-60\left( \frac{\lambda ^{VVV}}{M}\right)
^{2}\lambda _{\infty } \\
Z_{R} &=&Z_{R}^{r}({\mu })+\frac{40}{9}\left( \frac{M}{F}\right) ^{2}\frac{1%
}{M^{2}}(12d_{1}(d_{3}+d_{4})-d_{3}^{2}-9d_{4}^{2}+6d_{3}d_{4})\lambda
_{\infty } \\
&&+80\left( \frac{\lambda ^{VVV}}{M}\right) ^{2}\frac{1}{M^{2}}\lambda
_{\infty } \\
Y_{R} &=&Y_{R}^{r}({\mu })+\frac{40}{9}\left( \frac{M}{F}\right) ^{2}\frac{1%
}{M^{2}}(6d_{1}^{2}-12d_{1}(d_{3}+d_{4})+5d_{3}^{2}+9d_{4}^{2}-6d_{3}d_{4})%
\lambda _{\infty } \\
&&-40\left( \frac{\lambda ^{VVV}}{M}\right) ^{2}\frac{1}{M^{2}}\lambda
_{\infty } \\
X_{R} &=&X_{R}^{r}({\mu })+\frac{40}{9}\left( \frac{M}{F}\right) ^{2}\frac{1%
}{M^{2}}(d_{3}^{2}-6d_{3}d_{4}+5d_{4}^{2})\lambda _{\infty }-\left( \frac{%
G_{V}}{F}\right) ^{2}\frac{1}{M^{2}}\lambda _{\infty } \\
W_{R} &=&W_{R}^{r}({\mu })+\frac{40}{9}\left( \frac{M}{F}\right) ^{2}\frac{1%
}{M^{4}}(d_{3}^{2}+6d_{3}d_{4}-5d_{4}^{2})\lambda _{\infty }-10\left( \frac{%
\lambda ^{VVV}}{M}\right) ^{2}\frac{1}{M^{4}}\lambda _{\infty }
\end{eqnarray*}%
where
\begin{eqnarray*}
X_{R}^{r}({\mu }) &=&X_{R1}^{r}({\mu })+X_{R2}^{r}({\mu })+X_{R3}^{r}({\mu })
\\
W_{R}^{r}({\mu }) &=&W_{R1}^{r}({\mu })+W_{R2}^{r}({\mu }).
\end{eqnarray*}%
An explicit calculation gives for the renormalized self-energies (in the
following formulae $x=s/M^{2}$)
\begin{eqnarray*}
\Sigma _{L}^{r}(s) &=&M^{2}\left( \frac{M}{4\pi F}\right) ^{2}\left[
\sum_{i=0}^{3}{\alpha }_{i}x^{i}-\left( \frac{1}{2}\left( \frac{G_{V}}{F}%
\right) ^{2}x^{2}\widehat{B}(x)+\frac{40}{9}d_{3}^{2}(x^{2}-1)^{2}\widehat{J}%
(x)\right) \right] \\
&&-5\left( \frac{\lambda ^{VVV}}{4\pi }\right) ^{2}(x-4)(x+2)\overline{J}(x)
\\
\Sigma _{T}^{r}(s) &=&M^{2}\left( \frac{M}{4\pi F}\right) ^{2}\left[
\sum_{i=0}^{3}{\beta }_{i}x^{i}+\frac{20}{9}\left(
2d_{3}^{2}+(d_{3}^{2}+6d_{3}d_{4}+d_{4}^{2})x+2d_{4}^{2}x^{2}\right)
(x-1)^{2}\widehat{J}(x)\right] \\
&&+5\left( \frac{\lambda ^{VVV}}{4\pi }\right) ^{2}(x^{2}-2x+4)\overline{J}%
(x).
\end{eqnarray*}%
Here the functions $\widehat{B}(x)$ and $\widehat{J}(x)$ are same as in the
previous subsection and $\overline{J}(x)$ is given on the physical sheet by
\begin{equation*}
\overline{J}(x)=\overline{J}^{I}(x)=2+\sqrt{1-\frac{4}{x}}\ln \frac{\sqrt{1-%
\frac{4}{x}}-1}{\sqrt{1-\frac{4}{x}}+1}.
\end{equation*}%
with the same branch of the logarithm as before. On the second sheet we have
$\overline{J}^{II}(x-\mathrm{i}0)=\overline{J}^{I}(x+\mathrm{i}0)=\overline{J%
}^{I}(x-\mathrm{i}0)+2\mathrm{i}\pi \sqrt{1-4/x}$ and therefore
\begin{equation*}
\overline{J}^{II}(x)=\overline{J}^{I}(x)+2\mathrm{i}\pi \sqrt{1-\frac{4}{x}}.
\end{equation*}%
The explicit dependence of the renormalization scale invariant polynomial
parameters ${\alpha }_{i}$ and ${\beta }_{i}$ on the counterterm couplings
and $\chi $logs are given in the Appendix \ref{appendix tensor}.

In order to simplify the following discussion we put $\lambda ^{VVV}=0$ in
the rest of this subsection. This is in accord with the fact, that the
corresponding trilinear interaction term can be effectively removed by
resonance field redefinition \cite{Cirigliano:2006hb}. Also, the
two-resonance cut starts at $x=4$ which is far from the region we are
interested in. Here the effect of the resonance bubble can be effectively
absorbed to the polynomial part of the self-energies.

The equation for the propagator poles in the $1^{--}$ channel
\begin{equation*}
s-M^{2}-\Sigma _{L}(s)=0
\end{equation*}%
has an approximative perturbative solution corresponding to the original $%
1^{--}$ vector resonance, which develops a mass correction and a finite
width of the order $O(1/N_{C})$ due to the loops. This solution can be
written in the form $\overline{s}=M_{\mathrm{phys}}^{2}-\mathrm{i}M_{\mathrm{%
phys}}\Gamma _{\mathrm{phys}}$ where
\begin{eqnarray*}
M_{\mathrm{phys}}^{2} &=&M^{2}+\mathrm{Re}\Sigma _{L}(M^{2})=M^{2}\left[
1+\left( \frac{M}{4\pi F}\right) ^{2}\left( \sum_{i=0}^{3}{\alpha }_{i}-%
\frac{1}{2}\left( \frac{G_{V}}{F}\right) ^{2}\right) \right] \\
M_{\mathrm{phys}}\Gamma _{\mathrm{phys}} &=&-\mathrm{Im}\Sigma
_{L}(M^{2})=M^{2}\left( \frac{M}{4\pi F}\right) ^{2}\frac{1}{2}\left( \frac{%
G_{V}}{F}\right) ^{2}\pi \mathrm{,}
\end{eqnarray*}%
which gives a constraint on the values of ${\alpha }_{i}$'s
\begin{equation*}
M_{\mathrm{phys}}^{2}+\frac{1}{\pi }M_{\mathrm{phys}}\Gamma _{\mathrm{phys}%
}=M^{2}\left( 1+\frac{1}{(4\pi )^{2}}\left( \frac{M}{F}\right)
^{2}\sum_{i=0}^{3}{\alpha }_{i}\right) .
\end{equation*}%
This allows us to re-parameterize perturbatively $\Sigma _{L}(s)$ in terms of $%
M_{\mathrm{phys}}$ and $\Gamma _{\mathrm{phys}}$ as
\begin{eqnarray*}
\Sigma _{L}^{r}(s) &=&M_{\mathrm{phys}}^{2}\left( \frac{M_{\mathrm{phys}}}{%
4\pi F}\right) ^{2}\left[ \sum_{i=0}^{3}{\alpha }_{i}x^{i}-\frac{40}{9}%
d_{3}^{2}(x^{2}-1)^{2}\widehat{J}(x)\right] -\frac{1}{\pi }\Gamma _{\mathrm{%
phys}}M_{\mathrm{phys}}x^{2}\widehat{B}(x) \\
\Sigma _{T}^{r}(s) &=&M_{\mathrm{phys}}^{2}\left( \frac{M_{\mathrm{phys}}}{%
4\pi F}\right) ^{2}\left[ \sum_{i=0}^{3}{\beta }_{i}x^{i}+\frac{20}{9}\left(
2d_{3}^{2}+(d_{3}^{2}+6d_{3}d_{4}+d_{4}^{2})x+2d_{4}^{2}x^{2}\right)
(x-1)^{2}\widehat{J}(x)\right] .
\end{eqnarray*}

As for the Proca field case, within the on shell renormalization
prescription $M^{2}=M_{\mathrm{phys}}^{2}$ and we get a constraint
\begin{equation}
\frac{1}{\pi }\frac{\Gamma _{\mathrm{phys}}}{M_{\mathrm{phys}}}=\left( \frac{%
M_{\mathrm{phys}}}{4\pi F_{\pi }}\right) ^{2}\sum_{i=0}^{3}{\alpha }_{i}.
\label{alpha_size}
\end{equation}%
As a result, we can re-write the self-energy (in the units of $M_{\mathrm{%
phys}}^{2}$, \emph{i.e.} as in the previous section $\sigma _{T,L}^{r}(x)=M_{%
\mathrm{phys}}^{-2}\Sigma _{T,L}^{r}(M_{\mathrm{phys}}^{2}x)$ in what
follows) in the form
\begin{equation*}
{\sigma }_{L}^{r}(x)=\frac{1}{\pi }\frac{\Gamma _{\mathrm{phys}}}{M_{\mathrm{%
phys}}}\left[ 1-x^{2}\widehat{B}(x)+\sum_{i=1}^{3}a_{i}(x^{i}-1)\right] -%
\frac{40}{9}\left( \frac{M_{\mathrm{phys}}}{4\pi F_{\pi }}\right)
^{2}d_{3}^{2}(x^{2}-1)^{2}\widehat{J}(x)
\end{equation*}%
using the re-scaled parameters $a_{i}$ with a natural size $O(1)$
\begin{equation*}
a_{i}=\pi \frac{M_{\mathrm{phys}}}{\Gamma _{\mathrm{phys}}}\left( \frac{M_{%
\mathrm{phys}}}{4\pi F_{\pi }}\right) ^{2}{\alpha }_{i}\sim O(1).
\end{equation*}%
So that the $\Sigma _{L}^{r}(s)$ has four independent parameters ${\alpha }%
_{i}$, $i=1,2,3$ and $d_{3}$. Similarly, $\Sigma _{T}^{r}(s)$ can be written
in this scheme in terms of six independent dimensionless parameters $b_{i}$
, $d_{3}$ and $\gamma $
\begin{eqnarray*}
b_{i} &=&\pi \frac{M_{\mathrm{phys}}}{\Gamma _{\mathrm{phys}}}\left( \frac{%
M_{\mathrm{phys}}}{4\pi F_{\pi }}\right) ^{2}{\beta }_{i}\sim O(1) \\
\gamma &=&d_{4}/d_{3}\sim O(1)
\end{eqnarray*}%
as
\begin{equation*}
{\sigma }_{T}^{r}(x)=\frac{1}{\pi }\frac{\Gamma _{\mathrm{phys}}}{M_{\mathrm{%
phys}}}\sum_{i=0}^{3}b_{i}x^{i}+\frac{20}{9}\left( \frac{M_{\mathrm{phys}}}{%
4\pi F_{\pi }}\right) ^{2}d_{3}^{2}\left( 2+(1+6\gamma +\gamma
^{2})x+2\gamma ^{2}x^{2}\right) (x-1)^{2}\widehat{J}(x).
\end{equation*}%
In order to satisfy the OPE constraints for $VVP$ correlator \cite{VVP}%
, we have to put further (according to (\ref{d_3}))
\begin{equation}
d_{3}=-\frac{3}{4}\left( \frac{M_{\mathrm{phys}}}{4\pi F_{\pi }}\right)
^{2}\left( \frac{F_{\pi }}{F_{V}}\right) ^{2}\left[ 1-\frac{1}{6}\left(
\frac{4\pi F_{\pi }}{M_{\mathrm{phys}}}\right) ^{2}\right]  \label{d_3_1}
\end{equation}%
which reduces the number of the independent parameters for ${\sigma }%
_{L}^{r}(x)$ and ${\sigma }_{T}^{r}(x)$ to three and five respectively.

\subsection{The first order formalism}

\begin{figure}[t]
\par
\begin{center}
\epsfig{width=0.5\textwidth,file=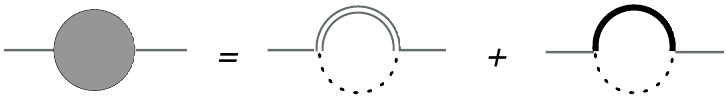}
\end{center}
\caption{The extra one-loop graphs contributing to the vector field
self-energy of in the first order formalism. The dotted and double
lines corresponds to the Goldstone boson and antisymmetric tensor
field propagators respectively, the thick line stay symbolically for
the ``mixed'' propagator.} \label{tensor_graphs vector}
\end{figure}

\begin{figure}[h]
\par
\begin{center}
\epsfig{width=0.5\textwidth,file=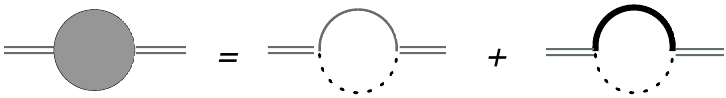}
\end{center}
\caption{The extra one-loop graphs contributing to the antisymmetric
tensor field self-energy in the first order formalism. The meaning
of the various types of lines is the same as in the previous
figures.} \label{tensor_graphs tensor}
\end{figure}

In this case, the interaction part of the Lagrangian describing $1^{--}$
resonances collects all the terms from the previous two formalisms. It
contains also one extra term which mixes the the fields $R_{\mu \nu }$ and $%
V_{\alpha }$
\begin{eqnarray*}
\mathcal{L}_{RV} &=&\frac{1}{2}M^{2}\langle V_{\mu }V^{\mu }\rangle +\frac{1%
}{4}M^{2}\langle R^{\mu \nu }R_{\mu \nu }\rangle -\frac{1}{2}\langle R^{\mu
\nu }\widehat{V}_{\mu \nu }\rangle \\
&&-\frac{\mathrm{i}}{2\sqrt{2}}g_{V}\langle \widehat{V}^{\mu \nu }[u_{\mu
},u_{\nu }]\rangle +\frac{1}{2}\sigma _{V}\varepsilon _{\alpha \beta \mu \nu
}\langle \{V^{\alpha },\widehat{V}^{\mu \nu }\}u^{\beta }\rangle \\
&&+\frac{iG_{V}}{2\sqrt{2}}\langle R^{\mu \nu }[u_{\mu },u_{\nu }]\rangle
+d_{1}\varepsilon _{\mu \nu \alpha \sigma }\langle D_{\beta }u^{\sigma
}\{R^{\mu \nu },R^{\alpha \beta }\}\rangle \\
&&+d_{3}\varepsilon _{\rho \sigma \mu \lambda }\langle u^{\lambda }\{D_{\nu
}R^{\mu \nu },R^{\rho \sigma }\}\rangle +d_{4}\varepsilon _{\rho \sigma \mu
\alpha }\langle u_{\nu }\{D^{\alpha }R^{\mu \nu },R^{\rho \sigma }\}\rangle
\\
&&+\frac{1}{2}M\sigma _{RV}\varepsilon _{\alpha \beta \mu \nu }\langle
\{V^{\alpha },R^{\mu \nu }\}u^{\beta }\rangle +\mathrm{i}\lambda
^{VVV}\langle R_{\mu \nu }R^{\mu \rho }R^{\nu \sigma }\rangle +\ldots
\end{eqnarray*}%
Because the free diagonal propagators are the same as in the pure
Proca or antisymmetric tensor cases, all the graphs depicted in the
Figs. \ref{vectorg}, \ref{tensor_graphs} contribute also here to the
diagonal self-energies $\Sigma _{RR}$ and $\Sigma _{VV}$. The mixed
vertex and mixed
propagator generate additional graphs contributing to $\Sigma _{RR}$, $%
\Sigma _{VV}$ and $\Sigma _{RV}$ which are depicted in the Figs.
\ref{tensor_graphs vector}, \ref{tensor_graphs tensor} and
\ref{tensor_graphs mixed} respectively (in the latter case also the
GB bubble contributes).

\begin{figure}[h]
\par
\begin{center}
\epsfig{width=1\textwidth,file=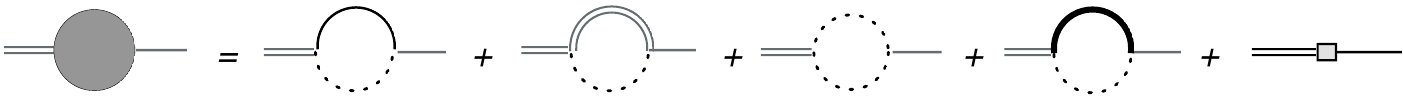}
\end{center}
\caption{The one-loop graphs contributing to the ``mixed''
self-energy  in the first order formalism.} \label{tensor_graphs
mixed}
\end{figure}

Similarly, the set of counterterms necessary to renormalize the infinities
includes all the terms (\ref{ctV}) and (\ref{ctR}) and additional mixed
terms
\begin{eqnarray*}
\mathcal{L}_{RV}^{ct} &=&\frac{1}{2}M^{2}Z_{MV}\langle V_{\mu }V^{\mu
}\rangle \mathcal{+}\frac{Z_{V}}{4}\langle \hat{V}_{\mu \nu }\hat{V}^{\mu
\nu }\rangle -\frac{Y_{V}}{2}\langle (D_{\mu }V^{\mu })^{2}\rangle \\
&&+\frac{X_{V1}}{4}\langle \{D_{\alpha },D_{\beta }\}V_{\mu }\{D^{\alpha
},D^{\beta }\}V^{\mu }\rangle +\frac{X_{V2}}{4}\langle \{D_{\alpha
},D_{\beta }\}V_{\mu }\{D^{\alpha },D^{\mu }\}V^{\beta }\rangle \\
&&+\frac{X_{V3}}{4}\langle \{D_{\alpha },D_{\beta }\}V^{\beta }\{D^{\alpha
},D^{\mu }\}V_{\mu }\rangle +\frac{X_{V4}}{2}\langle D^{2}V_{\mu }\{D^{\mu
},D^{\beta }\}V_{\beta }\rangle +X_{V5}\langle D^{2}V_{\mu }D^{2}V^{\mu
}\rangle \\
&&+\frac{1}{4}M^{2}Z_{MR}\langle R^{\mu \nu }R_{\mu \nu }\rangle +\frac{1}{2}%
Z_{R}\langle D_{\alpha }R^{{\alpha }{\mu }}D^{\beta }R_{{\beta }{\mu }%
}\rangle +\frac{1}{4}Y_{R}\langle D_{\alpha }R^{{\mu }{\nu }}D^{\alpha }R_{{%
\mu }{\nu }}\rangle \\
&&+\frac{1}{4}X_{R1}\langle D^{2}R^{{\mu }{\nu }}\{D_{\nu },D^{\sigma }\}R_{{%
\mu }{\sigma }}\rangle +\frac{1}{8}X_{R2}\langle \{D_{\nu },D_{\alpha }\}R^{{%
\mu }{\nu }}\{D^{\sigma },D^{\alpha }\}R_{{\mu }{\sigma }}\rangle \\
&&+\frac{1}{8}X_{R3}\langle \{D^{\sigma },D^{\alpha }\}R^{{\mu }{\nu }%
}\{D_{\nu },D_{\alpha }\}R_{{\mu }{\sigma }}\rangle \\
&&+\frac{1}{4}W_{R1}\langle D^{2}R^{{\mu }{\nu }}D^{2}R_{{\mu }{\nu }%
}\rangle +\frac{1}{16}W_{R2}\langle \{D^{{\alpha }},D^{\beta }\}R^{{\mu }{%
\nu }}\{D_{{\alpha }},D_{\beta }\}R_{{\mu }{\nu }} \\
&&-\frac{1}{2}Z_{RV}M\langle R^{\mu \nu }\widehat{V}_{\mu \nu }\rangle +%
\frac{1}{2}X_{RV1}M\langle D^{\alpha }R^{\mu \nu }D_{\alpha }\widehat{V}%
_{\mu \nu }\rangle +\frac{1}{2}X_{RV2}M\langle D_{\mu }R^{\mu \nu }D^{\sigma
}\widehat{V}_{\sigma \nu } \\
&&+\mathcal{L}_{RV}^{ct(6)}\label{ctRV}
\end{eqnarray*}
Now the infinite parts of the bare couplings have to be fixed as follows
\begin{eqnarray*}
Z_{RV} &=&Z_{RV}^{r}(\mu )-\frac{20}{3}\left( \frac{M}{F}\right) ^{2}(\sigma
_{RV}+2\sigma _{V})(2d_{1}-\sigma _{RV})\lambda _{\infty } \\
X_{RV} &=&X_{RV}^{r}(\mu )-\frac{20}{9}\left( \frac{M}{F}\right) ^{2}\frac{1%
}{M^{2}}(\sigma _{RV}+2\sigma _{V})(4d_{3}+\sigma _{RV})\lambda _{\infty } \\
Z_{MV} &=&Z_{MV}^{r}(\mu ) \\
Z_{V} &=&Z_{V}^{r}(\mu )+\frac{20}{3}\left( \frac{M}{F}\right) ^{2}\left(
\sigma _{RV}(\sigma _{RV}+2\sigma _{V})+4\sigma _{V}^{2}\right) \lambda
_{\infty } \\
X_{V} &=&X_{V}^{r}(\mu )-\frac{20}{9}\left( \frac{M}{F}\right) ^{2}\frac{1}{%
M^{2}}\left( \sigma _{RV}(\sigma _{RV}+2\sigma _{V})+4\sigma _{V}^{2}\right)
\lambda _{\infty } \\
Y_{V} &=&Y_{V}^{r}(\mu ) \\
X_{V}^{^{\prime }} &=&X_{V}^{^{\prime }r}(\mu ) \\
Z_{MR} &=&Z_{MR}^{r}({\mu })+\frac{20}{3}\left( \frac{M}{F}\right)
^{2}(4d_{1}^{2}-\sigma _{RV}(\sigma _{RV}-2d_{1}))\lambda _{\infty } \\
Z_{R} &=&Z_{R}^{r}({\mu })+\frac{40}{9}\left( \frac{M}{F}\right)
^{2}(12d_{1}(d_{3}+d_{4})-d_{3}^{2}-9d_{4}^{2}+6d_{3}d_{4})\lambda _{\infty
}+\frac{10}{9}\sigma _{RV}(10d_{3}+18d_{4}+\sigma _{RV})\lambda _{\infty } \\
Y_{R} &=&Y_{R}^{r}({\mu })+\frac{10}{9}\left( \frac{M}{F}\right) ^{2}\frac{1%
}{M^{2}}(24d_{1}^{2}-48d_{1}(d_{3}+d_{4})+20d_{3}^{2}+36d_{4}^{2} \\
&&-24d_{3}d_{4}-\sigma _{RV}^{2}+2\sigma _{RV}(d_{3}+3d_{4}))\lambda
_{\infty } \\
X_{R} &=&X_{R}^{r}({\mu })+\frac{40}{9}\left( \frac{M}{F}\right) ^{2}\frac{1%
}{M^{2}}(d_{3}^{2}-6d_{3}d_{4}+5d_{4}^{2})\lambda _{\infty } \\
&&-\frac{10}{9}\frac{1}{M^{2}}\sigma _{RV}(6(d_{3}+d_{4})-\sigma
_{RV})\lambda _{\infty }-\left( \frac{G_{V}}{F}\right) ^{2}\frac{1}{M^{2}}%
\lambda _{\infty } \\
W_{R} &=&W_{R}^{r}({\mu })+\frac{40}{9}\left( \frac{M}{F}\right) ^{2}\frac{1%
}{M^{4}}(d_{3}^{2}+6d_{3}d_{4}-5d_{4}^{2})\lambda _{\infty }+\frac{10}{9}%
\sigma _{RV}(\sigma _{RV}-2(d_{3}+3d_{4}))\lambda _{\infty }
\end{eqnarray*}%
where
\begin{eqnarray*}
X_{V}^{r}(\mu ) &=&X_{V1}^{r}(\mu )+X_{V5}^{r}(\mu ) \\
X_{V}^{^{\prime }r}(\mu ) &=&X_{V1}^{r}(\mu )+X_{V2}^{r}(\mu
)+X_{V3}^{r}(\mu )+X_{V4}^{r}(\mu )+X_{V5}^{r}(\mu ) \\
X_{R}^{r}({\mu }) &=&X_{R1}^{r}({\mu })+X_{R2}^{r}({\mu })+X_{R3}^{r}({\mu })
\\
W_{R}^{r}({\mu }) &=&W_{R1}^{r}({\mu })+W_{R2}^{r}({\mu }) \\
X_{RV}^{r}({\mu }) &=&X_{RV1}^{r}({\mu })+X_{RV2}^{r}({\mu })
\end{eqnarray*}%
The renormalized self-energies can be then written in the form
\begin{eqnarray*}
\Sigma _{RV}(s)^{r} &=&M\left( \frac{M}{4\pi F}\right) ^{2}\left[
\sum_{i=0}^{2}{\alpha }_{i}^{RV}x^{i}+\frac{1}{2}\frac{g_{V}G_{V}}{M}\left(
\frac{M}{F}\right) ^{2}x^{2}\widehat{B}(x)\right. \\
&&\left. +\frac{10}{9}(\sigma _{RV}+2\sigma _{V})(2d_{3}x+2d_{3}-\sigma
_{RV})(x-1)^{2}\widehat{J}(x)\right] =\Sigma _{VR}(s)^{r} \\
\Sigma _{VV}^{T}(s)^{r} &=&M^{2}\left( \frac{M}{4\pi F}\right) ^{2}\left[
\sum_{i=0}^{3}{\alpha }_{i}^{VV}x^{i}-\frac{1}{2}g_{V}^{2}\left( \frac{M}{F}%
\right) ^{2}x^{3}\widehat{B}(x)\right. \\
&&\left. -\frac{10}{9}\left( \sigma _{RV}(\sigma _{RV}+2\sigma _{V})+4\sigma
_{V}^{2}\right) (x-1)^{2}x\widehat{J}(x)\right] \\
\Sigma _{VV}^{L}(s)^{r} &=&M^{2}\left( \frac{M}{4\pi F}\right)
^{2}\sum_{i=0}^{3}{\beta }_{i}^{VV}x^{i} \\
\Sigma _{RR}^{L}(s)^{r} &=&M^{2}\left( \frac{M}{4\pi F}\right) ^{2}\left[
\sum_{i=0}^{3}{\alpha }_{i}^{RR}x^{i}-\frac{1}{2}\left( \frac{G_{V}}{F}%
\right) ^{2}x^{2}\widehat{B}(x)\right. \\
&&\left. -\frac{10}{9}(4d_{3}^{2}(x+1)^{2}-2d_{3}\sigma _{RV}(x+1)+\sigma
_{RV}^{2})(x-1)^{2}\widehat{J}(x)\right] \\
\Sigma _{RR}^{T}(s)^{r} &=&M^{2}\left( \frac{M}{4\pi F}\right) ^{2}\left[
\sum_{i=0}^{3}{\beta }_{i}^{RR}x^{i}+\frac{5}{9}(8d_{3}^{2}-4\sigma
_{RV}d_{3}+2\sigma _{RV}^{2}\right. \\
&&\left. +(4d_{3}^{2}-2\sigma _{RV}d_{3}+\sigma
_{RV}^{2}+24d_{3}d_{4}+4d_{4}^{2}-6\sigma
_{RV}d_{4})x+8d_{4}^{2}x^{2})(x-1)^{2}\widehat{J}(x)\right] .
\end{eqnarray*}%
Here again the renormalization scale independent coefficients of the
polynomial parts of the self-energies are expressed in terms of the
couplings and chiral logs; the explicit formulae can be found in the
Appendix \ref{appendix first order}.

The equation for the poles in the $1^{--}$ channel
\begin{equation*}
D(s)=(M^2+\Sigma _{RR}^L(s))(M^2+\Sigma _{VV}^T(s))-s(M+\Sigma
_{RV}(s))(M+\Sigma _{VR}(s))=0
\end{equation*}
can be solved perturbatively writing the solution in the form $\overline{s}%
=M_{\mathrm{phys}}^2-\mathrm{i}M_{\mathrm{phys}}\Gamma _{\mathrm{phys}%
}=M^2+\Delta $. To the first order in $\Delta $ and the self-energies we get
then
\begin{equation*}
\overline{s}=M^2+\Sigma _{RR}^L(M^2)+\Sigma _{VV}^T(M^2)-M(\Sigma
_{RV}(M^2)+\Sigma _{VR}(M^2))
\end{equation*}
and therefore
\begin{eqnarray*}
M_{\mathrm{phys}}^2 &=&M^2+\mathrm{Re}\left[ \Sigma _{RR}^L(M^2)+\Sigma
_{VV}^T(M^2)-M(\Sigma _{RV}(M^2)+\Sigma _{VR}(M^2))\right] \\
&=&M^2\left[ 1+\left( \frac M{4\pi F}\right) ^2\left( \sum_{i=0}^3({\alpha }%
_i^{RR}+{\alpha }_i^{VV})-2\sum_{i=0}^2{\alpha }_i^{RV}-\frac 12\left( \frac
MF\right) ^2\left( g_V+\frac{G_V}M\right) ^2\right) \right] \\
M_{\mathrm{phys}}\Gamma _{\mathrm{phys}} &=&-\mathrm{Im}\left[ \Sigma
_{RR}^L(M^2)+\Sigma _{VV}^T(M^2)-M(\Sigma _{RV}(M^2)+\Sigma _{VR}(M^2))%
\right] \\
&=&\pi M^2\left( \frac M{4\pi F}\right) ^2\frac 12\left( \frac MF\right)
^2\left( g_V+\frac{G_V}M\right) ^2
\end{eqnarray*}
which yield the constraint
\begin{equation*}
M_{\mathrm{phys}}^2+\frac 1\pi M_{\mathrm{phys}}\Gamma _{\mathrm{phys}%
}=M^2\left( 1+\left( \frac M{4\pi F}\right) ^2\left( \sum_{i=0}^3({\alpha }%
_i^{RR}+{\alpha }_i^{VV})-2\sum_{i=0}^2{\alpha }_i^{RV}\right) \right) .
\end{equation*}
In the on-shell scheme $M^2=M_{\mathrm{phys}}^2$we get further
\begin{equation*}
\frac 1\pi \frac{\Gamma _{\mathrm{phys}}}{M_{\mathrm{phys}}}=\left( \frac{M_{%
\mathrm{phys}}}{4\pi F_\pi }\right) ^2\left( \sum_{i=0}^3({\alpha }_i^{RR}+{%
\alpha }_i^{VV})-2\sum_{i=0}^2{\alpha }_i^{RV}\right)
\end{equation*}
On the contrary to the previous two cases, this allows to exclude both the
constants $g_V$ and $G_V$ in favor of the physical observables only for the
combination
\begin{eqnarray*}
\sigma (x) &\equiv &x\sigma _{RR}^L(x)+\sigma _{VV}^T(x)-x(\sigma
_{RV}(x)+\sigma _{VR}(x)) \\
&=&\frac 1\pi \frac{\Gamma _{\mathrm{phys}}}{M_{\mathrm{phys}}}\left(
1+\sum_{i=0}^4a_i(x-1)^i-x^3\widehat{B}(x)\right) \\
&&-\frac{20}9\left( \frac{M_{\mathrm{phys}}}{4\pi F_\pi }\right) ^2x(x-1)^2%
\widehat{J}(x)\left[ d_3(x+1)(2d_3(x+1)+\sigma _{RV}+4\sigma _V)+\sigma
_V(\sigma _{RV}-2\sigma _V)\right]
\end{eqnarray*}
(here $\Sigma _{RR}^{Lr}=M^2\sigma _{RR}^L$, $\Sigma _{RV}^r=M\sigma _{RV}$
etc.), where
\begin{equation*}
a_i=\pi \frac{M_{\mathrm{phys}}}{\Gamma _{\mathrm{phys}}}\left( \frac{M_{%
\mathrm{phys}}}{4\pi F_\pi }\right) ^2\left( {\alpha }_{i-1}^{RR}+{\alpha }%
_i^{VV}-2{\alpha }_{i-1}^{RV}\right)
\end{equation*}
with ${\alpha }_{-1}^{RR}={\alpha }_{-1}^{RV}=0$ are parameters of order $%
O(1)$,

From the OPE constraints applied to $VVP$ correlator within the first order
formalism we get further
\begin{eqnarray*}
d_3 &=&-\frac{N_C}{64\pi ^2}\left( \frac M{F_V}\right) ^2+\frac 18\left(
\frac F{F_V}\right) ^2+\frac 12(\sigma _{RV}+\sigma _V) \\
&=&-\frac 34\left( \frac{M_{\mathrm{phys}}}{4\pi F_\pi }\right) ^2\left(
\frac{F_\pi }{F_V}\right) ^2\left[ 1-\frac 16\left( \frac{4\pi F_\pi }{M_{%
\mathrm{phys}}}\right) ^2\right] +\frac 12(\sigma _{RV}+\sigma _V)
\end{eqnarray*}

Using dimensionless variables, we can write the condition for the poles in
the form
\begin{equation*}
(1+\sigma _{RR}^L(x))(1+\sigma _{VV}^T(x))-x(1+\sigma _{RV}(x))(1+\sigma
_{VR}(x))=0
\end{equation*}
in the $1^{--}$ channel and
\begin{eqnarray*}
1+\sigma _{RR}^T(x) &=&0 \\
1+\sigma _{VV}^L(x) &=&0
\end{eqnarray*}
in the $1^{+-}$ and $0^{+-}$ channels respectively. Within the
on-shell scheme
\begin{eqnarray*}
\sigma _{RV}(s)^r &=&\frac 1\pi \frac{\Gamma _{\mathrm{phys}}}{M_{\mathrm{%
phys}}}\left( \sum_{i=0}^2a_i^{RV}x^i-(1-C)Cx^2\widehat{B}(x)\right) \\
&&+\frac{10}9\left( \frac{M_{\mathrm{phys}}}{4\pi F_\pi }\right) ^2\left[
(\sigma _{RV}+2\sigma _V)(2d_3x+2d_3-\sigma _{RV})(x-1)^2\widehat{J}(x)%
\right] \\
\sigma _{VV}^T(s)^r &=&\frac 1\pi \frac{\Gamma _{\mathrm{phys}}}{M_{\mathrm{%
phys}}}\left( \sum_{i=0}^3a_i^{VV}x^i+C^2x^3\widehat{B}(x)\right) \\
&&-\frac{10}9\left( \frac{M_{\mathrm{phys}}}{4\pi F_\pi }\right) ^2\left[
\left( \sigma _{RV}(\sigma _{RV}+2\sigma _V)+4\sigma _V^2\right) (x-1)^2x%
\widehat{J}(x)\right] \\
\sigma _{VV}^L(s)^r &=&\frac 1\pi \frac{\Gamma _{\mathrm{phys}}}{M_{\mathrm{%
phys}}}\sum_{i=0}^3b_i^{VV}x^i \\
\sigma _{RR}^L(s)^r &=&\frac 1\pi \frac{\Gamma _{\mathrm{phys}}}{M_{\mathrm{%
phys}}}\left( \sum_{i=0}^3a_i^{VV}x^i+(1-C)^2x^3\widehat{B}(x)\right) \\
&&-\frac{10}9\left( \frac{M_{\mathrm{phys}}}{4\pi F_\pi }\right) ^2\left[
(4d_3^2(x+1)^2-2d_3\sigma _{RV}(x+1)+\sigma _{RV}^2)(x-1)^2\widehat{J}(x)%
\right] \\
\sigma _{RR}^T(s)^r &=&\frac 1\pi \frac{\Gamma _{\mathrm{phys}}}{M_{\mathrm{%
phys}}}\sum_{i=0}^3b_i^{RR}x^i+\frac 59\left( \frac{M_{\mathrm{phys}}}{4\pi
F_\pi }\right) ^2\left[ (8d_3^2-4\sigma _{RV}d_3+2\sigma _{RV}^2\right. \\
&&\left. +(4d_3^2-2\sigma _{RV}d_3+\sigma _{RV}^2+24d_3d_4+4d_4^2-6\sigma
_{RV}d_4)x+8d_4^2x^2)(x-1)^2\widehat{J}(x)\right] .
\end{eqnarray*}
and
\begin{equation*}
\frac 1\pi \frac{\Gamma _{\mathrm{phys}}}{M_{\mathrm{phys}}}C^2=\frac
12g_V^2\left( \frac{M_{\mathrm{phys}}}{F_\pi }\right) ^2M_{\mathrm{phys}%
}\left( \frac{M_{\mathrm{phys}}}{4\pi F_\pi }\right) ^2
\end{equation*}
and the other parameters are of natural size $O(1)$ with the constraint
\begin{equation*}
\sum_{i=0}^3(a_i^{RR}+a_i^{VV})-2\sum_{i=0}^2a_i^{RV}=\sum_{i=0}^4a_i=1.
\end{equation*}

\subsection{Note on the counterterms}

Let us note, that the counterterm Lagrangians (\ref{ctV}),
(\ref{ctR}) and (\ref{ctRV}) might be further simplified using the
leading order equations of motion (EOM) in order to eliminate the
terms with more then two derivatives as it has been done \emph{e.g.}
in \cite{SS2}. However, this does not mean, that we do not need to
introduce such counterterms at all. As we have proved by means of
the above explicit calculations, without the higher derivative
counterterms (or equivalently without the couterterms proportional
to the EOM) we would not have the off-shell self-energies finite.

In fact, the infinities originating in the missing EOM-proportional
counterterms are not always dangerous. Note \emph{e.g.}, that such
infinities are in fact harmless, provided we restrict our treatment
to strict one-loop contribution to the GF of quark bilinears or to
the corresponding on-shell S-matrix elements. Namely, in this case,
the one-loop generating functional of the GF is obtained by means of
the Gaussian functional integration of the quantum fluctuations
around the solution of the lowest order EOM. As a result, the EOM
can be safely used to simplify the infinite part of the one-loop
generating functional. On the strict one-loop level the infinite
parts of the self-energy subgraphs corresponding to the missing
EOM-proportional counterterms cancel with similar infinities
stemming from the vertex corrections.

Nevertheless, already at the one-loop level these counterterms might
be necessary under some conditions. Namely, near the resonance poles
we can (and in fact have to) go beyond the strict one-loop expansion
\emph{e.g.} by means of the Dyson resumation of the one-loop
self-energy contributions to the propagator. This will generally
destroy such a compensation of infinities. This is the reason why we
keep the counterterm Lagrangian in the general form (\ref{ctV}),
(\ref{ctR}) and (\ref{ctRV}).

\section{From self-energies to propagators\label{Section_propagators}}

\bigskip

In the previous sections we have given the explicit form of the
self-energies in a given approximation within all three formalisms for the
description of the spin-1 resonances. Here we would like to discuss
interpretation of these results and the construction of the corresponding
propagators. We will concentrate on the most frequently used antisymmetric
tensor representation, where all the characteristic features of other
approaches are visible without unsubstantial technical complications. The
remaining two cases can be discussed along the same lines with similar
results.

Let us remind the form of the self-energies for the antisymmetric tensor
case
\begin{eqnarray}
{\sigma }_{L}^{r}(x) &=&\frac{1}{\pi }\frac{\Gamma _{\mathrm{phys}}}{M_{%
\mathrm{phys}}}\left[ 1-x^{2}\widehat{B}(x)+\sum_{i=1}^{3}a_{i}(x^{i}-1)%
\right] -\frac{40}{9}\left( \frac{M_{\mathrm{phys}}}{4\pi F_{\pi }}\right)
^{2}d_{3}^{2}(x^{2}-1)^{2}\widehat{J}(x)  \label{selfL} \\
{\sigma }_{T}^{r}(x) &=&\frac{1}{\pi }\frac{\Gamma _{\mathrm{phys}}}{M_{%
\mathrm{phys}}}\sum_{i=0}^{3}b_{i}x^{i}+\frac{20}{9}\left( \frac{M_{\mathrm{%
phys}}}{4\pi F_{\pi }}\right) ^{2}d_{3}^{2}\left( 2+(1+6\gamma
+\gamma ^{2})x+2\gamma ^{2}x^{2}\right) (x-1)^{2}\widehat{J}(x),\nonumber \\
\label{selfT}
\end{eqnarray}
where $d_{3}$ is given by (\ref{d_3}) and where we have already
re-parametrized the general result in terms of the parameters of the
perturbative solution of the pole equation in the $1^{--}$ channel (which we
have identified with the original degree of freedom). In doing that we have
tacitly assumed the validity of the general relation between the
self-energies and the propagator (\ref{tensor_Delta_2}). The equations
determining the additional poles of the propagators are then
\begin{eqnarray}
f_{L}(x) &\equiv &x-1-{\sigma }_{L}^{r}(x)=0  \label{poleL} \\
f_{T}(x) &\equiv &1+{\sigma }_{T}^{r}(x)=0.  \label{poleT}
\end{eqnarray}
In what follows we shall discuss these equations in more detail.
We will find a lower and upper bound on the number of their
solutions and give a proof, that the corresponding lover bounds
are greater than one on both sheets. We will also briefly discuss
the compatibility of the relation (\ref{tensor_Delta_2})  with the
K\"{a}ll\'{e}n-Lehman representation and show, that at least one
of the roots of (\ref{poleL}) and (\ref{poleT}) corresponds
inevitably either to the negative norm ghost or the tachyon.

\subsection{The number of poles using Argument principle}

Let us first briefly discuss a determination of the number of
solution of the equations (\ref{poleL}) and (\ref{poleT}). This
can be made using the theorem known as Argument principle (see
\emph{e.g.} \cite{Ablowitz}). According to this theorem, for a
meromorphic function $f(z)$ with no zeros or poles on a simple
closed contour $C$, the difference between the number of zeros $N$
and poles $P$ (counted according to  their multiplicity) inside
$C$ is given as
\begin{equation}
N-P=\frac{1}{2\pi}[\arg f(z)]_{C}.
\end{equation}
Here $[\arg f(z)]_{C}$ is the change of the argument of $f(z)$
along $C$.
 Using this theorem we will show, that in both cases (\ref{poleL}) and
(\ref{poleT}) there is a nonzero lower bound on the number of
solutions on the first and the second
sheet, which correspond to the poles of the propagator (\ref{tensor_Delta_2}%
). We will also give conditions for the saturation of these lower bounds.

Let us start with (\ref{poleT}). The left hand side of the pole equation $%
f_{T}(z)=1+{\sigma }_{T}^{r}(z)$ is analytic on the first sheet (and
meromorphic on the second sheet) of the cut complex plane with cut from $z=1$
to $z=+\infty $. \ Let us choose contour $C=C_{+}+C_{R}-C_{-}+C_{\varepsilon
}$ which is usually used for the proof of the dispersive representation for
the self-energy, namely the one consisting of the infinitesimal circle $%
C_{\varepsilon }$ encircling the point $z=1$ clockwise, two straight lines $%
C_{\pm }$ infinitesimally above and bellow the real axis going from $z=1$ to
$z=R$ \ and a circle $C_{R}$ corresponding to $z=R\,\mathrm{e}^{\mathrm{i}%
\theta }$, $0<\theta <2\pi $, and take the limit with $\varepsilon
\rightarrow 0$, $R\rightarrow \infty $ in the end. According to the
argument
principle, the total change of the phase of $\ $\ the function $%
f_{T}^{I,II}(z)$ along this contour gives the number of zeros (with their
multiplicities) $n^{I}$ of $f(z)$ on the first sheet and $n^{II}-2$, where $%
n^{II}$ is the number of zeros of $f(z)$ on the second sheet (note that $%
f_{T}^{II}(z)$ has pole of the second order at $z=0$) lying inside the
contour $C$, i.e.
\begin{eqnarray*}
n^{I} &=&\frac{1}{2\pi }[\arg f_{T}^{I}(z)]_{C} \\
n^{II} &=&\frac{1}{2\pi }[\arg f_{T}^{II}(z)]_{C}+2.
\end{eqnarray*}%

Let us assume the contour  $C_{\varepsilon }$ first. Suppose that
$x=1$ is not a solution of the equation $f_{T}(z)=0$. As a
consequence, $[\arg f_{T}^{I,II}(z)]_{C_{\varepsilon }}$ vanishes
\footnote{%
In the case $f_{T}^{I,II}(x)\rightarrow 0$ for $x\rightarrow 1$ when $%
f_{T}^{I,II}(x)=(x-1)^{k}$ $g_{T}^{I,II}(x)$ where $k\leq 3$ and when $%
g_{T}^{I,II}(x)$ (which has the branching point at $x=1$)  has a finite nonzero limit at $%
x=1$) we get $[\arg f_{T}^{I,II}(z)]_{C_{\varepsilon }}=-2\pi
k$.\label{footnote_zero}}.

On the contour $C_R$, {\em{i.e.}} for $%
z=R\,\mathrm{e}^{\mathrm{i}\theta }$ we get for $b_{3}\neq 0$%
\begin{equation*}
f_{T}^{I,II}(R\,\mathrm{e}^{\mathrm{i}\theta })=R^{3}\mathrm{e}^{3\mathrm{i}%
\theta }\left( \frac{1}{\pi }\frac{\Gamma _{\mathrm{phys}}}{M_{\mathrm{phys}}%
}b_{3}+\frac{20}{9}\left( \frac{M_{\mathrm{phys}}}{4\pi F_{\pi }}\right)
^{2}d_{3}^{2}2\gamma ^{2}\left[ 1-\ln R+\mathrm{i}(2\pi -\theta \mp \pi )%
\right] +O\left( \frac{1}{R},\frac{\ln R}{R}\right) \right)
\end{equation*}%
and therefore, for $R\rightarrow \infty $, $[\arg
f_{T}^{I,II}(z)]_{C_{R}}\rightarrow 6\pi $. The same is valid also for $%
b_{3}=0$ with $\gamma \neq 0$. However, for $b_{3}=\gamma =0$ we get
\begin{equation*}
f_{T}^{I,II}(R\,\mathrm{e}^{\mathrm{i}\theta })=R^{2}\mathrm{e}^{2\mathrm{i}%
\theta }\left( \frac{1}{\pi }\frac{\Gamma _{\mathrm{phys}}}{M_{\mathrm{phys}}%
}b_{2}+\frac{20}{9}\left( \frac{M_{\mathrm{phys}}}{4\pi F_{\pi }}\right)
^{2}d_{3}^{2}\left[ 1-\ln R+\mathrm{i}(2\pi -\theta \mp \pi )\right]
+O\left( \frac{1}{R},\frac{\ln R}{R}\right) \right) .
\end{equation*}%
In this case $[\arg f_{T}^{I,II}(z)]_{C_{R}}\rightarrow 4\pi $ and because $%
d_{3}\neq 0$ (unless we are in a conflict with OPE for the tree level $VVP$
correlator\footnote{%
Note however, that the requirement that the tree level conditions for OPE
are satisfied might be modified by loop corrections.}), this gives also the
lower bound for $[\arg f_{T}^{I,II}(z)]_{C_{R}}$.

Finally let us discuss the lines $C_{\pm}$. Because $\mathrm{%
Im}$ $f_{T}^{I}(x\pm \mathrm{i}0)=\mathrm{Im}{\sigma }_{T}^{r}(x\pm \mathrm{i%
}0)\gtrless 0$ (and $f_{T}^{I}$ is real analytic), $\mathrm{Im}$ $%
f_{T}^{II}(x\pm \mathrm{i}0)>0$ for $x>1$, and $\mathrm{Re}f_{T}^{I,II}(R\pm
\mathrm{i}0\,)\rightarrow -\infty $ for $R\rightarrow \infty $, we can
easily conclude that in this limit $[\arg f_{T}^{I,II}(z)]_{C_{+}}=0$ unless
$f_{T}^{I,II}(1)>0$, in the latter case $[\arg f_{T}^{I,II}(z)]_{C_{+}}=\pi $
and in both cases $[\arg f_{T}^{I,II}(z)]_{C_{-}}=\pm \lbrack \arg
f_{T}^{I,II}(z)]_{C_{+}}$.

Putting all pieces together we get under the assumption
$f_{T}^{I,II}(1)\neq 0$ the following bound
\begin{equation*}
\lbrack \arg f_{T}^{I,II}(z)]_{C}\geq 4\pi
\end{equation*}%
and therefore for the number of zeros in the cut complex plane we get%
\begin{eqnarray}
2 &\leq &n^{I}\leq 4 \\
4 &\leq &n^{II}\leq 5
\end{eqnarray}%
where the lower bound is saturated for $f_{T}^{I,II}(1)<0$, $b_{3}=\gamma =0$
and the upper bound for $f_{T}^{I,II}(1)>0$ and either $b_{3}\neq 0$ or $%
\gamma \neq 0$. For $f_{T}^{I,II}(1)=0$ (provided we include also
this zero with its multiplicity into $n^{I,II}$) the  these bounds
are valid too\footnote{In this case the point $x=1$ is solution of
$f_{T}^{I,II}(x)=0$ and provided $f_{T}^{I,II}(x)=(x-1)^{k}$
$g_{T}^{I,II}(x)$ (zero with multiplicity $k\le 3$) we have
according to the footnote \ref{footnote_zero}  the phase deficit
$-2\pi k$ ({\em i.e.} the number of the poles different from $z=1$
is then reduced by $k$) in comparison with the case
$f_{T}^{I,II}(x)\neq 0$.}.

An analogous simple analysis for $f_{L}(z)=z-1-$ ${\sigma }_{L}^{r}(z)$ in
the cut complex plane with the cut from $z=0$ to $z=+\infty $ gives
\footnote{%
Note, that in this case,
\begin{equation*}
f_{L}^{I,II}(R\,\mathrm{e}^{\mathrm{i}\theta })=R^{3}\mathrm{e}^{3\mathrm{i}%
\theta }\left( -\frac{1}{\pi }\frac{\Gamma _{\mathrm{phys}}}{M_{\mathrm{phys}%
}}a_{3}+\frac{40}{9}\left( \frac{M_{\mathrm{phys}}}{4\pi F_{\pi }}\right)
^{2}d_{3}^{2}\left[ 1-\ln R+\mathrm{i}(2\pi -\theta \mp \pi )\right]
+O\left( \frac{1}{R},\frac{\ln R}{R}\right) \right)
\end{equation*}%
and therefore $[\arg f_{L}^{I,II}(z)]_{C_{R}}=6\pi $.} for $%
f_{L}^{I,II}(0)\neq 0$
\begin{eqnarray}
3 &\leq &n^{I}\leq 4 \\
n^{II} &=&5
\end{eqnarray}%
where the lower bounds are saturated for $f_{L}^{I,II}(0)<0$
otherwise $n^{I}$ equals to the upper bound.

We can not therefore avoid in any way the generation of the additional poles
(some of them might even be of the higher order) in both $1^{--}$ and $1^{+-%
\text{ }}$channels of the propagator only by means of an appropriate choice
of the free parameters $a_{i}$, $b_{i}$ and $\gamma $. The minimal number of
the additional poles (with their orders) on the second sheet is the same for
both channels (note that, one pole in $1^{--}$ channel has to correspond to
the perturbative solution describing the original degrees of freedom we have
started with). The conditions for the saturation of the lower bounds in the $%
1^{--}$ and $1^{+-}$ channels are
\begin{equation}
-\frac{1}{\pi }\frac{\Gamma _{\mathrm{phys}}}{M_{\mathrm{phys}}}\left[
1-\sum_{i=1}^{3}a_{i}\right] +\frac{20}{9}\left( \frac{M_{\mathrm{phys}}}{%
4\pi F_{\pi }}\right) ^{2}d_{3}^{2}<1
\end{equation}%
and
\begin{eqnarray}
b_{3} &=&\gamma =0 \\
-\frac{1}{\pi }\frac{\Gamma _{\mathrm{phys}}}{M_{\mathrm{phys}}}%
\sum_{i=0}^{3}b_{i} &>&1
\end{eqnarray}%
respectively. Note that, while the first condition is in accord with the
large $N_{C}$ counting, the last one is not. \bigskip Let us now discuss the
physical relevance of such additional poles.

\subsection{The K\"{a}ll\'{e}n-Lehman representation and nature of the poles}

In this subsection, we will show that the the propagator (\ref%
{tensor_Delta_2}) with self-energies (\ref{selfL}) and (\ref{selfT}) is
incompatible with the K\"{a}ll\'{e}n-Lehman representation with the positive
spectral function. Moreover, at least one of the solutions of both equations
(\ref{poleL}) and (\ref{poleT}) is pathological and corresponds to the
negative norm ghost or the tachyonic pole.

Let us first briefly remind the K\"{a}ll\'{e}n-Lehman representation of the
antisymmetric tensor field propagator. According to the Lorentz structure we
can write the following spectral representation of the full propagator
(modulo generally non-covariant contact terms)
\begin{equation*}
\Delta _{\mu \nu \alpha \beta }(p)=p^{2}\Pi _{\mu \nu \alpha \beta
}^{T}(p)\Delta _{T}(p^{2})-p^{2}\Pi _{\mu \nu \alpha \beta }^{L}(p)\Delta
_{L}(p^{2})+\Delta _{\mu \nu \alpha \beta }^{\mathrm{contact}}(p)
\end{equation*}%
where (up to the necessary subtractions)
\begin{equation}
\Delta _{L,T}(p^{2})=\int_{0}^{\infty }\mathrm{d}\mu ^{2}\frac{\rho
_{L,T}(\mu ^{2})}{p^{2}-\mu ^{2}+\mathrm{i}0}
\end{equation}%
and where the spectral functions $\rho _{T,L}(p^{2})$ are given in terms of
the sum over the intermediate states as
\begin{equation}
(2\pi )^{-3}\theta (p^{0})\left[ \rho _{T}(p^{2})p^{2}\Pi _{\mu \nu \alpha
\beta }^{T}(p)-\rho _{L}(p^{2})p^{2}\Pi _{\mu \nu \alpha \beta }^{L}(p)%
\right] =\sum\limits_{N}\delta ^{(4)}(p-p_{N})\langle 0|R_{\mu \nu
}(0)|N\rangle \langle N|R_{\alpha \beta }(0)|0\rangle .  \label{spectral}
\end{equation}%
Note that, in the above formula we assume all the states $|N\rangle $ to
have a positive norm; the spectral functions $\rho _{L,T}(p^{2})$ are then
positive (for the proof see the Appendix \ref{Appendix_positivity}). For the
one particle spin-one bound stated states $|p,\lambda \rangle $ with mass $M$
either
\begin{equation}
\langle 0|R_{\mu \nu }(0)|p,\lambda \rangle =Z_{L}^{1/2}u_{\mu \nu
}^{(\lambda )}(p)
\end{equation}%
or
\begin{equation}
\langle 0|R_{\mu \nu }(0)|p,\lambda \rangle =Z_{T}^{1/2}w_{\mu \nu
}^{(\lambda )}(p)
\end{equation}%
according to its parity (cf. (\ref{u_function}) and (\ref{w_function})).
Therefore (using the formulae from the Appendix \ref{Appendix_positivity}),
the corresponding one particle contribution to $\rho _{L,T}(\mu ^{2})$ is
\begin{equation}
\rho _{L,T}^{\text{one-particle}}(\mu ^{2})=\frac{2}{M^{2}}Z_{L,T}\delta
(\mu ^{2}-M^{2}).
\end{equation}%
Positivity $\rho _{L,T}(\mu ^{2})$ implies $Z_{L,T}>0$ in the above
one-particle contributions.

For free fields with mass $M$ we get
\begin{eqnarray}
\rho _{L}^{free}(\mu ^{2}) &=&\frac{2}{M^{2}}\left( \delta (\mu
^{2}-M^{2})-\delta (\mu ^{2})\right)  \notag \\
\rho _{T}^{free}(\mu ^{2}) &=&\frac{2}{M^{2}}\delta (\mu ^{2}).
\label{free_spectral}
\end{eqnarray}%
Note the kinematical poles in $\Delta _{L,T}(p^{2})$ at $\ p^{2}=0$, which
do not correspond to any one-particle intermediate state and which sum up to
the contact terms of the form
\begin{equation}
\Delta _{\mu \nu \alpha \beta }^{free,\mathrm{contact}}(p)=\frac{1}{M^{2}}%
\left( g_{\mu \alpha }g_{\beta \nu }-g_{\mu \beta }g_{\nu \alpha }\right) .
\end{equation}

Let us now define for complex $z$ by means of the analytic continuation (up
to the possible subtractions)
\begin{equation}
\Delta _{L,T}(z)=\int_{0}^{\infty }\mathrm{d}s\frac{\rho _{L,T}(s)}{s-z}~,~~~
\label{dispersiv_kallan}
\end{equation}%
Within the perturbation theory however, the primary quantities are the
self-energies, which we define as (cf. (\ref{tensor_Delta_2}))
\begin{eqnarray}
\Delta _{T}(s) &=&\frac{1}{s}\frac{2}{M^{2}+\Sigma _{T}(s)}  \notag \\
\Delta _{L}(s) &=&\frac{1}{s}\frac{2}{s-M^{2}-\Sigma _{L}(s^{2})}.
\label{simas_definition}
\end{eqnarray}%
The poles at $s=0$ are of the kinematical origin and in analogy with the
free propagator they sum up into the contact terms provided $\Sigma
_{T}(0)=\Sigma _{L}(0)$. The formulae (\ref{simas_definition}) can be
understood as the Dyson re-summation of the $1PI$ self-energy insertions to
the propagator or as an inversion of the $1PI$ two-point function. Due to
the positivity of $\rho _{L,T}(s)$, we get for the imaginary parts of $\
\Sigma _{L,T}$ the following positivity (negativity) constraints:
\begin{eqnarray}
\mathrm{Im}\Sigma _{L}(s+i0) &=&\frac{1}{2}\theta (s)s~\mathrm{Im}\Delta
_{L}(s+i0)|s-M^{2}-\Sigma _{L}(s+i0)|^{2}\leq 0  \notag \\
\mathrm{Im}\Sigma _{T}(s+i0) &=&-\frac{1}{2}\theta (s)s~\mathrm{Im}\Delta
_{T}(s+i0)|M^{2}+\Sigma _{L}(s+i0)|^{2}\geq 0.  \label{positiv_negativ}
\end{eqnarray}

Let us now turn to the $R\chi T$ -like effective theories and try to
demonstrate their possible limitations. In such a framework the
self-energies $\Sigma _{L,T}$ are given by a sum of the 1PI graphs organized
according to some counting rule (for $R\chi T$ \emph{e.g.} by the index $%
i_{\Gamma \text{ }}$, cf. (\ref{power_counting})). Up to a fixed given order
(which we assume to be fixed from now on) we have the asymptotic behavior $%
\Sigma _{L,T}(z)=O(z^{n}\ln ^{k}z))$ for $z\rightarrow -\infty $ according
to the Weinberg theorem. \ Here $n$ corresponds to the maximal degree of
divergence of the contributing (sub)graphs and \ therefore, it grows with
the number of loops as well as with the index of the vertices (cf. (\ref%
{index_O}) ). \

Such a grow of the inverse propagator is known to lead to problems. Suppose
\emph{e.g.}, that we can organize the result of the calculation of the 1PI
graphs in the form of a dispersive representation for the functions $\Sigma
_{L,T}(z)$ on the first sheet\footnote{%
Here we do not assume the existence of any CDD poles
\cite{Castillejo} for simplicity. In general case, provided the
spectral representation of $\Delta _{L,T}$ is valid in the form
(\ref{dispersiv_kallan}), and $\mathrm{Im}\Delta
_{L,T}^{-1}(s)=O(s^{n})$ for $s\rightarrow \infty $ we formally get
\begin{equation*}
\Delta _{L,T}^{-1}(z)=P_{n}(z)+Q_{n+1}^{L,T}(z)\left( \frac{1}{\pi }%
\int_{x_{t}}^{\infty }\frac{dx}{Q_{n+1}^{L,T}(x)}\frac{\mathrm{Im}\Delta
_{L,T}^{-1}(x)}{x-z}-\sum_{i}\frac{C_{i}}{z-z_{0i}}\right)
\end{equation*}%
where $C_{i}>0$ and $0<z_{0i}<x_{t}$ correspond to the CDD poles.}
\begin{equation}
\Sigma _{L,T}^{I}(z)=P_{n}^{L,T}(z)+\frac{Q_{n+1}^{L,T}(z)}{\pi }%
\int_{x_{t}}^{\infty }\frac{dx}{Q_{n+1}^{L,T}(x)}\frac{\mathrm{Im}\Sigma
_{L,T}(x+i0)}{x-z}  \label{dispersiv_sigma}
\end{equation}%
where $x_{t}$ $\geq 0$ is the lowest multi-particle threshold, $%
P_{n}^{L,T}(z)$ and $Q_{n+1}^{L,T}(z)$ (we suppose $Q_{n+1}^{L,T}(x)>0$ for $%
x>0$) are renormalization scale independent real polynomials of the order $n$
and $n+1$ respectively and $\mathrm{Im}\Sigma _{L,T}(x+i0)$ can be obtained
using the Cutkosky rules. The contributions to $P_{n}^{L,T}(z)$ stem from
the counterterms necessary to renormalize the superficial divergences of the
contributing 1PI graphs as well as from the loops ($\chi $logs)\footnote{%
In what follows we give such an representation of our one-loop $i_{\Gamma
}\leq 6$ result explicitly.}.

As a consequence, the functions $z^{k}\Delta _{L,T}(z)$ where $0\leq k\leq n$
and where $\Delta _{L,T}(s)$ is naively defined by (\ref{simas_definition})
are analytic (up to the finite number of complex poles $z_{j}$ generally
different for $\Delta _{L}$ and $\Delta _{T}$ and a kinematical pole at $z=0$
- see bellow) in the cut complex plane. As far as the number of poles $z_{j}$
are concerned, provided $\mathrm{Im}\Sigma _{L,T}(x+i0)\lessgtr 0$ as
suggested by (\ref{positiv_negativ}), we can almost literally repeat the
analysis from the previous subsection based on the argument principle. The
change of a phase of the inverse propagator along the path $C_{R}$ is now $%
[\arg \Delta ^{-1}_{L,T}(z)]_{C_{R}}\rightarrow 2\pi n$ (for $R\rightarrow \infty $%
), while the absolute value of the $[\arg \Delta
^{-1}_{L,T}(z)]_{C_{\pm }}$ is bounded by $\pi ~\ $due to the
positivity (negativity) of $\mathrm{Im}\Sigma _{L,T}(x\pm i0)$.
Provided $\Delta ^{-1}_{L,T}(x_{t})\neq 0$, we can therefore
conclude
\begin{eqnarray}
n-1 &\leq &n^{I} \\
n &\leq &n^{II}-p^{II}
\end{eqnarray}%
where $n^{I,II}$ is the number of the solutions of the equation $\Delta
^{-1}_{L,T}(z)=0$ on the first and second sheet respectively and $p^{II}$ is the number of the poles (weighted with their order) of $%
\Sigma _{L,T}(z)$ on the second sheet\footnote{%
Note that, the case $n=1$ is in some sense exceptional. In this case it is
possible to get a realistic resonance propagator compatible with the K\"{a}ll%
\'{e}n-Lehman representation with no pole on the first sheet and one pole on
the unphysical sheet. Such a propagator has been obtained in \cite%
{Achasov:2004uq} for scalar resonances. Cf. also \cite{Giacosa:2007bn}.}.

Therefore, because $z^{k}\Delta _{L,T}(z)=O(z^{k-n-1})$, we can write for $%
0<k\leq n$ an unsubtracted dispersion relation (cf. (\ref{dispersiv_kallan}%
), we will omit the subscript $L,~T$ in the following formulae for brevity
and write simply $\Delta (z)$, $\rho (s)$ \emph{etc.})
\begin{equation*}
z^{k}\Delta (z)=\sum_{j>0}\frac{R_{j}z_{j}^{k}}{z-z_{j}}+\frac{1}{\pi }%
\int_{x_{t}}^{\infty }\mathrm{d}x\frac{x^{k}\mathrm{disc}\Delta (x)}{x-z}
\end{equation*}%
or
\begin{equation}
\Delta (z)=\frac{1}{z^{k}}\sum_{j>0}\frac{R_{j}z_{j}^{k}}{z-z_{j}}+\frac{1}{%
\pi z^{k}}\int_{x_{t}}^{\infty }\mathrm{d}x\frac{x^{k}\mathrm{disc}\Delta (x)%
}{x-z}.
\end{equation}%
and for $k=0$ (note the kinematical pole at $z=0$)
\begin{equation}
\Delta (z)=\frac{R_{0}}{z}+\sum_{j>0}\frac{R_{j}}{z-z_{j}}+\frac{1}{\pi }%
\int_{x_{t}}^{\infty }\mathrm{d}x\frac{\mathrm{disc}\Delta (x)}{x-z}
\end{equation}%
Due to the asymptotic fall off $\Delta (z)=O(z^{-n-1})$ the discontinuity $%
\mathrm{disc}\Delta (x)$ has to satisfy the following sum rules
\begin{eqnarray}
-\frac{1}{\pi }\int_{x_{t}}^{\infty }\mathrm{d}xx^{k}\mathrm{disc}\Delta
(x)+\sum_{j}R_{j}z_{j}^{k} &=&0,~~~0<k\leq n-1.  \label{sum_rule} \\
-\frac{1}{\pi }\int_{x_{t}}^{\infty }\mathrm{d}x\mathrm{disc}\Delta
(x)+\sum_{j}R_{j}+R_{0} &=&0
\end{eqnarray}%
Suppose on the other hand validity of the dispersive representation (\ref%
{dispersiv_kallan}). Then all the poles have to be real, and we can identify
\begin{equation}
\rho (s)=-\frac{1}{\pi }\mathrm{disc}\Delta (s)+\sum_{j}R_{j}\delta
(s-z_{j})+R_{0}\delta (s).
\end{equation}

However, the sum rules (\ref{sum_rule}) are generally inconsistent with the
spectral representation (\ref{dispersiv_kallan}). The validity of some of
them might require either an appearance of the states with the negative norm
in the spectrum, \emph{i.e.} we are in a conflict with the positivity of the
spectral function $\rho (s)\geq 0$ or an appearance of physically
non-acceptable tachyon poles leading to the acausality. For instance,
suppose $\mathrm{disc}\Delta (s)\leq 0$, then for $R_{0}\geq 0$ at least one
of the poles has to correspond to a negative norm one-particle state
(ghost). On the other hand, for $\mathrm{disc}\Delta (s)\leq 0$, $R_{j}>0$
we can still satisfy the $k=0$ sum rule with negative $R_{0}$, however, from
the $k=1$ sum rule we need at least one pole to be negative (tachyon) (in
this case, however, the sum rules with even $k$ cannot be satisfied)%
\footnote{%
An analogous discussion can be done for the second sheet. Concrete examples
of various types of poles will be given in the next section.}. These
considerations illustrate the known fact that the representation of the
propagator based on the formulas (\ref{simas_definition}) has limited range
of validity within the fixed order of the perturbation theory and has to be
taken with some care.

One point of view might be that the range of applicability of the formulae (%
\ref{simas_definition}) is $|z|<\Lambda _{\mathrm{max}}=\mathrm{min}%
\{|z_{j}|\}$ where $\{z_{j}\}$ is the set of unwanted poles. Provided there
exists a genuine expansion parameter $\alpha $ applicable to the
organization of the perturbative series, according to which $\Sigma _{L,T}$ $%
=\sum_{i>0}\alpha ^{i}\Sigma _{L,T}^{(i)}$ (\emph{e.g.} expanding in powers
of $\alpha =1/N_{C}$ in $R\chi T$), one can expect the additional (generally
pathological) poles of $\Delta _{L,T}(z)$ to decouple ( \emph{i.e.} $\Lambda
_{\mathrm{max}}\rightarrow \infty $ for $\alpha \rightarrow 0$). In such a
case we could argue that they are in fact harmless. However, the size of $%
\Lambda _{\mathrm{max}}$ for actual value of $\ \alpha $ need not to be far
from $M$ which could invalidate this approach to the theory in the region
for which it was originally designed.

Alternatively, instead of using the (partial) Dyson re-summation, we can
expand directly $\Delta _{\mu \nu \alpha \beta }(p)$ to the fixed finite
order $n$ which leads to
\begin{eqnarray*}
\Delta _{L}(s) &=&\frac{2}{s}\left( \frac{1}{s-M^{2}}+\alpha \frac{1}{s-M^{2}%
}\Sigma _{L}^{(1)}(s^{2})\frac{1}{s-M^{2}}+\ldots +\alpha ^{n}\frac{1}{%
s-M^{2}}\Sigma _{L}^{(n)}(s^{2})\frac{1}{s-M^{2}}\right) \\
\Delta _{T}(s) &=&\frac{2}{s}\left( \frac{1}{M^{2}}+\alpha \frac{1}{M^{2}}%
\Sigma _{T}^{(1)}(s^{2})\frac{1}{M^{2}}+\ldots +\alpha ^{n}\frac{1}{M^{2}}%
\Sigma _{T}^{(n)}(s^{2})\frac{1}{M^{2}}\right) .
\end{eqnarray*}%
This expansion (which does not give rise to the additional poles of the
propagator) might be useful for $s\ll M^{2}$, however, in this case a
higher-order pole at $s=M^{2}$ is generated, which is not correct physically
in the resonance region $s\sim M^{2}$. Here we instead expect a single pole
on the second sheet of $\Delta _{L}(z)$, where $z=M_{\mathrm{phys}}^{2}-iM_{%
\mathrm{phys}}\Gamma _{\mathrm{phys}}$ (where the mass $M_{\mathrm{phys}%
}^{2}=M^{2}+O(\alpha )$ and the width $\Gamma _{\mathrm{phys}}=O(\alpha )$)
corresponding to the original degree of freedom of the free Lagrangian.
Therefore, the Dyson re-summation (\emph{i.e}. the application of the
formulae (\ref{simas_definition})) suplemented with some other more
sophisticated approaches (\emph{e.g.} the Redmond and Bogolyubov method \cite%
{Redmond, Bogolyubov} consisting of the subtraction\footnote{%
Note that, in order to perform this on the lagrangian level, nonperturbative
and nonlocal counterterms would have to be added to the theory. However the
status of such a counterterms is not clear, cf. \cite{Caro:1996ex}.} of the
additional unwanted poles from the propagator, or diagonal Pad\'{e}
approximation method \cite{Lambert1973}) seems to be inevitable for $s\sim
M^{2}$.

However, in the concrete case of our calculations of the antisymmetric
tensor field propagator, the plain Dyson re-summation might produce various
types of poles some of which we illustrate in the next subsection.

\bigskip

\subsection{Examples of the poles}

\begin{figure}[h]
\par
\begin{center}
\epsfig{width=0.35\textwidth,file=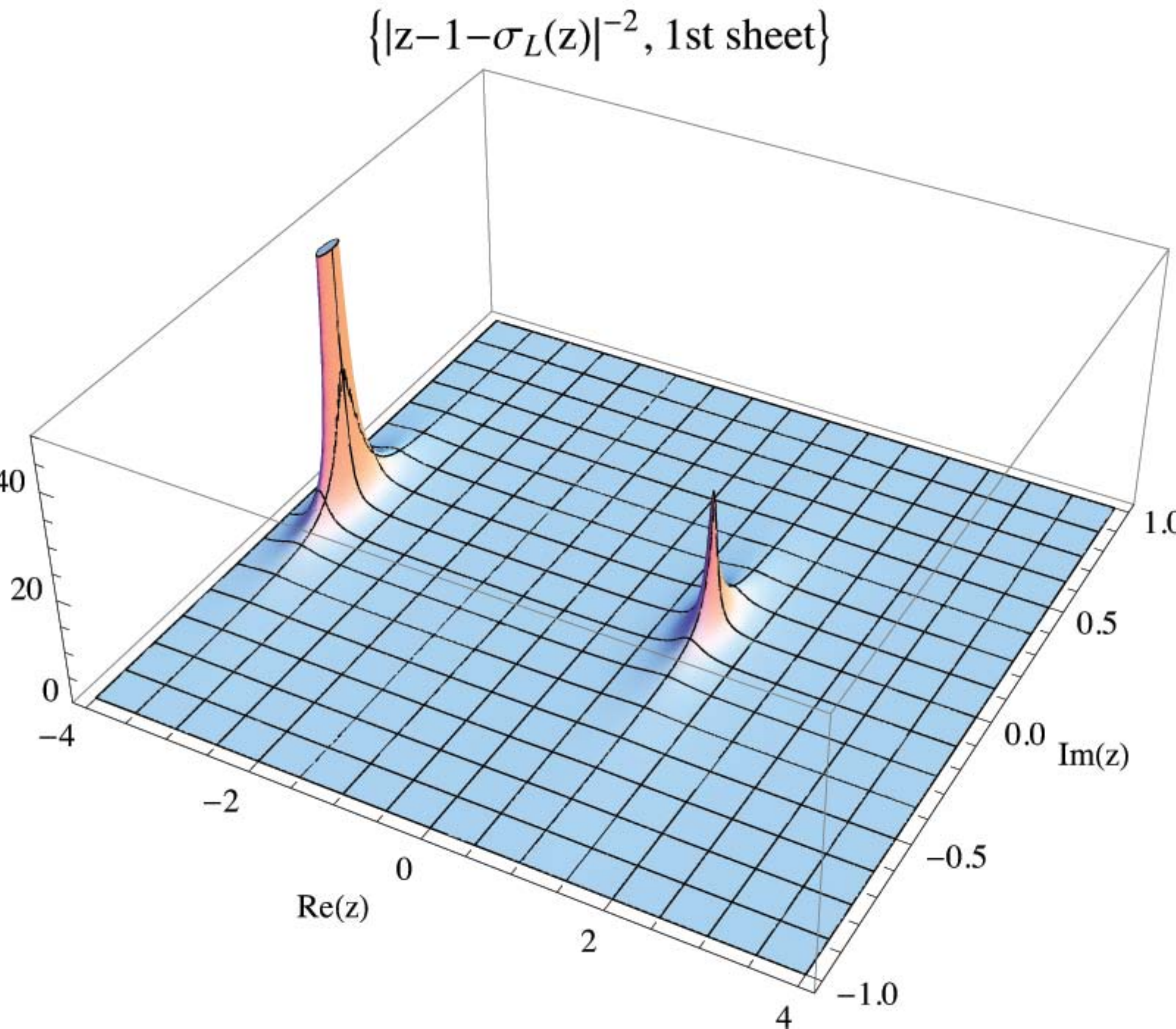} \hskip2cm
\epsfig{width=0.35\textwidth,file=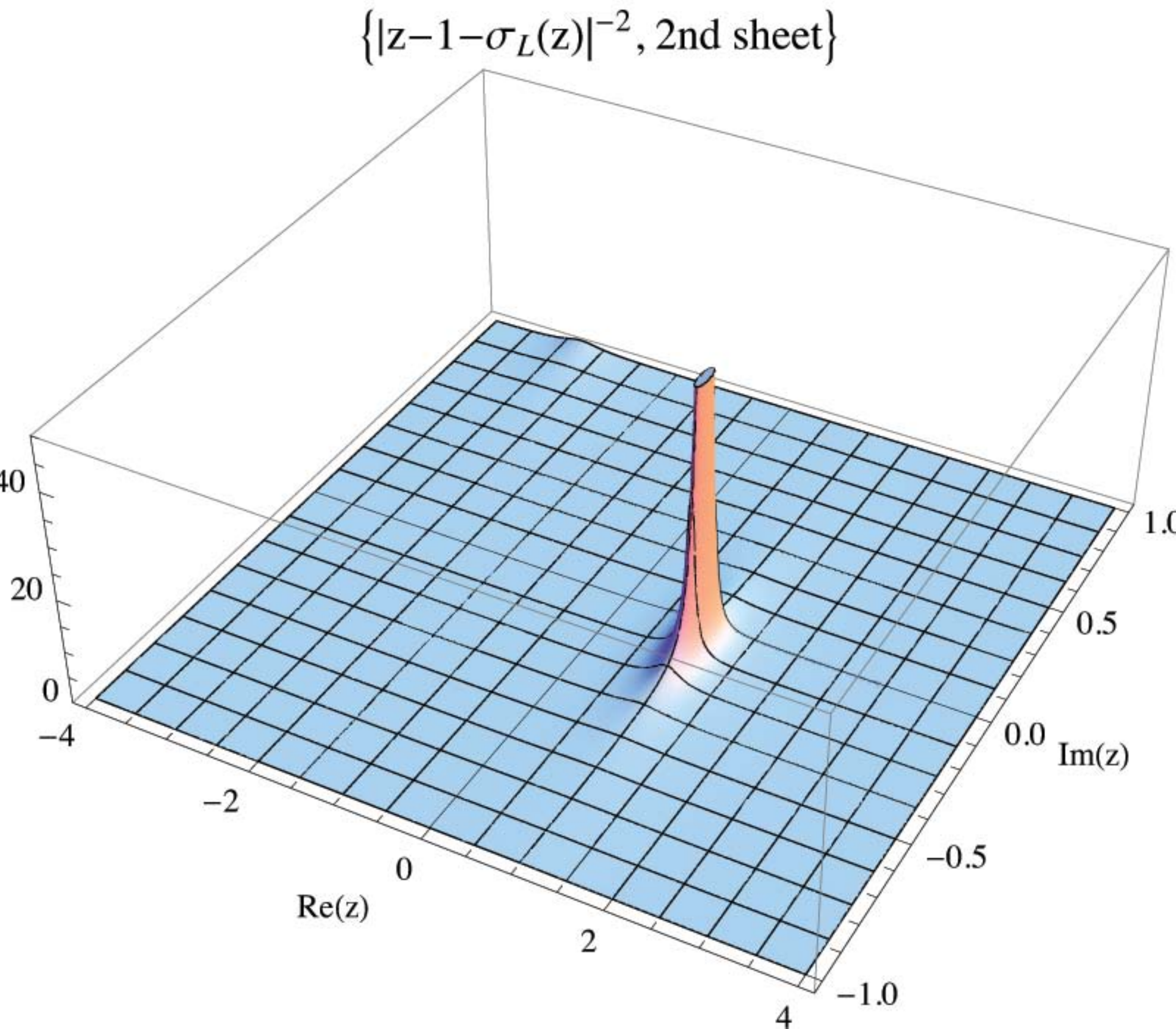}
\end{center}
\caption{The plot of the the square of the modulus of the propagator
function $|z-1-\sigma_L(z)|^{-2}$ on the first and the second sheet
for $a_0=a_1=a_2=10$, $a_3=0$. The additional pole on the first
sheet is a tachyon.} \label{poleLT}
\end{figure}

The additional poles of the propagator can have different nature.
Let us assume the $1^{--}$ channel first. By construction for any
values of the constants $a_i$ we have one pole on the second sheet
(which is directly accessible from the physical sheet by means of
the crossing of the cut for $0<z<1$) which corresponds to the
physical resonance ($\rho$ meson) we have started with at the tree
level. On the first sheet we get then a typical resonance  peak.
These two structures are illustrated in the Fig. \ref{poleL}, where
the square of the modulus of the propagator function, namely {\em
i.e.} $|z-1-\sigma_L(z)|^{-2}$, is plotted\footnote{We have used the
following numerical inputs: $M_{\rm{phys}}=770MeV$,
$\Gamma_{\rm{phys}}=150MeV$, $F=93.2MeV$, $F_V=154MeV$.} on the
first and the second sheet for $a_i=0$. In this case, no additional
pole appears in the region of assumed applicability of $R\chi T$.
However, for another set of parameters we can get also pathological
poles not far from this region ({\em e.g.} tachyon as it is
illustrated in analogous Fig. \ref{poleLT}, now for
$a_0=a_1=a_2=10$, $a_3=0$).

\begin{figure}[t]
\par
\begin{center}
\epsfig{width=0.35\textwidth,file=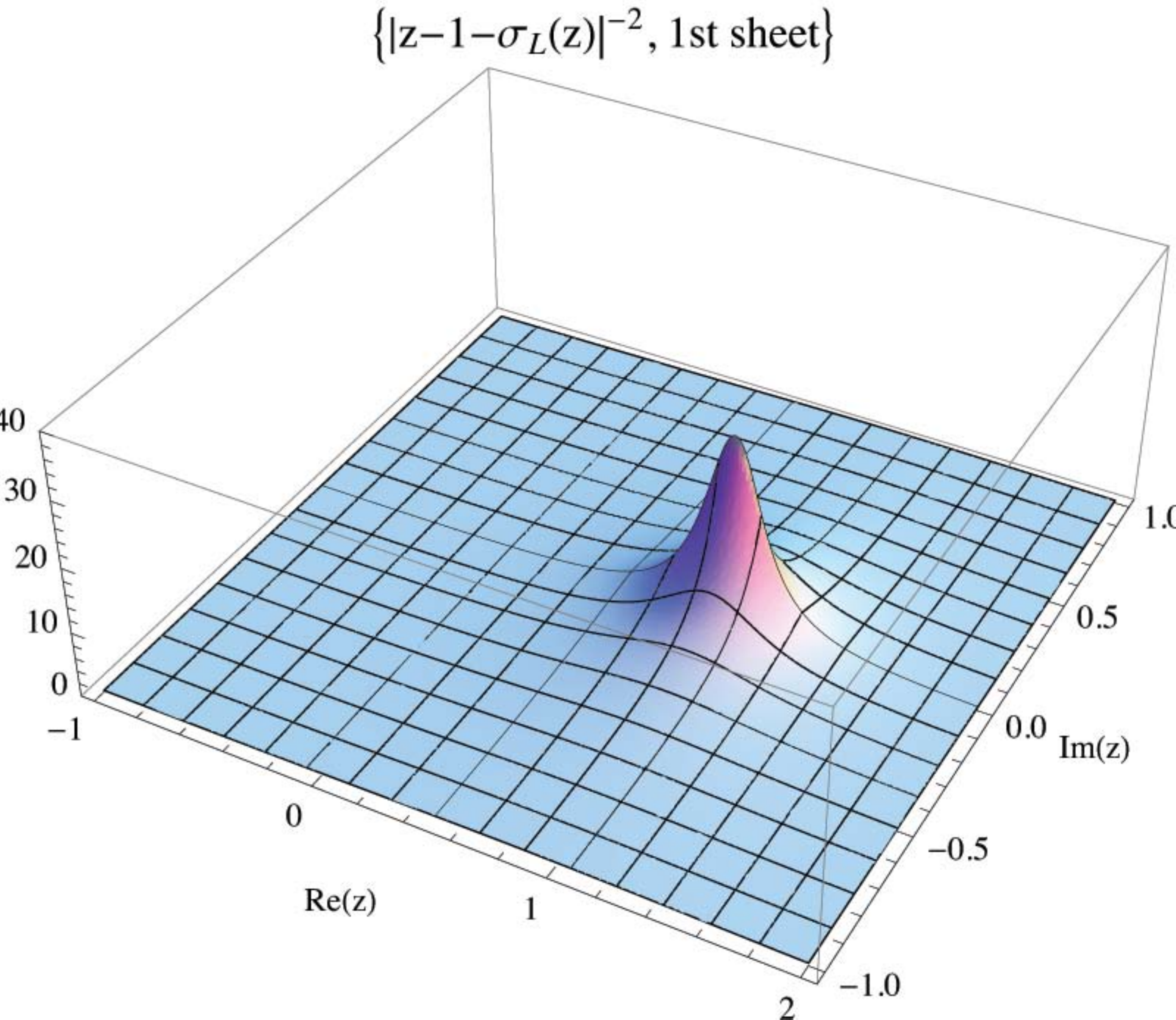} \hskip2cm
\epsfig{width=0.35\textwidth,file=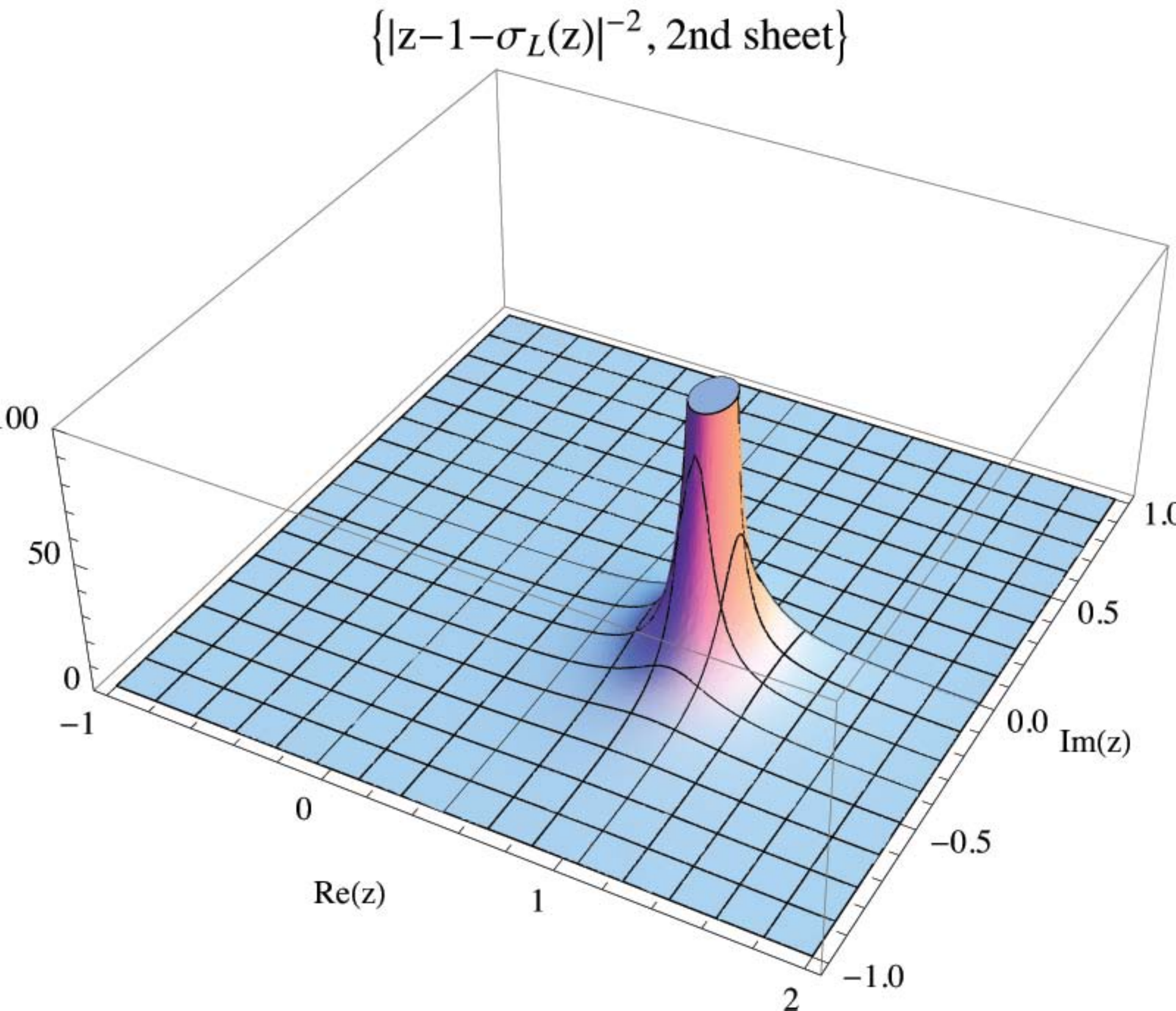}
\end{center}
\caption{The plot of the the square of the modulus of the propagator
function $|z-1-\sigma_L(z)|^{-2}$ on the first and the second sheet
for $a_i=0$. The pole on the second sheet and the peak on the first
sheet correspond to the $\rho(770)$.} \label{poleL}
\end{figure}

In the $1^{+-}$ channel, there is no tree-level pole in the
propagator. The structure of the poles of the Dyson resumed
propagator is strongly dependent on the parameters $b_i$ and
$\gamma$ in this case. Let us illustrate this briefly. Note
\emph{e.g.} that, the equation (\ref{poleT}) can have (exact)
solution $x=1$ on the first sheet provided the parameters $b_{i}$
satisfy the following constraint
\begin{equation}
\sum_{i=0}^{3}b_{i}=-\pi \frac{M_{\mathrm{phys}}}{\Gamma _{\mathrm{phys}}}%
\sim -16  \label{bconstraint1}
\end{equation}%
where the numerical estimate corresponds to $(M_{\mathrm{phys}},\Gamma _{%
\mathrm{phys}})\sim (M_{\rho },\Gamma _{\rho })$. In order to interpret this
solution as a $1^{+-\text{ }}$bound state pole we need the residuum $Z_{A}$
at this pole to be positive,\emph{\ i.e.}
\begin{equation}
Z_{A}^{-1}={\sigma }_{T}^{r}(x)^{^{\prime }}|_{x=1}=\frac{1}{\pi }\frac{%
\Gamma _{\mathrm{phys}}}{M_{\mathrm{phys}}}\sum_{j=1}^{3}jb_{j}>0
\label{bconstraint2}
\end{equation}%
otherwise the pole is a negative norm ghost state. Of course, from the
phenomenological point of view, both these possibilities are meaningless.
Note also that, the constraints (\ref{bconstraint1}) and (\ref{bconstraint2}%
) require unnatural large values of the parameters $b_{i}$ and it is also in
a conflict with the large $N_{C}$ counting\footnote{%
While $b_{i}=O(1)$ in the large $N_{C}$ limit, the right hand side of (\ref%
{bconstraint1}) begaves as $O(N_{C})$.}.

For $\gamma =0$, a pathological tachyonic solution of (\ref{poleT}) exists
for $x=-2$ provided
\begin{equation*}
\sum_{i=0}^{3}(-2)^{i}b_{i}=-\pi \frac{M_{\mathrm{phys}}}{\Gamma _{\mathrm{%
phys}}}
\end{equation*}%
which might be satisfied with more reasonable values of the parameters $%
b_{i} $ than in the previous case. More generally, we can have pathological
poles $x=x_{\gamma }$ where $x_{\gamma }$ is a solution of
\begin{equation*}
2+(1+6\gamma +\gamma ^{2})x_{\gamma }+2\gamma ^{2}x_{\gamma }^{2}=0.
\end{equation*}%
This $x_{\gamma }$ is a pole of the propagator on both physical and
unphysical sheets under the conditions that the following constraint on the
parameters $b_{i}$%
\begin{equation*}
\sum_{i=0}^{3}x_{\gamma }^{i}b_{i}=-\pi \frac{M_{\mathrm{phys}}}{\Gamma _{%
\mathrm{phys}}}
\end{equation*}%
is satisfied. Here $x_{\gamma }$ is real (and negative) for $|\gamma +5|>2%
\sqrt{6}$ and it represents therefore a physically unacceptable tachyonic
pole. Outside of this region of $\gamma $ we get pair of complex conjugate
poles on the physical sheet with $\mathrm{Re}x_{\gamma }>0$ when $-3+2\sqrt{2%
}>\gamma >-3-2\sqrt{2}$.

\begin{figure}[t]
\par
\begin{center}
\epsfig{width=0.35\textwidth,file=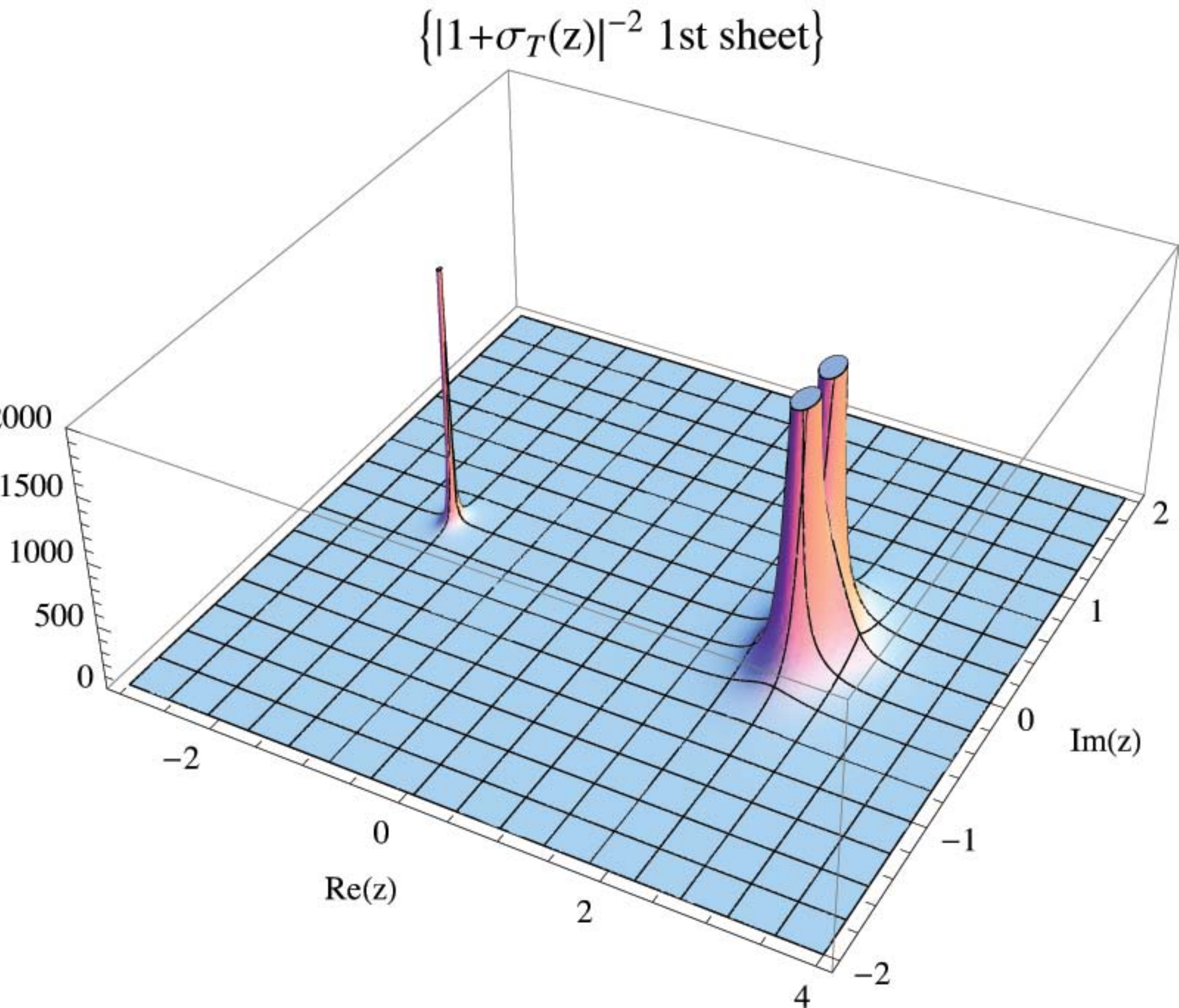} \hskip2cm
\epsfig{width=0.35\textwidth,file=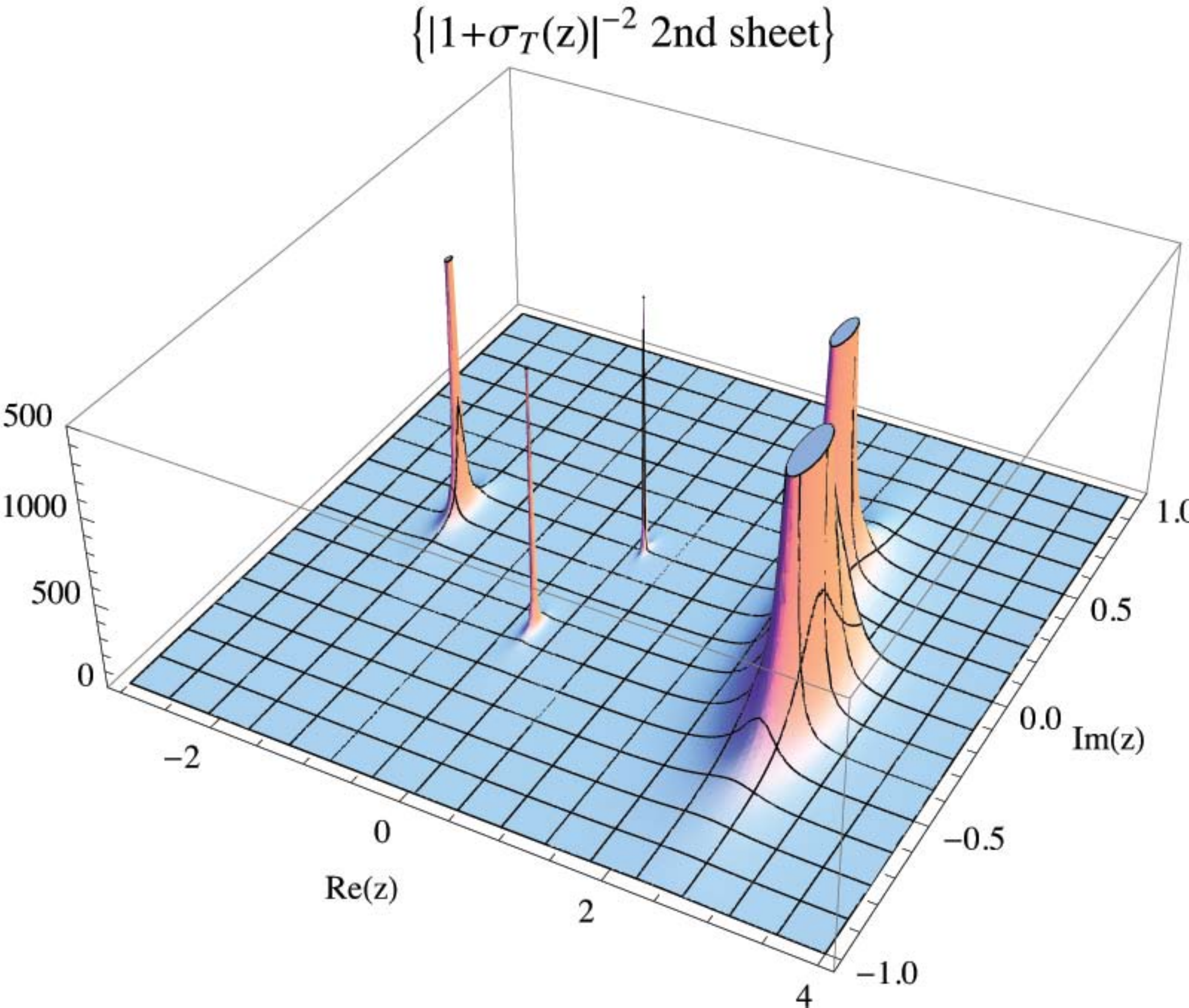}
\end{center}
\caption{The plot of the the square of the modulus of the propagator
function $|1+\sigma_T(z)|^{-2}$ on the first and the second sheet
for $b_0=-2.16$, $b_1=-3.66$, $b_2=-4.45$, $b_3=1.47$ and
$\gamma=0$. Along the desired $b_1(1235)$ pole on the 2nd sheet
($z=2.552-0.295{{\rm i}}$) and peak on the 1st sheet, additional
structures appear.} \label{pole_b1}
\end{figure}

However, we can easily get a more realistic situation and ensure that the
position of the complex pole $z_{R}=x_{R}-\mathrm{i}y_{R}$ on the second
sheet in the $1^{+-\text{ }}$channel corresponds \emph{e.g.} to a resonance $%
b_{1}(1235)$. In this case, two conditions for $b_{i}$, and
$\gamma $ have to be satisfied, which correspond to the real and
imaginary part of the pole equation $1+{\sigma
}_{T}^{r}(z_{R})=0$. This allows us to eliminate two of the five
independent parameters in favor of the mass and the width of the
desired resonance\footnote{Similar conditions we get in the
$1^{--}$ channel, provided we demand to generate \emph{e.g.} $\rho
(1450)$ dynamically.}. However, it might be difficult to eliminate
additional pathological poles in the assumed region of
applicability of $R\chi T$. We illustrate this in the Fig.
\ref{pole_b1}, where the the square of the modulus of the
propagator function $|1+\sigma_T(z)|^{-2}$ on the first and the
second sheet for $b_0=-2.16$, $b_1=-3.66$, $b_2=-4.45$, $b_3=1.47$
and $\gamma=0$ is plotted on the first and the second sheet. In
addition to the desired $b_1(1235)$  pole on the second sheet we
get also four additional poles on the second sheet which is
difficult to interpret physically as well as two additional
structures the first sheet one of which can be interpreted as an
tachyonic pole.

In general it is not so straightforward to formulate the
conditions for $a_{i}$, $b_{i}$, and $\gamma $ under which there are \emph{no%
} additional poles on the real axis in the antisymmetric tensor field
propagator. Because $\mathrm{Im}\sigma _{L}^{r}(x+i0)$ is negative for $x>0$
(and similarly $\mathrm{Im}\sigma _{T}^{r}(x+i0)$ is positive for $x>1$), we
can clearly conclude, that there is no real pole in these regions on the
first and the second sheet. As far as the regions of $x<0$ (for $\sigma
_{L}^{r}$) and $x<1$ (for $\sigma _{T}^{r}$) are concerned, we can proceed
as follows. Note, that we can write for the functions $\widehat{J}(x)$ and $%
\widehat{B}(x)$ the following dispersive representation
\begin{eqnarray*}
\widehat{B}(x) &=&2+x+(x+1)^{2}\int_{0}^{\infty }\frac{dx^{^{\prime }}}{%
(x^{^{\prime }}+1)^{2}}\frac{1}{x^{^{\prime }}-x} \\
&\equiv &2+x+b(x) \\
\widehat{J}(x) &=&\int_{1}^{\infty }\frac{dx^{^{\prime }}}{x^{^{\prime }}}%
\left( 1-\frac{1}{x^{^{\prime }}}\right) \frac{1}{x^{^{\prime }}-x},
\end{eqnarray*}%
from which the representation (\ref{dispersiv_sigma}) for $\Sigma _{L}$ with
desired properties easily follows. From this we can see that on the first
sheet $b(x)$, $\widehat{J}(x)>0$ for $x<0$ and $x<1$ respectively.
Similarly, for $\Sigma _{T}$ we can write
\begin{equation*}
\left( 2+(1+6\gamma +\gamma ^{2})x+2\gamma ^{2}x^{2}\right) \widehat{J}(x)=1+%
\frac{1}{6}\left( 3\gamma ^{2}+18\gamma +5\right) x+j(x)
\end{equation*}%
where
\begin{equation*}
j(x)=x^{2}\int_{1}^{\infty }\frac{dx^{^{\prime }}}{x^{^{\prime }3}}\left( 1-%
\frac{1}{x^{^{\prime }}}\right) \left( 2+(1+6\gamma +\gamma ^{2})x^{^{\prime
}}+2\gamma ^{2}x^{^{\prime }2}\right) \frac{1}{x^{^{\prime }}-x}
\end{equation*}%
and $j(x)>0$ for $x<1$. The equations (\ref{P_L}) and (\ref{poleT}) have
therefore the following structure
\begin{eqnarray}
p_{L}(x) &=&-\frac{1}{\pi }\frac{\Gamma _{\mathrm{phys}}}{M_{\mathrm{phys}}}%
x^{2}\left( b(x)+2\right) -\frac{40}{9}\left( \frac{M_{\mathrm{phys}}}{4\pi
F_{\pi }}\right) ^{2}d_{3}^{2}(x^{2}-1)^{2}\widehat{J}(x)  \label{eq_L} \\
p_{T}(x) &=&-\frac{20}{9}\left( \frac{M_{\mathrm{phys}}}{4\pi F_{\pi }}%
\right) ^{2}d_{3}^{2}(x-1)^{2}\left( j(x)+1\right),  \label{eq_T}
\end{eqnarray}%
where $p_{L,T}(x)$ are the following polynomials of the third order
\begin{eqnarray*}
p_{L}(x) &=&x-1-\frac{1}{\pi }\frac{\Gamma _{\mathrm{phys}}}{M_{\mathrm{phys}%
}}\left[ 1-x^{3}+\sum_{i=1}^{3}a_{i}(x^{i}-1)\right] =(x-1)q_{L}(x) \\
p_{T}(x) &=&1+\frac{1}{\pi }\frac{\Gamma _{\mathrm{phys}}}{M_{\mathrm{phys}}}%
\sum_{i=0}^{3}b_{i}x^{i}+\frac{10}{27}\left( \frac{M_{\mathrm{phys}}}{4\pi
F_{\pi }}\right) ^{2}d_{3}^{2}\left( 3\gamma ^{2}+18\gamma +5\right)
x(x-1)^{2}.
\end{eqnarray*}%
where
\begin{equation*}
q_{L}(x)=1-\frac{1}{\pi }\frac{\Gamma _{\mathrm{phys}}}{M_{\mathrm{phys}}}%
\left( (1+x+x^{2})(a_{3}-1)+a_{1}+a_{2}(x+1)\right)
\end{equation*}%
Because the right hand sides of the equations (\ref{eq_L}) and (\ref{eq_T})
are negative in the regions of interest, the sufficient (but not necessary)
condition of the absence of the poles in these regions is $q_{L}(x)<0$ for $%
x<0$ and $p_{T}(x)>0$ for $x<1$. For $q_{L}(x)$ this can be achieved in many
ways, \emph{e.g. }for
\begin{eqnarray*}
a_{3} &\geq &1 \\
q_{L}(0) &=&1-\frac{1}{\pi }\frac{\Gamma _{\mathrm{phys}}}{M_{\mathrm{phys}}}%
\left( a_{1}+a_{2}+a_{3}-1\right) <0 \\
q_{L}^{^{\prime }}(0) &=&-\frac{1}{\pi }\frac{\Gamma _{\mathrm{phys}}}{M_{%
\mathrm{phys}}}(a_{3}+a_{2}-1)>0
\end{eqnarray*}%
\emph{i.e}.
\begin{equation*}
a_{1}>\pi \frac{M_{\mathrm{phys}}}{\Gamma _{\mathrm{phys}}}%
,\,\,a_{2}<0,\,\,a_{3}\geq 1.\,\,
\end{equation*}%
Note however, that such a condition for $a_{1}$ requires unnatural value for
this parameter and is in a conflict with the large $N_{C}$ counting.
Similarly, the condition $p_{T}(x)>0$ can be ensured \emph{e.g.} when the
coefficients at the third power of $x$ vanish identically, \emph{i.e.}
\begin{equation*}
b_{3}=-\frac{10}{27}\left( \frac{M_{\mathrm{phys}}}{4\pi F_{\pi }}\right)
^{2}\pi \frac{M_{\mathrm{phys}}}{\Gamma _{\mathrm{phys}}}d_{3}^{2}\left(
3\gamma ^{2}+18\gamma +5\right) ,
\end{equation*}%
the coefficients at the second power of $x$ are positive, \emph{i.e}
\begin{equation*}
b_{2}>\frac{20}{27}\left( \frac{M_{\mathrm{phys}}}{4\pi F_{\pi }}\right)
^{2}\pi \frac{M_{\mathrm{phys}}}{\Gamma _{\mathrm{phys}}}d_{3}^{2}\left(
3\gamma ^{2}+18\gamma +5\right) ,
\end{equation*}%
and
\begin{eqnarray*}
p_{T}(1) &=&1+\frac{1}{\pi }\frac{\Gamma _{\mathrm{phys}}}{M_{\mathrm{phys}}}%
\sum_{i=0}^{3}b_{i}>0 \\
p_{T}^{^{\prime }}(1) &=&\frac{1}{\pi }\frac{\Gamma _{\mathrm{phys}}}{M_{%
\mathrm{phys}}}\sum_{j=0}^{3}b_{j}j>0.
\end{eqnarray*}%
On the contrary to the previous case, these conditions respect the large $%
N_{C}$ counting. Therefore without any detailed information about the actual
value of the $a_{i}$ and $b_{i}$ it seems to be quite natural to have
tachyonic pole in the $1^{--}$ channel and no bound states or tachyon poles
in the $1^{+-}$ channel of the propagator.

\section{Summary and discussion\label{Section_summary}}

In this paper we have studied and illustrated various aspects of the
renormalization procedure of \ the Resonance Chiral Theory using the
spin-one resonance self-energy and the corresponding propagator as a
concrete example. The explicit calculation of the one-loop self-energies
within three possible formalisms for the description of the spin-one
resonances, namely the Proca filed, antisymmetric tensor field and the first
order formalism is the main result of our article. Because the theory is
non-renormalizable and the loop corrections break the ordinary chiral power
counting, we had presumed an accurence of problems of several types which
have proved to be true within our explicit example.

The first sort of problems concerned the technical aspects of the process of
renormalization, namely the organization of the loop corrections and the
counterterms and the mixing of the ordinary chiral orders by the loops. In
order to organize our calculations we have proposed a self-consistent scheme
for classification of the one-particle irreducible graphs $\Gamma $ and
corresponding counterterms $\mathcal{O}_{i}$ which renormalize its
superficial divergences. The classification is according to the indices $%
i_{\Gamma }$ and $i_{\mathcal{O}_{i}}$ assigned to graph $\Gamma $ and
operator $\mathcal{O}_{i}$ respectively. Though the scheme based on $i_{%
\mathcal{O}}$ restricts both the chiral order of the chiral building blocs
(number of derivatives and external sources) as well as the number of
resonance fields in the operators in the $R\chi T$ Lagrangian at each fixed
order and can be understood as a combination of the chiral and $1/N_{C}$
counting, it is however not possible to assign to $i_{\Gamma }$ a clear
physical meaning connected with the infrared characteristics of the graphs $%
\Gamma $. Nevertheless the scheme works at least formally and can be used
for the proof of the renormalizability of $R\chi T$ to given order $%
i_{\Gamma },i_{\mathcal{O}_{i}}\leq $ $i_{\max }$. We have used it at the
level $i_{\max }\leq 6$ and proved that the complete set of counterterms
from zero up to six derivatives is necessary to renormalize the divergences
of the one-loop self-energies in the contrary to the naive expectations
based on the usual chiral powercounting.

The last aspect, namely that the complete set of counterterms
including also those with two derivatives (\emph{i.e.} the kinetic
terms) is necessary, is connected to the second sort of problems.
The tree level Lagrangian is constructed using just one of such a
kinetic term in order to ensure the propagation of just three
degrees of freedom corresponding to the spin-one particle state. If
we would include all possible kinetic terms with two derivatives
into the free Lagrangian, we would get (according to the formalism
used) additional poles in the free propagator corresponding to the
additional one-particle states some of them being necessarily either
negative norm ghost or tachyon. This was the first signal of the
problems with unphysical degrees of freedom connected with the
one-loop corrections to the self-energies. The higher derivative
kinetic terms further increase the number of these extra degrees of
freedom. We have studied this feature also using the path integral
representation and integrated in additional fields which appear to
be responsible for the additional propagator poles.

The problems with additional degrees of freedom are also connected
with the well known fact that the propagator obtained by means of
the Dyson re-summation of the perturbative one-particle irreducible
self-enery insertions might be \ incompatible with the
K\"{a}ll\'{e}n-Lehman spectral representation even in the case of
the renormalizable theories \cite{Symanzik}. As is well known, in
this case tachyonic or negative norm ghost state can appear as an
additional pole. Such an extra pole is usually harmless because it
is very far from the energy range where the theory is applicable. In
the power-counting non-renormalizable effective theories like $R\chi
T$ such problems are much stronger either because of the worse UV
behavior of the self-energies (which increases the number of
additional poles) or because the additional pathological poles might
lie near the region where the theory was assumed to be valid. The
nontrivial Lorentz structure of the fields describing spin-one
resonances further complicates this delineation because some of the
additional poles might have different quantum numbers than the
original tree-level degrees of freedom. As far as this type of poles
is concerned, we have demonstrated using the path integral formalism
that it can be eliminated by means of the requirement of additional
protective symmetry of the interaction Lagrangian, which is an
analog of $U(1)$ gauge transformation known for the Proca and
Rarita-Schwinger fields. However, these symmetries are in general in
conflict with chiral symmetry, though individual interaction
vertices can posses such a symmetry accidentally.

The results of our calculations proved to fit this general picture.
Using the explicit example of the one-loop antisymmetric tensor
self-energy we have shown that the Dyson re-summed propagator has
always (ie. irrespectively to the actual values of the couterterm
couplings) at least three additional poles on the first sheet in the
$1^{--}$ channel, just five such poles on the second sheet (one of
them corresponding to the original degree of freedom) and at least
two additional poles on the first sheet in the $1^{+-}$ channel and
at least four such poles on the second sheet. As we have seen in
explicit analysis of the pole equations, without any additional
information about the size of the counterterm couplings and
consequently about the actual values of the renormalization scale
invariant parameters entering the polynomial part of the
self-energies, a rich variety of poles in the propagator is
possible. Some of the poles might be unphysical (complex conjugated
pairs of poles on the first sheet and tachyonic or negative norm
ghosts on both sheets) and some of them even can be situated near or
inside the assumed applicability region of $R\chi T$.

It might be argued that the additional poles are just artifacts of
the inappropriate treatment of the theory and that the one-loop
one-particle irreducible self-energy insertion cannot be re-summed
in order to construct a reliable approximation of the full resonance
propagator. However, the mere truncation of the Dyson series keeping
only first two terms (corresponding to tree-level contribution and
to the strict one-loop correction to the propagator respectively)
generates double poles at $s=M^{2}$ on both sheets and is therefore
in contradiction with the expected analytic structure of the full
propagator. Though this might be an useful approximation of the full
propagator for $s\ll M^{2}$, it cannot be correct in the resonance
region. Therefore provided we would like to use $R\chi T$ at
one-loop also for $s\sim M^{2}$, the construction the propagator
using some sort of re-summation (\emph{i.e.} the Dyson one or its
modifications like \emph{e.g.} the Redmond and Bogolyubov procedure
or Pad\'{e} approximation) might be inevitable. The actual position
of the additional poles (if there are any within the chosen
procedure) might be then understood as a bound limiting the range of
applicability of the theory. In the most optimistic scenario all the
additional poles are far form the region of interest and $R\chi T$
can be treated as a consistent effective theory describing just the
degrees of freedom we start with at the tree level. The less
satisfactory case when only the pathological poles are far-distant,
we can either abandon the theory as inconsistent or alternatively we
can try to interpret the non-pathological poles as a prediction of
the theory corresponding to the dynamical generation of higher
resonances. Such a treatment was used in the case of scalar
resonances in \cite{Boglione:2002vv} (see also
\cite{Tornqvist:1995kr,Tornqvist:1995ay}). Eventually in the case
when all the additional poles lie near $s\sim M^{2}$, either the
approximative construction of the propagator or one-loop $R\chi T$
itself might be problematic. Which scenario actually turns up
depends on the values of the couplings in the $R\chi T$ Lagrangian.

\section*{Acknowledgement}

We would like to thank A.~Pich, J.~J.~Sanz-Cillero and M.~Zdrahal
for useful discussions and valuable comments on the manuscript. We
also thank M.~Chizhov for drawing our attention to an error in
previous version of the manuscript. This work is supported in part
by the Center for Particle Physics (Project no. LC 527), GAUK
(Project no.6908; 114-10/258002), GACR [202/07/P249] and the EU
Contract No. MRTN-CT-2006-035482, "FLAVIAnet". J.~T. is also
supported by the U.S. Department of State (International Fulbright
S\&T award).

\appendix

\section{Additional degrees of freedom in the path integral - the Proca
field \label{PI_Proca}}

Suppose that the interaction Lagrangian has the form
\begin{equation}
\mathcal{L}_{int}=\mathcal{L}_{ct}+\mathcal{L}_{int}^{^{\prime }}
\end{equation}
where $\mathcal{L}_{ct}$ is the toy interaction Lagrangian (\ref{L_V_toy}).
Our aim will be to transform $Z[J]$ to the form of the path integral with
all the additional degrees of freedom represented explicitly in the
Lagrangian and the integration measure. In terms of the transverse and
longitudinal degrees of freedom we get
\begin{eqnarray}
\mathcal{L}_{int}(V_{\perp }-\partial \Lambda ,J,\ldots ) &=&\mathcal{L}%
_{ct}(V_{\perp }-\partial \Lambda ,J,\ldots )+\mathcal{L}_{int}^{^{\prime
}}(V_{\perp }-\partial \Lambda ,J,\ldots )  \notag \\
&=&\frac{\alpha }{2}V_{\perp }^{\mu }\Box V_{\perp \mu }-\frac{\beta }{2}%
(\Box \Lambda )^{2}+\frac{\gamma }{2M^{2}}(\Box V_{\perp }^{\mu })(\Box
V_{\perp \mu })+\frac{\delta }{2M^{2}}(\partial _{\mu }\square \Lambda
)(\partial ^{\mu }\Box \Lambda )  \notag \\
&&+\mathcal{L}_{int}^{^{\prime }}(V_{\perp }-\partial \Lambda ,J\ldots ).
\end{eqnarray}
In order to lower the number of derivatives in the kinetic terms we
integrate in auxiliary scalar fields $\chi $, $\rho $, $\pi $, $\sigma $ and
auxiliary transverse vector field $B_{\perp \mu }$. Writing
\begin{equation}
\exp \left( -i\int \mathrm{d}^{4}x\frac{\beta }{2}(\Box \Lambda )^{2}\right)
=\int \mathcal{D}\chi \exp \left( i\int \mathrm{d}^{4}x\left( \frac{1}{%
2\beta }\chi ^{2}-\partial _{\mu }\chi \partial ^{\mu }\Lambda \right)
\right)
\end{equation}
and similarly for other higher derivative terms we can finally formulate the
theory as
\begin{equation}
Z[J]=\int \mathcal{D}V_{\perp }\mathcal{D}B_{\perp }\mathcal{D}\Lambda
\mathcal{D}\chi \mathcal{D}\rho \mathcal{D}\sigma \mathcal{D}\pi \exp \left(
i\int \mathrm{d}^{4}x\mathcal{L}(V_{\perp },B_{\perp },\Lambda ,\chi ,\rho
,\sigma ,\pi ,J,\ldots )\right)
\end{equation}
with
\begin{eqnarray}
\mathcal{D}B_{\perp } &=&\mathcal{D}B\delta (\partial _{\mu }B^{\mu }) \\
B_{\perp }^{\mu } &=&\left( g^{\mu \nu }-\frac{\partial ^{\mu }\partial
^{\nu }}{\square }\right) B_{\nu }.
\end{eqnarray}
and
\begin{eqnarray}
\mathcal{L}(V_{\perp },B_{\perp },\Lambda ,\chi ,\rho ,\sigma ,\pi ,J,\ldots
) &=&\frac{1}{2}(1+\alpha )V_{\perp }^{\mu }\square V_{\perp \mu }+\frac{1}{2%
}M^{2}V_{\perp }^{\mu }V_{\perp \mu }-\frac{1}{2\gamma }M^{2}B_{\perp }^{\mu
}B_{\perp \mu }-B_{\perp }^{\mu }\square V_{\perp }^{\mu }  \notag \\
&&+\frac{1}{2}M^{2}\partial _{\mu }\Lambda \partial ^{\mu }\Lambda +\frac{1}{%
2\beta }\chi ^{2}-\partial _{\mu }\chi \partial ^{\mu }\Lambda  \notag \\
&&-\frac{1}{2\delta }M^{2}\partial _{\mu }\rho \partial ^{\mu }\rho
-\partial _{\mu }\rho \partial ^{\mu }\sigma -\partial _{\mu }\pi \partial
^{\mu }\Lambda -\pi \sigma  \notag \\
&&+\mathcal{L}_{int}^{^{\prime }}(V_{\perp }-\partial \Lambda ,J,\ldots )
\label{trans_L_V}
\end{eqnarray}
In this formulation the kinetic terms have no more than two derivatives,
however, the number of fields is higher than the actual number of degrees of
freedom. We therefore have to integrate out the redundant variables. As a
first step we diagonalize the kinetic terms performing the shifts
\begin{eqnarray}
V_{\perp }^{\mu } &\rightarrow &V_{\perp }^{\mu }+\frac{1}{1+\alpha }%
B_{\perp }^{\mu }  \notag \\
\Lambda &\rightarrow &\Lambda +\frac{1}{M^{2}}\chi +\frac{1}{M^{2}}\pi
\notag \\
\rho &\rightarrow &\rho -\frac{\delta }{M^{2}}\sigma  \notag \\
\chi &\rightarrow &\chi -\pi  \label{shift_L_V}
\end{eqnarray}
respectively to the form
\begin{eqnarray}
\mathcal{L}(V_{\perp },B_{\perp },\Lambda ,\chi ,\rho ,\sigma ,\pi ,J,\ldots
) &=&\frac{1}{2}(1+\alpha )V_{\perp }^{\mu }\square V_{\perp \mu }+\frac{1}{2%
}M^{2}V_{\perp }^{\mu }V_{\perp \mu }  \notag \\
&&-\frac{1}{2}(1+\alpha )^{-1}B_{\perp }^{\mu }\square B_{\perp }^{\mu }+%
\frac{1}{2}M^{2}\left( (1+\alpha )^{-2}-\gamma ^{-1}\right) B_{\perp }^{\mu
}B_{\perp \mu }  \notag \\
&&+M^{2}(1+\alpha )^{-1}V_{\perp }^{\mu }B_{\perp \mu }  \notag \\
&&+\frac{1}{2}M^{2}\partial _{\mu }\Lambda \partial ^{\mu }\Lambda -\frac{1}{%
2M^{2}}\partial _{\mu }\chi \partial ^{\mu }\chi +\frac{1}{2\beta }(\chi
-\pi )^{2}  \notag \\
&&-\frac{1}{2\delta }M^{2}\partial _{\mu }\rho \partial ^{\mu }\rho +\frac{%
\delta }{2M^{2}}\partial _{\mu }\sigma \partial ^{\mu }\sigma -\pi \sigma
\notag \\
&&+\mathcal{L}_{int}^{^{\prime }}(\overline{V},J,\ldots ).
\label{shifted_L_V}
\end{eqnarray}
where
\begin{equation}
\overline{V}=V_{\perp }+\frac{1}{1+\alpha }B_{\perp }-\partial \Lambda -%
\frac{1}{M^{2}}\partial \chi
\end{equation}
Now the superfluous degrees of freedom are easily identified. Namely, the
fields $\rho $ and $\sigma $ decouple and moreover $\pi $ has no kinetic
term. Both of them can be therefore easily integrated out. As a result of
the gaussian integration we get
\begin{equation}
Z[J]=\int \mathcal{D}V_{\perp }\mathcal{D}B_{\perp }\mathcal{D}\Lambda
\mathcal{D}\chi \mathcal{D}\sigma \exp \left( i\int \mathrm{d}^{4}x\mathcal{L%
}(V_{\perp },B_{\perp },\Lambda ,\chi ,\sigma ,J,\ldots )\right)
\end{equation}
where
\begin{eqnarray}
\mathcal{L}(V_{\perp },B_{\perp },\Lambda ,\chi ,\sigma ,J,\ldots ) &=&\frac{%
1}{2}(1+\alpha )V_{\perp }^{\mu }\square V_{\perp \mu }+\frac{1}{2}%
M^{2}V_{\perp }^{\mu }V_{\perp \mu }  \notag \\
&&-\frac{1}{2}(1+\alpha )^{-1}B_{\perp }^{\mu }\square B_{\perp }^{\mu }+%
\frac{1}{2}M^{2}\left( (1+\alpha )^{-2}-\gamma ^{-1}\right) B_{\perp }^{\mu
}B_{\perp \mu }  \notag \\
&&+M^{2}(1+\alpha )^{-1}V_{\perp }^{\mu }B_{\perp \mu }  \notag \\
&&-\frac{1}{2M^{2}}\partial _{\mu }\chi \partial ^{\mu }\chi +\frac{\delta }{%
2M^{2}}\partial _{\mu }\sigma \partial ^{\mu }\sigma -\frac{1}{2}\beta
\sigma ^{2}-\chi \sigma  \notag \\
&&+\frac{1}{2}M^{2}\partial _{\mu }\Lambda \partial ^{\mu }\Lambda  \notag \\
&&+\mathcal{L}_{int}^{^{\prime }}(\overline{V},J,\ldots ).
\label{integrated_L_V}
\end{eqnarray}
Let us assume $\alpha >-1\,$and $\delta >0$ in what follows. Note that, in
this case the fields $B_{\perp }^{\mu }$ and $\chi $ have opposite minus
sign at their kinetic terms. This is a signal of the appearence of the
negative norm ghosts in the spectrum of the theory. The \textquotedblright
dangerous\textquotedblright\ fields $B_{\perp }^{\mu }$ and $\chi $ mix with
the fields $V_{\perp }^{\mu }$ and $\sigma $ respectively. In order to
identify the mass eigenstates we further rescale the fields
\begin{eqnarray}
V_{\perp }^{\mu } &\rightarrow &(1+\alpha )^{-1/2}V_{\perp }^{\mu }  \notag
\\
B_{\perp }^{\mu } &\rightarrow &(1+\alpha )^{1/2}B_{\perp }^{\mu }  \notag \\
\chi &\rightarrow &M\chi  \notag \\
\sigma &\rightarrow &\delta ^{-1/2}M\sigma
\end{eqnarray}
and afterwards we diagonalize the mass terms
\begin{eqnarray}
\mathcal{L}_{mass} &=&\frac{1}{2}\frac{M^{2}}{1+\alpha }\left( V_{\perp
}^{\mu }V_{\perp \mu }+\left( 1-\frac{(1+\alpha )^{2}}{\gamma }\right)
B_{\perp }^{\mu }B_{\perp \mu }\right)  \notag \\
&&-\frac{1}{2}M^{2}\left( \beta \sigma ^{2}+\delta ^{-1/2}\chi \sigma \right)
\label{L_V_mass}
\end{eqnarray}
by means of an appropriate $Sp(2)$ symplectic rotation of the fields $%
V_{\perp }^{\mu }$, $B_{\perp }^{\mu }$ and $\chi $, $\sigma $
\begin{eqnarray}
V_{\perp }^{\mu } &\rightarrow &V_{\perp }^{\mu }\cosh \theta _{V}+B_{\perp
}^{\mu }\sinh \theta _{V}  \notag \\
B_{\perp }^{\mu } &\rightarrow &V_{\perp }^{\mu }\sinh \theta _{V}+B_{\perp
}^{\mu }\cosh \theta _{V}  \notag \\
\chi &\rightarrow &\chi \cosh \theta _{S}+\sigma \sinh \theta _{S}  \notag \\
\sigma &\rightarrow &\chi \sinh \theta _{S}+\sigma \cosh \theta _{S}.
\end{eqnarray}
This is possible for $(1+\alpha )^{2}>4\gamma $ and $\beta ^{2}>4\delta $,
when the off-diagonal elements of the mass matrix vanish for
\begin{eqnarray}
\tanh \theta _{V} &=&\frac{(1+\alpha )^{2}-2\gamma -(1+\alpha )\sqrt{%
(1+\alpha )^{2}-4\gamma }}{2\gamma }  \notag \\
\tanh \theta _{S} &=&\frac{\sqrt{\beta ^{2}-4\delta }-\beta }{2\delta ^{1/2}}%
.
\end{eqnarray}
We get finally for the generating functional
\begin{equation}
Z[J]=\int \mathcal{D}V_{\perp }\mathcal{D}B_{\perp }\mathcal{D}\Lambda
\mathcal{D}\chi \mathcal{D}\sigma \exp \left( i\int \mathrm{d}^{4}x\mathcal{L%
}(V_{\perp },B_{\perp },\Lambda ,\chi ,\sigma ,J,\ldots )\right)
\end{equation}
where
\begin{eqnarray}
\mathcal{L}(V_{\perp },B_{\perp },\Lambda ,\chi ,\sigma ,J,\ldots ) &=&\frac{%
1}{2}V_{\perp }^{\mu }\square V_{\perp \mu }+\frac{1}{2}M_{V+}^{2}V_{\perp
}^{\mu }V_{\perp \mu }-\frac{1}{2}B_{\perp }^{\mu }\square B_{\perp }^{\mu }+%
\frac{1}{2}M_{V-}^{2}B_{\perp }^{\mu }B_{\perp \mu }  \notag \\
&&+\frac{1}{2}\partial _{\mu }\sigma \partial ^{\mu }\sigma -\frac{1}{2}%
M_{S+}^{2}\sigma ^{2}-\frac{1}{2}\partial _{\mu }\chi \partial ^{\mu }\chi -%
\frac{1}{2}M_{S-}^{2}\chi ^{2}+\frac{1}{2}M^{2}\partial _{\mu }\Lambda
\partial ^{\mu }\Lambda  \notag \\
&&+\mathcal{L}_{int}^{^{\prime }}(\overline{V}^{(\theta )},J,\ldots ).
\notag \\
&&
\end{eqnarray}
where now
\begin{equation}
\overline{V}^{(\theta )}=\frac{\exp \theta _{V}}{(1+\alpha )^{1/2}}(V_{\perp
}+B_{\perp })-\partial \chi \cosh \theta _{S}-\partial \sigma \sinh \theta
_{S}-\partial \Lambda
\end{equation}
and where $M_{V\pm }^{2}$, $M_{S\pm }^{2}$ are the mass eigenvalues (\ref%
{MV}) and (\ref{MS}). The theory is now formulated in terms of two spin one
and two spin zero fields, whereas two of them, namely $B_{\perp }^{\mu }$
and $\chi $, are negative norm ghosts. The field $\Lambda $ do not
correspond to any dynamical degree of freedom, its role is merely to cancel
the spurious poles of the free propagators of the transverse fields $%
V_{\perp }$ and $B_{\perp }$ at $p^{2}=0$.

\section{The additional degrees of freedom in the path integral-the
antisymmetric tensor case\label{PI_antisymmetric}}

\bigskip We assume the interaction Lagrangian to be of the form
\begin{equation}
\mathcal{L}_{int}=\mathcal{L}_{ct}+\mathcal{L}_{int}^{^{\prime }},
\end{equation}
where $\mathcal{L}_{ct}$ is given by (\ref{L_R_toy}) and re-express it in
the terms of the longitudinal and transverse components of the original
field $R_{\mu \nu }$
\begin{equation}
\mathcal{L}_{int}(R_{\parallel }^{\mu \nu }-\frac{1}{2}\varepsilon ^{\mu \nu
\alpha \beta }\widehat{\Lambda }_{\alpha \beta },J,\ldots )=\mathcal{L}%
_{ct}(R_{\parallel }^{\mu \nu }-\frac{1}{2}\varepsilon ^{\mu \nu \alpha
\beta }\widehat{\Lambda }_{\alpha \beta },J,\ldots )+\mathcal{L}%
_{int}^{^{\prime }}(R_{\parallel }^{\mu \nu }-\frac{1}{2}\varepsilon ^{\mu
\nu \alpha \beta }\widehat{\Lambda }_{\alpha \beta },J,\ldots )
\end{equation}
where
\begin{eqnarray}
\mathcal{L}_{ct}(R^{\mu \nu }-\frac{1}{2}\varepsilon ^{\mu \nu \alpha \beta }%
\widehat{\Lambda }_{\alpha \beta }J,\ldots ) &=&\frac{\alpha }{4}%
R_{\parallel }^{\mu \nu }\square R_{\parallel \,\mu \nu }+\frac{\gamma }{%
4M^{2}}(\square R_{\parallel }^{\mu \nu })(\square R_{\parallel \,\mu \nu })
\notag \\
&&+\frac{\beta }{2}(\square \Lambda _{\perp }^{\mu })(\square \Lambda
_{\perp \mu })-\frac{\delta }{2M^{2}}(\partial ^{\alpha }\square \Lambda
_{\perp }^{\mu })(\partial _{\alpha }\square \Lambda _{\perp \mu }).
\end{eqnarray}
We can introduce the auxiliary (longitudinal) antisymmetric tensor field $%
B_{\parallel }^{\mu \nu }$ and (transverse) vector fields $\chi _{\perp
}^{\mu }$, $\rho _{\perp }^{\mu }$, $\sigma _{\perp }^{\mu }$ and $\pi
_{\perp }^{\mu }$ in order to avoid the higher derivative terms and write in
complete analogy with the Proca field case
\begin{equation}
Z[J]=\int \mathcal{D}R_{\parallel }\mathcal{D}B_{\parallel }\mathcal{D}%
\Lambda _{\perp }\mathcal{D}\chi _{\perp }\mathcal{D}\rho _{\perp }\mathcal{D%
}\sigma _{\perp }\mathcal{D}\pi _{\perp }\exp \left( \mathrm{i}\int \mathrm{d%
}^{4}x\mathcal{L}(R_{\parallel },B_{\parallel },\Lambda _{\perp },\chi
_{\perp },\rho _{\perp },\sigma _{\perp },\pi _{\perp },J,\ldots )\right)
\end{equation}
where the measures and fields are
\begin{eqnarray}
\mathcal{D}B_{\parallel } &=&\mathcal{D}B\delta (\partial _{\alpha }B_{\mu
\nu }+\partial _{\nu }B_{\alpha \mu }+\partial _{\mu }B_{\nu \alpha }) \\
B_{\parallel }^{\mu \nu } &=&-\frac{1}{2\square }(\partial ^{\mu }g^{\nu
\alpha }\partial ^{\beta }+\partial ^{\nu }g^{\mu \beta }\partial ^{\alpha
}-(\mu \leftrightarrow \nu ))B_{\alpha \beta }
\end{eqnarray}
and for $\phi ^{\mu }=\chi ^{\mu }$, $\rho ^{\mu }$, $\sigma ^{\mu }$ and $%
\pi ^{\mu }$
\begin{eqnarray}
\mathcal{D}\phi _{\perp } &=&\mathcal{D}\phi \delta (\partial _{\mu }\phi
^{\mu }) \\
\phi _{\perp }^{\mu } &=&\left( g^{\mu \nu }-\frac{\partial ^{\mu }\partial
^{\nu }}{\square }\right) \phi _{\perp \nu }.
\end{eqnarray}
The Lagrangian is then
\begin{eqnarray}
\mathcal{L} &=&\frac{1+\alpha }{4}R_{\parallel }^{\mu \nu }\square
R_{\parallel \,\mu \nu }+\frac{1}{4}M^{2}R_{\parallel }^{\mu \nu
}R_{\parallel \,\mu \nu }  \notag \\
&&-\frac{1}{\gamma }M^{2}B_{\parallel }^{\mu \nu }B_{\parallel \,\mu \nu
}+B_{\parallel }^{\mu \nu }\square R_{\parallel \,\mu \nu }  \notag \\
&&+\frac{1}{2}M^{2}\Lambda _{\perp }^{\mu }\square \Lambda _{\perp \mu }-%
\frac{1}{2\beta }\chi _{\perp }^{\mu }\chi _{\perp \mu }-\chi _{\perp }^{\mu
}\square \Lambda _{\perp \mu }  \notag \\
&&+\frac{1}{2\delta }M^{2}\partial ^{\alpha }\rho _{\perp }^{\mu }\partial
_{\alpha }\rho _{\perp \mu }-\partial ^{\alpha }\rho _{\perp }^{\mu
}\partial _{\alpha }\sigma _{\perp \mu }-\partial ^{\alpha }\Lambda _{\perp
}^{\mu }\partial _{\alpha }\pi _{\perp \mu }-\pi _{\perp }^{\mu }\sigma
_{\perp \mu }  \notag \\
&&+\mathcal{L}_{int}\left( R^{\mu \nu }-\frac{1}{2}\varepsilon ^{\mu \nu
\alpha \beta }\widehat{\Lambda }_{\alpha \beta },J,\ldots \right) .
\label{trans_L_R}
\end{eqnarray}
Note that, the fields $\chi ^{\mu }$, $\rho ^{\mu }$, $\sigma ^{\mu }$ and $%
\pi ^{\mu }$ mix with $\Lambda ^{\mu }$ and are therefore pseudovectors. The
Lagrangian (\ref{trans_L_R}) is completely analogical to (\ref{trans_L_V})
up to the more Lorentz indices, so will be brief in the next steps. First we
identify the redundant degrees of freedom diagonalizing the kinetic terms by
means of the following sequence of shifts (cf. (\ref{shift_L_V}))
\begin{eqnarray}
R_{\parallel }^{\mu \nu } &\rightarrow &R_{\parallel }^{\mu \nu }-2(1+\alpha
)^{-1}B_{\parallel }^{\mu \nu }  \notag \\
\Lambda _{\perp }^{\mu } &\rightarrow &\Lambda _{\perp }^{\mu }+\frac{1}{%
M^{2}}\chi _{\perp }^{\mu }-\frac{1}{M^{2}}\pi _{\perp }^{\mu }  \notag \\
\rho _{\perp }^{\mu } &\rightarrow &\rho _{\perp }^{\mu }+\frac{\delta }{%
M^{2}}\sigma _{\perp }^{\mu }  \notag \\
\chi _{\perp }^{\mu } &\rightarrow &\chi _{\perp }^{\mu }+\pi _{\perp }^{\mu
}.
\end{eqnarray}
As a result we get the Lagrangian in the form (cf. (\ref{shifted_L_V}))
\begin{eqnarray}
\mathcal{L} &=&\frac{1}{4}(1+\alpha )R_{\parallel }^{\mu \nu }\square
R_{\parallel \,\mu \nu }+\frac{1}{4}M^{2}R_{\parallel }^{\mu \nu
}R_{\parallel \,\mu \nu }  \notag \\
&&-(1+\alpha )^{-1}B_{\parallel }^{\mu \nu }\square B_{\parallel \,\mu \nu
}+(1+\alpha )^{-2}M^{2}B_{\parallel }^{\mu \nu }B_{\parallel \,\mu \nu }-%
\frac{1}{\gamma }M^{2}B_{\parallel }^{\mu \nu }B_{\parallel \,\mu \nu }
\notag \\
&&-(1+\alpha )^{-1}M^{2}R_{\parallel }^{\mu \nu }B_{\parallel \,\mu \nu }
\notag \\
&&+\frac{1}{2}M^{2}\Lambda _{\perp }^{\mu }\square \Lambda _{\perp \mu }-%
\frac{1}{2M^{2}}\chi _{\perp }^{\mu }\square \chi _{\perp \mu }-\frac{1}{%
2\beta }(\chi _{\perp }^{\mu }+\pi _{\perp }^{\mu })(\chi _{\perp \mu }+\pi
_{\perp \mu })  \notag \\
&&+\frac{1}{2\delta }M^{2}\partial ^{\alpha }\rho _{\perp }^{\mu }\partial
_{\alpha }\rho _{\perp \mu }-\frac{\delta }{2M^{2}}\partial ^{\alpha }\sigma
_{\perp }^{\mu }\partial _{\alpha }\sigma _{\perp \mu }-\pi _{\perp }^{\mu
}\sigma _{\perp \mu }  \notag \\
&&+\mathcal{L}_{int}\left( \overline{R},J,\ldots \right) ,
\end{eqnarray}
where
\begin{equation}
\overline{R}^{\mu \nu }=R_{\parallel }^{\mu \nu }-2(1+\alpha
)^{-1}B_{\parallel }^{\mu \nu }-\frac{1}{2}\varepsilon ^{\mu \nu \alpha
\beta }(\widehat{\Lambda }_{\alpha \beta }+\frac{1}{M^{2}}\widehat{\chi }%
_{\perp \alpha \beta }).
\end{equation}
Integrating out the superfluous fields $\rho _{\perp \mu }$ and $\pi _{\perp
\mu }$ which are decoupled from the interaction we get
\begin{equation}
Z[J]=\int \mathcal{D}R_{\parallel }\mathcal{D}B_{\parallel }\mathcal{D}%
\Lambda _{\perp }\mathcal{D}\chi _{\perp }\mathcal{D}\rho _{\perp }\mathcal{D%
}\sigma _{\perp }\mathcal{D}\pi _{\perp }\exp \left( \mathrm{i}\int \mathrm{d%
}^{4}x\mathcal{L}(R_{\parallel },B_{\parallel },\Lambda _{\perp },\chi
_{\perp },\rho _{\perp },\sigma _{\perp },\pi _{\perp },J,\ldots )\right)
\end{equation}
with (cf. (\ref{integrated_L_V}))
\begin{eqnarray}
\mathcal{L} &=&\frac{1}{4}(1+\alpha )R_{\parallel }^{\mu \nu }\square
R_{\parallel \,\mu \nu }+\frac{1}{4}M^{2}R_{\parallel }^{\mu \nu
}R_{\parallel \,\mu \nu }  \notag \\
&&-(1+\alpha )^{-1}B_{\parallel }^{\mu \nu }\square B_{\parallel \,\mu \nu
}+(1+\alpha )^{-2}M^{2}B_{\parallel }^{\mu \nu }B_{\parallel \,\mu \nu }-%
\frac{1}{\gamma }M^{2}B_{\parallel }^{\mu \nu }B_{\parallel \,\mu \nu }
\notag \\
&&-(1+\alpha )^{-1}M^{2}R_{\parallel }^{\mu \nu }B_{\parallel \,\mu \nu }
\notag \\
&&+\frac{1}{2}M^{2}\Lambda _{\perp }^{\mu }\square \Lambda _{\perp \mu }
\notag \\
&&-\frac{1}{2M^{2}}\chi _{\perp }^{\mu }\square \chi _{\perp \mu }+\frac{%
\delta }{2M^{2}}\sigma _{\perp }^{\mu }\square \sigma _{\perp \mu }+\frac{1}{%
2}\beta \sigma _{\perp }^{\mu }\sigma _{\perp \mu }+\chi _{\perp }^{\mu
}\sigma _{\perp \mu }  \notag \\
&&+\mathcal{L}_{int}(S,J,\ldots )
\end{eqnarray}
Again, assuming $\alpha >-1$ and $\delta >0$ we have two pairs of fields
with opposite signs of the kinetic terms, namely $(R_{\parallel }^{\mu \nu
},B_{\parallel }^{\mu \nu })$ and $(\chi _{\perp }^{\mu },\sigma _{\perp
}^{\mu })$ respectively.The fields within both of these these pairs mix.
After re-scaling
\begin{eqnarray}
R_{\parallel }^{\mu \nu } &\rightarrow &(1+\alpha )^{-1/2}R_{\parallel
}^{\mu \nu } \\
B_{\parallel }^{\mu \nu } &\rightarrow &\frac{1}{2}(1+\alpha
)^{1/2}B_{\parallel }^{\mu \nu }  \notag \\
\chi _{\perp }^{\mu } &\rightarrow &M\chi _{\perp }^{\mu }  \notag \\
\sigma _{\perp }^{\mu } &\rightarrow &\frac{M}{\sqrt{\delta }}\sigma _{\perp
}^{\mu }  \notag
\end{eqnarray}
the form of the mass matrix becomes identical to that of (\ref{L_V_mass})
(with obvious identifications) and we can therefore perform the same
symplectic rotations as in the Proca field case and under the same
assumptions to get diagonal mass terms corresponding to the eigenvalues (\ref%
{MV}, \ref{MS}). As a result we have found four spin-one states, two of them
being negative norm ghosts, namely $B_{\parallel }^{\mu \nu }$ and $\sigma
_{\perp }^{\mu }$ and two of them with opposite parity, namely $\chi _{\perp
}^{\mu }$ and $\sigma _{\perp }^{\mu }$. As in the Proca field case, the
field $\Lambda _{\perp }^{\mu }$ effectively compensates for the spurious $%
p^{2}=0$ poles in the $R_{\parallel }^{\mu \nu }$ and $B_{\parallel }^{\mu
\nu }$ propagators within Feynman graphs.\qquad \qquad

\section{Path integral formulation of the first order formalism\label%
{PI_first_order}}

Within the first order formalism, the path integral formulation is merely a
generalization of the previous two cases, so we will be as brief as possible
in what follows. Note that, now the kinetic term is invariant with respect
to the both transformations (\ref{V_gauge}) and (\ref{R_gauge}), therefore
the manifestation of the degrees of freedom within the the path integral
formalism can be done in analogy with the previous two cases. Using triple
Faddeev-Popov trick in the path integral
\begin{equation}
Z[J]=\int \mathcal{D}R\exp \left( \mathrm{i}\int \mathrm{d}^{4}x\left(
MV_{\nu }\partial _{\mu }R^{\mu \nu }+\frac{1}{2}M^{2}V_{\mu }V^{\mu }+\frac{%
1}{4}M^{2}R_{\mu \nu }R^{\mu \nu }+\mathcal{L}_{int}(V^{\alpha },R^{\mu \nu
},J,\ldots )\right) \right)
\end{equation}
we get
\begin{equation}
Z[J]=\int \mathcal{D}R_{\parallel }\mathcal{D}\Lambda _{\perp }\mathcal{D}%
V_{\perp }\mathcal{D}\Lambda \exp \left( \mathrm{i}\int \mathrm{d}^{4}x%
\mathcal{L}(R_{\parallel }^{\mu \nu },\Lambda _{\perp }^{\rho },V_{\perp
}^{\alpha }\ldots ,\Lambda ,J,\ldots )\right)
\end{equation}
where
\begin{eqnarray}
\mathcal{L}(R_{\parallel }^{\mu \nu },\Lambda _{\perp }^{\rho },V_{\perp
}^{\alpha }\ldots ,\Lambda ,J,\ldots ) &=&MV_{\perp \nu }\partial _{\mu
}R_{\parallel }^{\mu \nu }+\frac{1}{2}M^{2}V_{\perp \mu }V_{\perp }^{\mu }+%
\frac{1}{4}M^{2}R_{\parallel \mu \nu }R_{\parallel }^{\mu \nu }  \notag \\
&&+\frac{1}{2}M^{2}\Lambda _{\perp }^{\mu }\square \Lambda _{\perp \mu }+%
\frac{1}{2}M^{2}\partial _{\mu }\Lambda \partial ^{\mu }\Lambda  \notag \\
&&+\mathcal{L}_{int}(R_{\parallel }^{\mu \nu }-\frac{1}{2}\varepsilon ^{\mu
\nu \alpha \beta }\widehat{\Lambda }_{\alpha \beta },V_{\perp }^{\alpha
}-\partial ^{\alpha }\Lambda ,J,)  \label{FO_phys}
\end{eqnarray}
and, as in the previous subsections
\begin{eqnarray}
\mathcal{D}R_{\parallel } &=&\mathcal{D}R\delta (\partial_{\alpha } R_{\mu
\nu }+\partial _{\nu }R_{ \alpha \mu}+\partial _{\mu }R_{\nu \alpha })
\notag \\
\mathcal{D}\Lambda _{\perp } &=&\mathcal{D}\Lambda \delta (\partial _{\mu
}\Lambda ^{\mu })  \notag \\
\mathcal{D}V_{\perp } &=&\mathcal{D}V\delta (\partial _{\mu }\Lambda ^{\mu })
\notag \\
R_{\parallel }^{\mu \nu } &=&-\frac{1}{2\square }(\partial ^{\mu }g^{\nu
\alpha }\partial ^{\beta }+\partial ^{\nu }g^{\mu \beta }\partial ^{\alpha
}-(\mu \leftrightarrow \nu ))R_{\alpha \beta }  \notag \\
\Lambda _{\perp }^{\mu } &=&\left( g^{\mu \nu }-\frac{\partial ^{\mu
}\partial ^{\nu }}{\square }\right) \Lambda _{\nu }  \notag \\
V_{\perp }^{\mu } &=&\left( g^{\mu \nu }-\frac{\partial ^{\mu }\partial
^{\nu }}{\square }\right) V_{\nu }.
\end{eqnarray}
In order to diagonalize the kinetic terms we perform a shift
\begin{equation}
V_{\perp }^{\mu }\rightarrow V_{\perp }^{\mu }-\frac{1}{M}\partial _{\nu
}R_{\parallel }^{\nu \mu }
\end{equation}
and get
\begin{eqnarray}
\mathcal{L}(R_{\parallel }^{\mu \nu },\Lambda _{\perp }^{\rho },V_{\perp
}^{\alpha }\ldots ,\Lambda ,J,\ldots ) &=&\frac{1}{4}R_{\parallel }^{\mu \nu
}\square R_{\parallel \,\mu \nu }+\frac{1}{4}M^{2}R_{\parallel \mu \nu
}R_{\parallel }^{\mu \nu }+\frac{1}{2}M^{2}V_{\perp \mu }V_{\perp }^{\mu }
\notag \\
&&+\frac{1}{2}M^{2}\Lambda _{\perp }^{\mu }\square \Lambda _{\perp \mu }+%
\frac{1}{2}M^{2}\partial _{\mu }\Lambda \partial ^{\mu }\Lambda  \notag \\
&&+\mathcal{L}_{int}(R_{\parallel }^{\mu \nu }-\frac{1}{2}\varepsilon ^{\mu
\nu \alpha \beta }\widehat{\Lambda }_{\alpha \beta },V_{\perp }^{\alpha }-%
\frac{1}{M}\partial _{\nu }R_{\parallel }^{\nu \mu }-\partial ^{\alpha
}\Lambda ,J,).  \notag \\
&&
\end{eqnarray}
The discussion of the role of the field $R_{\parallel }^{\mu \nu }$ and the $%
\Lambda _{\perp }^{\mu }$ is the same as in the antisymmetric tensor case.
The extra fields $V_{\perp }^{\mu }$ and $\Lambda $ do not correspond to the
original degree of freedom, their free propagators are
\begin{eqnarray}
\Delta _{V\perp }^{\mu \nu }(p) &=&\frac{P^{T\,\mu \nu }}{M^{2}} \\
\Delta _{\Lambda }(p) &=&\frac{1}{M^{2}}\frac{1}{p^{2}}
\end{eqnarray}
with spurious poles at $p^{2}=0$. According to the form of the interaction,
only the combination with spurious poles cancelled, namely
\begin{equation}
\Delta _{V\perp }^{\mu \nu }(p)+p^{\mu }p^{\nu }\Delta _{\Lambda }(p)+\frac{1%
}{M^{2}}p_{\alpha }p_{\beta }\Delta _{\parallel }^{\alpha \mu \beta \nu
}(p)=-\frac{P^{T\,\mu \nu }}{p^{2}-M^{2}}+\frac{P^{L\,\,\mu \nu }}{M^{2}}
\end{equation}
enters the Feynman graphs.

Alternatively, we could make in (\ref{FO_phys}) the following shift
\begin{equation}
R_{\parallel }^{\mu \nu }\rightarrow R_{\parallel }^{\mu \nu }+\frac{1}{M}%
\left( \partial ^{\mu }V_{\perp }^{\nu }-\partial ^{\nu }V_{\perp }^{\mu
}\right)
\end{equation}
leading to
\begin{eqnarray}
\mathcal{L}(R_{\parallel }^{\mu \nu },\Lambda _{\perp }^{\rho },V_{\perp
}^{\alpha }\ldots ,\Lambda ,J,\ldots ) &=&\frac{1}{2}V_{\perp \mu }\square
V_{\perp }^{\mu }+\frac{1}{2}M^{2}V_{\perp \mu }V_{\perp }^{\mu }+\frac{1}{4}%
M^{2}R_{\parallel \mu \nu }R_{\parallel }^{\mu \nu }  \notag \\
&&+\frac{1}{2}M^{2}\Lambda _{\perp }^{\mu }\square \Lambda _{\perp \mu }+%
\frac{1}{2}M^{2}\partial _{\mu }\Lambda \partial ^{\mu }\Lambda  \notag \\
&&+\mathcal{L}_{int}(R_{\parallel }^{\mu \nu }+\frac{1}{M}\left( \partial
^{\mu }V_{\perp }^{\nu }-\partial ^{\nu }V_{\perp }^{\mu }\right) -\frac{1}{2%
}\varepsilon ^{\mu \nu \alpha \beta }\widehat{\Lambda }_{\alpha \beta
},V_{\perp }^{\alpha }-\partial ^{\alpha }\Lambda ,J,\ldots ).  \notag \\
&&
\end{eqnarray}
In this formulation, the role of the fields $V_{\perp }^{\mu }$ and the
field $\Lambda $ is the same as in the Proca field case. $R_{\parallel
}^{\mu \nu }$ does not correspond to the original degree of freedom and, as
in the previous formulation, it serves together with $\Lambda _{\perp \mu }$
to cancel the spurious $p^{2}=0$ poles.

Let us end up this subsection with the path integral treatment of the toy
quadratic interaction Lagrangian (\ref{L_RV_toy}). Using the same
transformations as before we get
\begin{eqnarray}
\mathcal{L}(R_{\parallel }^{\mu \nu },\Lambda _{\perp }^{\rho },V_{\perp
}^{\alpha }\ldots ,\Lambda ,J,\ldots ) &=&MV_{\perp \nu }\partial _{\mu
}R_{\parallel }^{\mu \nu }+\frac{1}{2}M^{2}V_{\perp \mu }V_{\perp }^{\mu }+%
\frac{1}{4}M^{2}R_{\parallel \mu \nu }R_{\parallel }^{\mu \nu }  \notag \\
&&+\frac{1}{2}M^{2}\Lambda _{\perp }^{\mu }\square \Lambda _{\perp \mu }+%
\frac{1}{2}M^{2}\partial _{\mu }\Lambda \partial ^{\mu }\Lambda  \notag \\
&&+\frac{\alpha _{V}}{2}V_{\perp \mu }\square V_{\perp }^{\mu }-\frac{\beta
_{V}}{2}(\square \Lambda )^{2}+\frac{\alpha _{R}}{4}R_{\parallel \mu \nu
}\square R_{\parallel }^{\mu \nu }+\frac{\beta _{R}}{4}\square \Lambda
_{\perp }^{\mu }\square \Lambda _{\perp \mu }  \notag \\
&&+\mathcal{L}_{int}^{^{\prime }}(R_{\parallel }^{\mu \nu }-\frac{1}{2}%
\varepsilon ^{\mu \nu \alpha \beta }\widehat{\Lambda }_{\perp \alpha \beta
},V_{\perp }^{\alpha }-\partial ^{\alpha }\Lambda ,J,\ldots )
\end{eqnarray}
Introducing the auxiliary fields analogous to the previous two examples, we
have
\begin{eqnarray}
\mathcal{L}(R_{\parallel }^{\mu \nu },\Lambda _{\perp }^{\mu },\chi ,\chi
_{\perp }^{\mu },\sigma _{\perp }^{\mu },\pi _{\perp }^{\mu },J,\ldots ) &=&%
\frac{\alpha _{V}}{2}V_{\perp \mu }\square V_{\perp }^{\mu }+\frac{1}{2}%
M^{2}V_{\perp \mu }V_{\perp }^{\mu }  \notag \\
&&+\frac{\alpha _{R}}{4}R_{\parallel \mu \nu }\square R_{\parallel }^{\mu
\nu }+\frac{1}{4}M^{2}R_{\parallel \mu \nu }R_{\parallel }^{\mu \nu }  \notag
\\
&&+MV_{\perp \nu }\partial _{\mu }R_{\parallel }^{\mu \nu }  \notag \\
&&+\frac{1}{2}M^{2}\Lambda _{\perp }^{\mu }\square \Lambda _{\perp \mu }-%
\frac{1}{2}M^{2}\Lambda \square \Lambda  \notag \\
&&+\frac{1}{2\beta _{V}}\chi ^{2}+\chi \square \Lambda -\frac{1}{2\beta _{R}}%
\chi _{\perp }^{\mu }\chi _{\perp \mu }-\chi _{\perp }^{\mu }\square \Lambda
_{\perp \mu }  \notag \\
&&+\mathcal{L}_{int}^{^{\prime }}(R_{\parallel }^{\mu \nu }-\frac{1}{2}%
\varepsilon ^{\mu \nu \alpha \beta }\widehat{\Lambda }_{\perp \alpha \beta
},V_{\perp }^{\alpha }-\partial ^{\alpha }\Lambda ,J,\ldots )
\end{eqnarray}
The kinetic terms can be diagonalized now by means of the shifts
\begin{eqnarray}
\Lambda _{\perp }^{\mu } &\rightarrow &\Lambda _{\perp }^{\mu }+\frac{1}{%
M^{2}}\chi _{\perp }^{\mu } \\
\Lambda &\rightarrow &\Lambda +\frac{1}{M^{2}}\chi
\end{eqnarray}
to the form
\begin{eqnarray}
\mathcal{L}(R_{\parallel }^{\mu \nu },\Lambda _{\perp }^{\mu },\chi ,\chi
_{\perp }^{\mu },J,\ldots ) &=&\frac{\alpha _{V}}{2}V_{\perp \mu }\square
V_{\perp }^{\mu }+\frac{1}{2}M^{2}V_{\perp \mu }V_{\perp }^{\mu }  \notag \\
&&+\frac{\alpha _{R}}{4}R_{\parallel \mu \nu }\square R_{\parallel }^{\mu
\nu }+\frac{1}{4}M^{2}R_{\parallel \mu \nu }R_{\parallel }^{\mu \nu }  \notag
\\
&&+MV_{\perp \nu }\partial _{\mu }R_{\parallel }^{\mu \nu }  \notag \\
&&+\frac{1}{2}M^{2}\Lambda _{\perp }^{\mu }\square \Lambda _{\perp \mu }-%
\frac{1}{2}M^{2}\Lambda \square \Lambda  \notag \\
&&-\frac{1}{2M^{2}}\chi _{\perp }^{\mu }\square \chi _{\perp \mu }-\frac{1}{%
2\beta _{R}}\chi _{\perp }^{\mu }\chi _{\perp \mu }  \notag \\
&&+\frac{1}{2M^{2}}\chi \square \chi +\frac{1}{2\beta _{V}}\chi ^{2}  \notag
\\
&&+\mathcal{L}_{int}^{^{\prime }}(S,W,J,\dots ),  \notag \\
&&  \label{L_FO_final}
\end{eqnarray}
where
\begin{eqnarray*}
\overline{R}^{\mu \nu } &=&R_{\parallel }^{\mu \nu }-\frac{1}{2}\varepsilon
^{\mu \nu \alpha \beta }\widehat{\Lambda }_{\perp \alpha \beta }-\frac{1}{%
2M^{2}}\varepsilon ^{\mu \nu \alpha \beta }\widehat{\chi }_{\perp \alpha
\beta } \\
\overline{V}^{\mu } &=&V_{\perp }^{\alpha }-\partial ^{\alpha }\Lambda -%
\frac{1}{M^{2}}\partial ^{\alpha }\chi .
\end{eqnarray*}
In the formula (\ref{L_FO_final}) the scalar and axial-vector ghost field as
well as two propagating dynamically mixed spin-1 degrees of freedom are
explicit.

\section{The parameters $\protect\alpha _{i}$ and $\protect\beta _{i}$ in
terms of LECs\label{Appendix_alpha_beta}}

In this appendix we present the expressions for the renormalization scale
independent polynomial parameters entering the self-energies (cf. Section %
\ref{ch4}).

\subsection{The Proca field case\label{appendix Proca}}

\begin{eqnarray*}
\alpha _0 &=&\left( \frac{4\pi F}M\right) ^2Z_M^r(\mu ) \\
\alpha _1 &=&\left( \frac{4\pi F}M\right) ^2Z_V^r(\mu )-\frac{40}3\sigma
_V^2\left( \ln \frac{M^2}{\mu ^2}+\frac 13\right) \\
\alpha _2 &=&\left( \frac{4\pi F}M\right) ^2M^2X_V^r(\mu )+\frac{40}9\sigma
_V^2\left( \ln \frac{M^2}{\mu ^2}+\frac 13\right) \\
\alpha _3 &=&\left( \frac{4\pi F}M\right) ^2M^4U_V^r(\mu )+g_V^2\left( \frac
MF\right) ^2\left( \ln \frac{M^2}{\mu ^2}-\frac 23\right) \\
\beta _0 &=&\left( \frac{4\pi F}M\right) ^2Z_M^r(\mu )=\alpha _0 \\
\beta _1 &=&\left( \frac{4\pi F}M\right) ^2Y_V^r(\mu ) \\
\beta _2 &=&\left( \frac{4\pi F}M\right) ^2M^2X_V^{^{\prime }r}(\mu ) \\
\beta _3 &=&\left( \frac{4\pi F}M\right) ^2M^4V_V^r(\mu ).
\end{eqnarray*}
Here $U_V$ and $V_V$ are certain linear combinations of the couplings of $%
\mathcal{L}_V^{ct(6)}$ renormalized as
\begin{eqnarray*}
U_V &=&U_V^r(\mu )-2g_V^2\left( \frac MF\right) ^4\frac 1{M^4}\lambda _\infty
\\
V_V &=&V_V^r(\mu )
\end{eqnarray*}

\subsection{The antisymmetric tensor case\label{appendix tensor}}

\begin{eqnarray*}
{\alpha }_0 &=&\left( \frac{4\pi F}M\right) ^2Z_M^r({\mu })-\frac{40}%
3d_1^2\ln \frac{M^2}{{\mu }^2}-\frac{20}9(3d_1^2-d_3^2)-5\left( \frac{%
\lambda ^{VVV}}M\right) ^2\left( \frac FM\right) ^2\left( 7-6\ln \frac{M^2}{{%
\mu }^2}\right) \\
{\alpha }_1 &=&\left( \frac{4\pi F}M\right) ^2(Z_R^r({\mu })+Y_R^r({\mu }))-%
\frac{40}9(3d_1^2+2d_3^2)\ln \frac{M^2}{{\mu }^2}-\frac{20}3\left(
d_1^2+\frac 19d_2^2\right) \\
&&+\frac{10}3\left( \frac{\lambda ^{VVV}}M\right) ^2\left( \frac FM\right)
^2\left( 7-6\ln \frac{M^2}{{\mu }^2}\right) \\
{\alpha }_2 &=&\left( \frac{4\pi F}M\right) ^2M^2(X_R^r({\mu })+W_R^r({\mu }%
))-\frac{40}9d_3^2\left( \ln \frac{M^2}{{\mu }^2}+\frac 13\right) +\frac
12\left( \frac{G_V}F\right) ^2\left( \ln \frac{M^2}{{\mu }^2}-\frac 23\right)
\\
&&-\frac 53\left( \frac{\lambda ^{VVV}}M\right) ^2\left( \frac FM\right)
^2\left( 2-3\ln \frac{M^2}{{\mu }^2}\right) \\
{\alpha }_3 &=&\left( \frac{4\pi F}M\right) ^2M^4U_R^r({\mu })+\frac{40}%
9d_3^2\left( \ln \frac{M^2}{{\mu }^2}+\frac 13\right) \\
{\beta }_0 &=&\left( \frac{4\pi F}M\right) ^2Z_M^r({\mu })-\frac{40}%
3d_1^2\ln \frac{M^2}{{\mu }^2}-\frac{20}9(3d_1^2+d_3^2)-\frac 53\left( \frac{%
\lambda ^{VVV}}M\right) ^2\left( \frac FM\right) ^2\left( 11-6\ln \frac{M^2}{%
{\mu }^2}\right) \\
{\beta }_1 &=&\left( \frac{4\pi F}M\right) ^2Y_R^r({\mu })-\frac{20}%
9(6d_1^2-12d_1(d_3+d_4)+5d_3^2+9d_4^2-6d_3d_4)\ln \frac{M^2}{{\mu }^2} \\
&&-\frac{20}{27}(9d_1^2-18d_1(d_3+d_4)-7d_3^2-12d_4^2+18d_3d_4) \\
&&\frac{20}3\left( \frac{\lambda ^{VVV}}M\right) ^2\left( \frac FM\right)
^2\left( 7+3\ln \frac{M^2}{{\mu }^2}\right) \\
{\beta }_2 &=&\left( \frac{4\pi F}M\right) ^2M^2W_R^r({\mu })-\frac{20}%
9(d_3^2+6d_3d_4-5d_4^2)\ln \frac{M^2}{{\mu }^2}-\frac{80}{27}(d_3^2+4d_4^2)
\\
&&-\frac 53\left( \frac{\lambda ^{VVV}}M\right) ^2\left( \frac FM\right)
^2\left( 4-3\ln \frac{M^2}{{\mu }^2}\right) \\
{\beta }_3 &=&\left( \frac{4\pi F}M\right) ^2M^4V_R^r({\mu })-\frac{40}%
9d_4^2\left( \ln \frac{M^2}{{\mu }^2}-\frac 23\right) .
\end{eqnarray*}
Here $U_R$ and $V_R$ are certain linear combinations of the couplings of $%
\mathcal{L}_R^{ct(6)}$ with the infinite parts fixed as
\begin{eqnarray*}
U_R &=&U_R^r({\mu })-\frac{80}9\left( \frac MF\right) ^2\frac
1{M^4}d_3^2\lambda _\infty \\
V_R &=&V_R^r({\mu })+\frac{80}9\left( \frac MF\right) ^2\frac
1{M^4}d_4^2\lambda _\infty .
\end{eqnarray*}

\subsection{The first order formalism\label{appendix first order}}

\begin{eqnarray*}
{\alpha }_{0}^{RV} &=&\left( \frac{4\pi F}{M}\right) ^{2}Z_{RV}^{r}(\mu )+%
\frac{10}{9}(\sigma _{RV}+2\sigma _{V})\left[ (d_{1}-d_{3})+3(2d_{1}-\sigma
_{RV})\left( \ln \frac{M^{2}}{\mu ^{2}}+\frac{1}{3}\right) \right] \\
{\alpha }_{1}^{RV} &=&\left( \frac{4\pi F}{M}\right) ^{2}M^{2}X_{RV}^{r}(\mu
)+\frac{10}{9}(\sigma _{RV}+2\sigma _{V})(4d_{3}+\sigma _{RV})\left( \ln
\frac{M^{2}}{\mu ^{2}}+\frac{1}{3}\right) \\
{\alpha }_{2}^{RV} &=&\left( \frac{4\pi F}{M}\right) ^{2}M^{4}Y_{RV}^{r}(\mu
)-\frac{20}{9}(\sigma _{RV}+2\sigma _{V})d_{3}\left( \ln \frac{M^{2}}{\mu
^{2}}+\frac{1}{3}\right) \\
&&-\frac{1}{2}\frac{g_{V}G_{V}}{M}\left( \frac{M}{F}\right) ^{2}\left( \ln
\frac{M^{2}}{\mu ^{2}}-\frac{2}{3}\right) \\
{\alpha }_{0}^{VV} &=&\left( \frac{4\pi F}{M}\right) ^{2}Z_{MV}^{r}(\mu ) \\
{\alpha }_{1}^{VV} &=&\left( \frac{4\pi F}{M}\right) ^{22}Z_{V}^{r}(\mu )-%
\frac{10}{3}\left( \sigma _{RV}(\sigma _{RV}+2\sigma _{V})+4\sigma
_{V}^{2}\right) \left( \ln \frac{M^{2}}{\mu ^{2}}+\frac{1}{3}\right) \\
{\alpha }_{2}^{VV} &=&\left( \frac{4\pi F}{M}\right) ^{2}M^{2}X_{V}^{r}(\mu
)+\frac{10}{9}\left( \sigma _{RV}(\sigma _{RV}+2\sigma _{V})+4\sigma
_{V}^{2}\right) \left( \ln \frac{M^{2}}{\mu ^{2}}+\frac{1}{3}\right) \\
{\alpha }_{3}^{VV} &=&\left( \frac{4\pi F}{M}\right) ^{2}M^{4}U_{V}^{r}(\mu
)+g_{V}^{2}\left( \frac{M}{F}\right) ^{2}\left( \ln \frac{M^{2}}{\mu ^{2}}-%
\frac{2}{3}\right) \\
\beta _{0}^{VV} &=&\left( \frac{4\pi F}{M}\right) ^{2}Z_{MV}^{r}(\mu
)=\alpha _{0}^{VV} \\
\beta _{1}^{VV} &=&\left( \frac{4\pi F}{M}\right) ^{2}Y_{V}^{r}(\mu ) \\
\beta _{2}^{VV} &=&\left( \frac{4\pi F}{M}\right) ^{2}M^{2}X_{V}^{^{\prime
}r}(\mu ) \\
\beta _{3}^{VV} &=&\left( \frac{4\pi F}{M}\right) ^{2}M^{4}V_{V}^{r}(\mu ) \\
{\alpha }_{0}^{RR} &=&\left( \frac{4\pi F}{M}\right) ^{2}Z_{MR}^{r}({\mu })+%
\frac{10}{3}\left( \sigma _{RV}(2d_{1}-\sigma _{RV})-4d_{1}^{2}\right)
\left( \ln \frac{M^{2}}{{\mu }^{2}}+\frac{1}{3}\right) \\
&&-\frac{10}{9}\left( d_{1}-d_{3}\right) \left( 2d_{1}+2d_{3}-\sigma
_{RV}\right) \\
{\alpha }_{1}^{RR} &=&\left( \frac{4\pi F}{M}\right) ^{2}(Z_{R}^{r}({\mu }%
)+Y_{R}^{r}({\mu }))-\frac{40}{9}(3d_{1}^{2}+2d_{3}^{2})\ln \frac{M^{2}}{{%
\mu }^{2}}-\frac{20}{3}\left( d_{1}^{2}+\frac{1}{9}d_{2}^{2}\right) \\
&&+\frac{10}{9}\left( \ln \frac{M^{2}}{{\mu }^{2}}+\frac{1}{3}\right) \sigma
_{RV}(4d_{3}+\sigma _{RV}) \\
{\alpha }_{2}^{RR} &=&\left( \frac{4\pi F}{M}\right) ^{2}M^{2}(X_{R}^{r}({%
\mu })+W_{R}^{r}({\mu }))-\frac{20}{9}d_{3}(2d_{3}+\sigma _{RV})\left( \ln
\frac{M^{2}}{{\mu }^{2}}+\frac{1}{3}\right) \\
&&+\frac{1}{2}\left( \frac{G_{V}}{F}\right) ^{2}\left( \ln \frac{M^{2}}{{\mu
}^{2}}-\frac{2}{3}\right) \\
{\alpha }_{3}^{RR} &=&\left( \frac{4\pi F}{M}\right) ^{2}M^{4}U_{R}^{r}({\mu
})+\frac{40}{9}d_{3}^{2}\left( \ln \frac{M^{2}}{{\mu }^{2}}+\frac{1}{3}%
\right) \\
{\beta }_{0}^{RR} &=&\left( \frac{4\pi F}{M}\right) ^{2}Z_{MR}^{r}({\mu })+%
\frac{10}{3}(\sigma _{RV}(2d_{1}-\sigma _{RV})-4d_{1}^{2})\left( \ln \frac{%
M^{2}}{{\mu }^{2}}+\frac{1}{3}\right) \\
&&-\frac{20}{9}(d_{1}^{2}+d_{3}^{2})+\frac{10}{9}\sigma _{RV}\left(
d_{1}+d_{3}-\sigma _{RV}\right) \\
&=&{\alpha }_{0}^{RR}-\frac{40}{9}d_{3}^{2}+\frac{10}{9}\sigma _{RV}\left(
2d_{3}-\sigma _{RV}\right) \\
{\beta }_{1}^{RR} &=&\left( \frac{4\pi F}{M}\right) ^{2}Y_{R}^{r}({\mu })-%
\frac{20}{9}%
(6d_{1}^{2}-12d_{1}(d_{3}+d_{4})+5d_{3}^{2}+9d_{4}^{2}-6d_{3}d_{4})\ln \frac{%
M^{2}}{{\mu }^{2}} \\
&&-\frac{20}{27}%
(9d_{1}^{2}-18d_{1}(d_{3}+d_{4})-7d_{3}^{2}-12d_{4}^{2}+18d_{3}d_{4}) \\
&&-\frac{5}{27}\sigma _{RV}\left( 32d_{3}+6(d_{3}+9d_{4})\ln \frac{M^{2}}{{%
\mu }^{2}}-3\sigma _{RV}\left( \ln \frac{M^{2}}{{\mu }^{2}}-\frac{2}{3}%
\right) \right) \\
{\beta }_{2}^{RR} &=&\left( \frac{4\pi F}{M}\right) ^{2}M^{2}W_{R}^{r}({\mu }%
)-\frac{20}{9}(d_{3}^{2}+6d_{3}d_{4}-5d_{4}^{2})\ln \frac{M^{2}}{{\mu }^{2}}-%
\frac{80}{27}(d_{3}^{2}+4d_{4}^{2}) \\
&&+\frac{5}{27}\sigma _{RV}\left( 8d_{3}+6(d_{3}+3d_{4})\ln \frac{M^{2}}{{%
\mu }^{2}}-3\sigma _{RV}\left( \ln \frac{M^{2}}{{\mu }^{2}}-\frac{2}{3}%
\right) \right) \\
{\beta }_{3}^{RR} &=&\left( \frac{4\pi F}{M}\right) ^{2}M^{4}V_{R}^{r}({\mu }%
)-\frac{40}{9}d_{4}^{2}\left( \ln \frac{M^{2}}{{\mu }^{2}}-\frac{2}{3}%
\right) .
\end{eqnarray*}
Here $U_{V}$, $V_{V}$, $U_{R}$, $V_{R}$ and $Y_{RV}$ are certain linear
combination of the couplings from $\mathcal{L}_{RV}^{ct(6)}$ with infinite
parts fixed according to
\begin{eqnarray*}
U_{V} &=&U_{V}^{r}(\mu )-2g_{V}^{2}\left( \frac{M}{F}\right) ^{4}\frac{1}{%
M^{4}}\lambda _{\infty } \\
V_{V} &=&V_{V}^{r}(\mu ) \\
U_{R} &=&U_{R}^{r}({\mu })-\frac{80}{9}\left( \frac{M}{F}\right) ^{2}\frac{1%
}{M^{4}}d_{3}^{2}\lambda _{\infty } \\
V_{R} &=&V_{R}^{r}({\mu })+\frac{80}{9}\left( \frac{M}{F}\right) ^{2}\frac{1%
}{M^{4}}d_{4}^{2}\lambda _{\infty } \\
Y_{RV} &=&Y_{RV}^{r}(\mu )+\frac{40}{9}\left( \frac{M}{F}\right) ^{2}\frac{1%
}{M^{4}}(\sigma _{RV}+2\sigma _{V})d_{3}\lambda _{\infty }+\frac{g_{V}G_{V}}{%
M}\left( \frac{M}{F}\right) ^{2}\frac{1}{M^{4}}\lambda _{\infty }
\end{eqnarray*}

\section{Proof of the positivity of the spectral functions\label%
{Appendix_positivity}}

\bigskip

Here we prove the positivity of the spectral functios $\rho _{L,T}(\mu ^{2})$
defined as
\begin{equation}
(2\pi )^{-3}\theta (p^{0})\left[ \rho _{T}(p^{2})p^{2}\Pi _{\mu \nu \alpha
\beta }^{T}(p)-\rho _{L}(p^{2})p^{2}\Pi _{\mu \nu \alpha \beta }^{L}(p)%
\right] =\sum\limits_{N}\delta ^{(4)}(p-p_{N})\langle 0|R_{\mu \nu
}(0)|N\rangle \langle N|R_{\alpha \beta }(0)|0\rangle .
\end{equation}
Let us define for $p^{2}>0$%
\begin{eqnarray*}
u_{\mu \nu }^{(\lambda )}(p) &=&\frac{\mathrm{i}}{\sqrt{p^{2}}}\left( p_{\mu
}\varepsilon _{\nu }^{(\lambda )}(p)-p_{\nu }\varepsilon _{\mu }^{(\lambda
)}(p)\right) \\
w_{\mu \nu }^{(\lambda )}(p) &=&\frac{1}{2}\varepsilon _{\mu \nu }^{%
\phantom{\mu\nu}\alpha \beta }u_{\alpha \beta }^{(\lambda )}(p)
\end{eqnarray*}
where $\varepsilon _{\mu }^{(\lambda )}(p)$ are the usual spin-one
polarization vectors corresponding to the mass $\sqrt{p^{2}}$. Then for $%
p^{2}>0$ we get the following orthogonality relations
\begin{eqnarray*}
u_{\mu \nu }^{(\lambda )}(p)u^{(\lambda ^{^{\prime }})\mu \nu }(p)^{\ast }
&=&-2\delta ^{\lambda \lambda ^{^{\prime }}} \\
w_{\mu \nu }^{(\lambda )}(p)w^{(\lambda ^{^{\prime }})\mu \nu }(p)^{\ast }
&=&2\delta ^{\lambda \lambda ^{^{\prime }}} \\
u_{\mu \nu }^{(\lambda )}(p)w^{(\lambda ^{^{\prime }})\mu \nu }(p)^{\ast }
&=&0
\end{eqnarray*}
and the projectors can be written for $p^2>0$ in terms of the polarization
sums as
\begin{eqnarray*}
\Pi _{\mu \nu \alpha \beta }^{L}(p) &=&-\frac{1}{2}\sum_{\lambda }u_{\mu \nu
}^{(\lambda )}(p)u_{\alpha \beta }^{(\lambda )}(p)^{\ast } \\
\Pi _{\mu \nu \alpha \beta }^{T}(p) &=&\frac{1}{2}\sum_{\lambda }w_{\mu \nu
}^{(\lambda )}(p)w_{\alpha \beta }^{(\lambda )}(p)^{\ast }.
\end{eqnarray*}
Multiplying (\ref{spectral}) by $u_{\mu \nu }^{(\lambda )}(p)^{\ast
}u_{\alpha \beta }^{(\lambda )}(p)$ and $w_{\mu \nu }^{(\lambda )}(p)^{\ast
}w_{\alpha \beta }^{(\lambda )}(p)$ respectively we get the positivity
constraints for the spectral functions
\begin{eqnarray*}
0 &\leq &\sum\limits_{N}\delta ^{(4)}(p-p_{N})|\langle 0|R_{\mu \nu
}(0)|N\rangle u^{(\lambda ^{^{\prime }})\mu \nu }(p)^{\ast }|^{2}=2(2\pi
)^{-3}\theta (p^{0})\rho _{L}(p^{2})p^{2} \\
0 &\leq &\sum\limits_{N}\delta ^{(4)}(p-p_{N})|\langle 0|R_{\mu \nu
}(0)|N\rangle w^{(\lambda ^{^{\prime }})\mu \nu }(p)^{\ast }|^{2}=2(2\pi
)^{-3}\theta (p^{0})\rho _{T}(p^{2})p^{2}.
\end{eqnarray*}

\end{document}